\documentclass[
reprint,
superscriptaddress,
longbibliography,
bibnotes,
amsmath,amssymb,
aps,
prb,
]{revtex4-2}
\usepackage[utf8]{inputenc}
\usepackage[english]{babel}
\usepackage[plainpages = false, pdfpagelabels, 
                 bookmarks,
                 bookmarksopen = true,
                 bookmarksnumbered = true,
                 breaklinks = true,
                 linktocpage,
                 colorlinks = true,
                 linkcolor = blue,
                 urlcolor  = blue,
                 citecolor = blue,
                 anchorcolor = green,
                 hyperindex = true,
                 hyperfigures
                 ]{hyperref} 
\usepackage{enumerate}
\usepackage{tabularx}
\usepackage{makecell, multirow}
\usepackage{tikz}
\usetikzlibrary{patterns}
\usepackage{slashed}
\usepackage{physics}
\usepackage{verbatim}
\usepackage{svg}
\usepackage{mathtools}
\usepackage{cancel}
\usepackage{bbold}
\usepackage{bm}
\usepackage{soul}
\usepackage{booktabs}
\usepackage[caption=false,singlelinecheck=off]{subfig}
\usepackage{blindtext}
\usepackage{bbm}
\usepackage{graphicx}
\usepackage{dcolumn}
\usepackage{bm}
\usepackage[normalem]{ulem}
\usepackage{xcolor}
\renewcommand{\vec}[1]{\bm{#1}}
\usepackage[percent]{overpic}
\usepackage{subfig}
\begin{document}
\title{\texorpdfstring{Non-equilibrium bosonization of fractional quantum Hall edges}{}}

\author{Christian Sp{\r a}nsl{\"a}tt}
\affiliation{\mbox{Department of Engineering and Physics, Karlstad University, Karlstad, Sweden}}

\author{Jinhong Park}
\email{jinhongpark@konkuk.ac.kr}
\affiliation{\mbox{Department of Physics, Konkuk University, Seoul 05029, Republic of Korea}}
\affiliation{\mbox{Institute for Quantum Materials and Technologies, Karlsruhe Institute of Technology, 76131 Karlsruhe, Germany}}

\author{Alexander D. Mirlin}
\affiliation{\mbox{Institute for Quantum Materials and Technologies, Karlsruhe Institute of Technology, 76131 Karlsruhe, Germany}}
\affiliation{\mbox{Institut f{\"u}r Theorie der Kondensierten Materie, Karlsruhe Institute of Technology, 76131 Karlsruhe, Germany}}

\date{\today}
\begin{abstract}
Edge transport serves as a powerful probe of remarkable low-energy properties of fractional quantum Hall states, including the anyonic character of their excitations. Here, we develop a theory of fractional quantum Hall edges driven out of equilibrium, which is based on the Keldysh action for the bosonized chiral Luttinger liquid. With this non-equilibrium FQH bosonization framework, we first consider a single-mode Laughlin edge and analyze the full counting statistics of charge, the quasiparticle Green's functions, and tunneling transport properties through a quantum point contact, allowing for generic edge excitations. We then extend the formalism to multi-mode edges with inter-mode interactions, and explore, with focus on the $\nu=4/3$ and $\nu=2/3$ edges as paradigmatic examples, how interaction-induced fractionalization of anyons modifies the edge dynamics and the associated transport observables. While the full counting statistics probes the fractionalized charge of the excitations, the Green's functions and tunneling transport are governed by mutual braiding phases of fractionalized excitations and tunneling quasiparticles. We emphasize in particular the effect of interaction-induced fractionalization on the Fano factor $F$ and the differential Fano factor $F_d$, observables that can be measured experimentally. Our formalism, which provides a unified framework for non-equilibrium transport in FQH edges and Luttinger liquids, permits extracting anyonic braiding information from non-equilibrium edge-transport experiments, and paves the way to various extensions, including more involved experimental geometries and edge structures. 
\end{abstract}
\maketitle
\section{Introduction}\label{sec:Introduction}
The fractional quantum Hall (FQH) effect~\cite{Tsui_fqh_1982,Laughlin_fqh_1983, Haldane_fqh_1983,Halperin1984statistics,Halperin_FQHE_2020}  is a prototypical platform for exploring the physics of a remarkable variety of strongly correlated, topological states of matter. The topological order~\cite{Wen_topological_1990,Wen_Quantum_2004} of FQH states manifests itself in the unconventional properties of their excitations, including  fractional charge and fractional (anyonic) quantum statistics. The bulk topology is further reflected in the properties of the FQH edge~\cite{Wen_CLL_edge_1990,Wen_Electrodynamical_edge_1990,Wen_Topological_1995}, making it possible to explore the topological order and, in particular, the nature of the excitations, by studying FQH edge transport in appropriately tailored geometries; see Refs.~\cite{Heiblum2020edge, feldman2021fractional} for recent reviews. A paradigmatic example is the detection of the fractional charge of excitations from shot noise measurements~\cite{Kane1994Jan,Picciotto_fractional_charge_1998, Saminadayar_fractional_charge_1997}.

The field theory resulting from bosonization of an Abelian FQH edge~\cite{Wen_Topological_1995} is a chiral Luttinger liquid theory (in general, with multiple, interacting edge modes). In equilibrium, and in the absence of inter-mode particle tunneling, this is a Gaussian bosonic field theory, which has served as a basis for major theoretical progress in the field. Realistic FQH systems with multiple modes are further characterized by disorder-induced inter-mode tunneling, which crucially affects the properties of FQH edges with counter-propagating modes \cite{Johnson_Composite_edge_1991, Kane_Randomness_1994,Kane_Impurity_1995}. In recent years, a growing body of works have theoretically explored and experimentally demonstrated various manifestations of topological order in the electric and thermal transport along Abelian and non-Abelian FQH edges~\cite{Rosenow_signatures_2010,Protopopov_transport_2_3_2017,Nosiglia2018,Park_Noise_2019,Spanslatt_Noise_2019,Spanslatt_condplateau,Spanslatt2021,Park_noise_2020,Park2024Jun,Spanslatt_Noise_2022,Spanslatt_binding_2023,Yutushui_Identifying_2022,Yutushui2023Dec,Yutushui2024,Yutushui2025May,Banerjee_Observed_2017, Banerjee_observation_2018,Melcer_Absent_2022, Kumar2022Jan,Kumar2024Jun,Dutta_Isolated_2022,Srivastav2021May,
Srivastav2022Sep,Breton2022,Grivnin2014, Wang2021, Hashisaka2021, Dutta_novel_2022, Hashisaka2023,Ronen2018, Cohen2019,Cohen2023,Ronen2021,Deprez2021}.
However, unambiguous experimental determination of the anyonic statistics of the FQH quasiparticles has turned out to be particularly challenging. This endeavor has attracted growing attention, with important theoretical and experimental progress in recent years. One approach is to use Fabry-Pérot or Mach-Zehnder interferometers based on FQH edges~\cite{Chamon1997, Law2006, Halperin2011,  Nakamura2020direct, nakamura2022impact, Nakamura2023, kundu2023anyonic, Batra2023}. Another class of setups, on which we focus here, involves measuring the current correlations generated by quasiparticle tunneling between FQH edges driven out of equilibrium via voltage-biased quantum point contacts~\cite{ Rosenow2016Apr,
Levkivskyi2016Apr,
Han2016topological,Campagnano2016,Lee2019anyonbraiding, Bartolomei2020Apr,Lee2020,
Morel2022fractionalization,Lee2022non-abelian,Taktak2022two-particle,
Schiller2023anyon,Ruelle2023Mar,Glidic2023Mar,Lee2023May, Zhang2024, Thamm2024effect,Iyer2024finite,ruelle2025time, Zhang2025Mar,Zhang2025fractional,De2025electronic,zhang2025effectivelinearresponsenonequilibrium, Ronetti2025, safi2025timedomainbraidinganyons, latyshev2025hongoumandelinterferometryfractionalexcitations}.
These conditions emphasize the need to develop a generic theory for FQH edges driven out of equilibrium, which is the goal of the present work. 

A non-equilibrium bosonization framework for ``conventional'' (non-chiral), one-dimensional (1D), and interacting fermionic systems was developed in Ref.~\cite{Gutman2010}; see also related works \cite{Gutman2008nonequilibrium,Kovrizhin2009exactly,Levkivskyi2009noise-induced}. This framework, based on the Keldysh functional-integral formalism to describe non-thermal distribution functions of fermions, has been used to analyze various physical properties of this class of systems, including tunneling spectral density, full counting statistics of charge, long-range density correlations, non-equilibrium interferometry, and energy relaxation~\cite{Gutman2010full,Gutman2011non-equilibrium,Protopopov2011many-particle,Kovrizhin2012relaxation,Levkivskyi2012energy,Protopopov2013correlations,Milletari2013shot-noise,NgoDinh2013analytically,
Schneider2017transient}.

The key point of the non-equilibrium bosonization formalism of Ref.~\cite{Gutman2010} is the formulation of the Keldysh action characterizing a non-equilibrium state of an interacting 1D fermionic system (Luttinger liquid) in terms of single-particle Fredholm determinants on Toeplitz form. This feature permits  expressing various observables in terms of singular Toeplitz determinants, which are analogous to those appearing in the theory of free-fermion full counting statistics of charge~\cite{Levitov1993Aug,Levitov1996Oct,Ivanov1997Sep,Klich2003,Abanov2011quantum}. Importantly, the asymptotic (infrared) properties of this class of determinants can be obtained by using the Szeg\H{o} theorems and the generalized Fisher-Hartwig conjecture, see Refs.~\cite{Gutman2011non-equilibrium,Abanov2011quantum,deift2011asymptotics} and references therein. It is also possible to efficiently evaluate these determinants numerically. 

In this work, we expand the scope of the non-equilibrium bosonization formalism to encompass Abelian FQH edges (described by the chiral Luttinger liquid model) driven out of equilibrium \footnote{Some ideas of extending the non-equilibrium bosonization to FQH systems, for a certain class of states and observables, were put forward in Ref.~\cite{Levkivskyi2016Apr}.}.
The formalism that we develop reproduces, and makes more precise, many results obtained by different approaches in the last few years. Furthermore, it allows exploration, in a consistent and systematic way, of a variety of setups, including complex (multi-mode) FQH edges with co-propagating or counter-propagating modes coupled by interactions and driven out of equilibrium by tunneling at voltage-biased quantum point contacts or by temperature differences.
In particular, our formalism allows us to study Green's functions of any anyonic excitations compatible with the considered FQH edge, 
which can be probed in tunneling experiments. We also show that our formalism determines the full counting statistics of charge (i.e., all statistical moments of charge fluctuations) for non-equilibrium FQH edges, which is very instructive for understanding the rich physics of this peculiar class of systems. 

The remainder of this paper is organized as follows. In Sec.~\ref{sec:Keldysh}, we develop and benchmark the non-equilibrium FQH bosonization framework, based on the Keldysh action for the chiral Luttinger liquid describing the Laughlin edge. In Sec.~\ref{sec:GF_Laughlin}, we analyze, by using this formalism as well as the Szeg\H{o} approximation and the generalized Fisher-Hartwig formula for Toeplitz determinants, the
Green's functions of generic quasiparticles on a non-equilibrium Laughlin edge. We use these results for the Green's functions in Sec.~\ref{sec:tunneling} to evaluate the current, noise, and Fano factors for tunneling through a quantum point contact between two Laughlin edges driven our of equilibrium by injection of various quasiparticles. In Sec.~\ref{sec:multiple-mode-edges}, we extend the non-equilibrium bosonization formalism to complex FQH edges with multiple, interacting, co-propagating or counter-propagating modes, focusing on $\nu=4/3$ and $\nu=2/3$ edges as paradigmatic examples  and calculate the full counting statistics, quasiparticle Green's functions, as well as the current noise, and Fano factors for the tunneling transport. Finally, Sec.~\ref{sec:Summary_Conclusions} contains a summary of our results and a discussion of their implications and further prospects.  

Throughout the paper, we use units where $k_{\rm B}=\hbar=1$.

\section{Keldysh action for the Laughlin edge}
\label{sec:Keldysh}
\subsection{Formalism}
\label{sec:formalism}
Our starting point is the FQH edge of a Laughlin state with filling factor $\nu=1/m$, where $m$ is an odd integer. The case $m=1$ (i.e., $\nu=1$) corresponds to a single chiral mode of free fermions (or, equivalently, to an integer quantum Hall edge), which will serve as the guiding example for our formalism. 

Within the bosonization formalism, the Keldysh action of the edge has the form of the chiral Luttinger liquid $S=S_0+S_V$, where
\begin{align}
&S_0 = \frac{1}{4\pi \nu}\int_c dt \int dx \,\partial_x \phi_\mu(x,t)\left[-\eta\partial_t-v  \partial_x\right] \phi_\mu(x,t), \label{eq:S0}\\
&S_V = \int_c dt \int dx \,V_\mu(x,t) \frac{\eta\partial_x \phi_\mu(x,t)}{2\pi}. \label{eq:SV}
\end{align}
Here $\phi$ is a compact bosonic field, $\eta=\pm 1$ is the mode chirality, $v>0$ is the propagation speed, and we have introduced a source term $V(x,t)$.
The time integration runs along the Keldysh contour
\begin{align}
    \int_c dt(\hdots) = \sum_{\mu=\pm 1}\mu \int_{-\infty}^{\infty} dt(\hdots), 
\end{align}
where the positive time branch, $\mu=+1$, runs from $t:-\infty\to \infty$ and the negative branch, $\mu=-1$, runs from  $t:+\infty\to -\infty)$ (see, e.g., Ref.~\cite{Kamenev2011Sep} for details on the Keldysh approach to non-equilibrium field theories).

From the action, one can read off the equal-time commutation relations
\begin{subequations}
\label{eq:commutators}
\begin{align}
    &\left[\phi(x),\phi(y) \right] = \eta i\pi \nu \text{sgn}(x-y),\\
     &\left[\rho(x),\rho(y) \right] = -\eta \nu\frac{i}{2\pi} \partial_x \delta(x-y),
\end{align}
\end{subequations}
where the edge density $\rho$ is related to $\phi$ as
\begin{align}
\label{eq:density_def}
    \rho(x,t)\equiv \frac{\eta}{2\pi} \partial_x\phi(x,t).
\end{align}
The Hamiltonian corresponding to the total action $S$, expressed in terms of the density~\eqref{eq:density_def}, reads
\begin{align}
\label{eq:Ham_Laughlin}
    H =\int\, dx \left[ \frac{\pi v}{\nu}\rho^2 - V\rho\right].
\end{align}
\begin{figure}[t!]
    \centering
\includegraphics[width=\columnwidth]{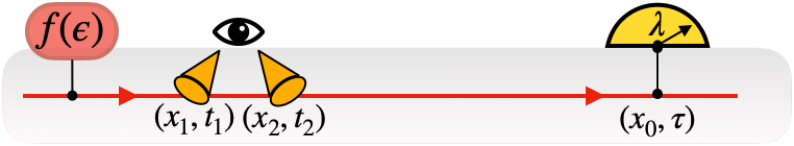}  
    \caption{A single-mode (Laughlin) FQH edge characterized by a non-equilibrium distribution function $f(\epsilon)$, see Sec.~\ref{sec:formalism} for description of the formalism. In this non-equilibrium state, we study the full counting statistics of charge (depicted by the yellow measurement pointer), Sec.~\ref{sec:noneq-Laughlin-FCS}, and Green's functions (depicted by the eye symbol),
Sec.~\ref{sec:Laughlin-equilibrium}.}
    \label{fig:Lauglin_setup}
\end{figure}
Furthermore, edge excitations are described by the vertex operators
\begin{align} 
\label{eq:vertex_operators}
    \psi_n^\dagger(x,t) = \frac{1}{\sqrt{2\pi a}} e^{-in\phi(x,t)}\,,
\end{align}
where $n$ is an integer and $a$ is a short-distance cut-off. 
In this work, we do not consider interferometer setups, so that 
the Klein factors~\cite{Guyon2002Klein} always appear in such a way that their product evaluates to unity. We thus ignore them in Eq.~\eqref{eq:vertex_operators} and in the rest of the paper. Combining the vertex operator~\eqref{eq:vertex_operators} with the density~\eqref{eq:density_def} and the commutators~\eqref{eq:commutators}, one finds that $\psi_1^\dagger(x,t)$ (i.e., taking $n=1$) creates excitations with fractional charge $e^*=e/m$ and exchange-statistics phase $\theta=\pi/m$. Likewise, $\psi_m^\dagger(x,t)$ ($n=\nu^{-1}=m$) creates electronic excitations, i.e., with charge $e$ and the electronic exchange phase $\theta=\pi$. More generally, we define the exchange phase of excitations~\eqref{eq:vertex_operators} as
\begin{align}
\label{eq:Laughlin_exchange_angle}
    \theta = \pi n^2 \nu = \frac{\pi n^2}{m}.
\end{align}
The corresponding \textit{braiding} phase, generated upon double exchange, is then $2\theta$. 
For two different types of quasiparticles, parametrized by integers $n_1$ and $n_2$, respectively, the exchange phase is not well defined~\cite{Wen_Topological_1995}, but their mutual braiding phase is and reads
\begin{align}
\label{eq:theta12_def}
    2\theta_{12} \equiv 2\pi \nu n_1 n_2=\frac{2\pi n_1 n_2}{m}.
\end{align}

Returning to the action $S$, it is convenient to make a rotation in the Keldysh branch space ($\mu=\pm$) and define classical and quantum components (where the latter is defined with a ``bar''):
\begin{subequations}
\label{eq:Keldysh_rotation}
\begin{align}
    \phi,\overline{\phi} &= (\phi_+\pm\phi_-)/\sqrt{2} ,\\
    \rho,\overline{\rho} &= (\rho_+\pm\rho_-)/\sqrt{2},\\
    V,\overline{V} &= (V_+\pm V_-)/\sqrt{2}.
\end{align}
\end{subequations}
With this notation, we can express the key quantity for the non-equilibrium description of the FQH edge, namely the partition function as a functional of the source fields, in the following way:
\begin{align}
\label{eq:Z_V}
    \mathcal{Z}[V,\overline{V}] =  \left\langle \exp \left\{i \left(V \overline{\rho} + \overline{V}\rho\right)\right\}\right\rangle_{S_0}.
\end{align}
Here and below, we use a short-hand notation  where e.g.,  $V\overline{\rho}\equiv \int dt dx V(x,t)\overline{\rho}(x,t)$. Implicit in Eq.~\eqref{eq:Z_V} is the dependence of the partition function $\mathcal{Z}[V,\overline{V}]$ on the state of the edge, expressed by the many-body density matrix $\rho_0$. In terms of the Hamiltonian $H$ in Eq.~\eqref{eq:Ham_Laughlin}, the generating function $\mathcal{Z}[V,\overline{V}]$ is given by
\begin{equation}
\mathcal{Z}[V,\overline{V}] = \text{Tr}\left[\rho_0 U_c \right], 
\end{equation}
 where $U_c$ is the time evolution operator along the Keldysh contour,
\begin{align}
    U_c = T_c \left\{e^{-i \int_c dt H}\right\},
\end{align}
and $T_c$ is the time ordering operator along this contour.

Generalizing Ref.~\cite{Gutman2010}, we analyze now the structure of the generating function $\mathcal{Z}[V,\overline{V}]$. A crucial point is that the bosonic theory is non-interacting (the Hamiltonian~\eqref{eq:Ham_Laughlin} is quadratic with respect to the density fields). It then follows that the generating function has the structure
\begin{align} \label{eq:Partition_function_general}
    \mathcal{Z}[V, \overline{V}] = \exp [-i \nu V \Pi_a \overline{V}] {\cal F}[\overline{V}].
    \end{align}
This structure reflects the properties of the correlation functions of the density fields in this non-interacting bosonic theory. The first factor in Eq.~\eqref{eq:Partition_function_general} encodes the two-point retarded and advanced density correlation functions, which are independent of the state of the system. This factor can be written solely in terms of 
$\Pi_a$, the advanced component of the polarization operator, given in energy-momentum space by 
\begin{align}
\label{eq:advanced_pol}
    \Pi_a (q,\omega) = \frac{1}{2\pi} \frac{\eta q}{\eta vq - \omega+i 0^+}.
\end{align}
The other non-zero (in general, multi-point) correlation functions involve only the classical density components $\rho(x,t)$. Correspondingly, 
the second factor in Eq.~\eqref{eq:Partition_function_general} depends on the quantum component $\overline{V}$ only. It follows that 
 ${\cal F}[\overline{V}]$ can be presented as
\begin{equation}
\label{eq:noise_functional}
{\cal F}[\overline{V}] = \exp \left(
 \sum_{n=2}^\infty\frac{i^n}{n!}\overline{V}^n {\cal S}_{n} \right),
 \end{equation}
where 
\begin{eqnarray}
\overline{V}^n{\cal S}_{n} &\equiv&
\int [dt][dx]
\overline{V}_{}(x_1,t_1)\dots \overline{V}_{}(x_n,t_n)\nonumber \\
& \times & {\cal S}_{n}(x_1,t_1;\dots;x_n,t_n)\,,
\end{eqnarray}
with $\int[dt][dx]$ denoting integration over all spatial and time coordinates.
Moreover, ${\cal S}_{n}(x_1,t_1;\ldots;x_n,t_n)$ are the irreducible noise cumulants,
\begin{eqnarray}
\label{mass-shell-xt}
{\cal S}_{n}(x_1,t_1,;\ldots;x_n,t_n) & = &
\langle\langle{\rho}(x_1,t_1)\ldots {\rho}(x_n,t_n)\rangle\rangle
\nonumber \\
& = & \langle\langle{\rho}(z_1,0)\ldots
{\rho}(z_n,0)\rangle\rangle, \nonumber \\
&&
\end{eqnarray}
where $z_i=x_i-\eta v t_i$. The generating function $\mathcal{Z}[V,\overline{V}]$ is thus fully governed by the set of noise cumulants, i.e., by the full counting statistics (FCS) of charge in the system. The noise cumulants in turn depend on the (in general, non-equilibrium) state of the system. 

It is now straightforward to transform the generating function to the $\rho$ representation, 
\begin{align}
\label{bosonic_action}
 e^{i S[\rho, \overline{\rho}]} =
\int {\cal D}V {\cal D}\overline{V}  {\mathcal{Z}}[V,\overline{V}]
e^{-iV_{}\overline{\rho}_{}-i\overline{V}_{}\rho_{}}\,.
\end{align}
Substituting here Eq.~\eqref{eq:Partition_function_general}, we get 
\begin{align}
\label{eq:partition_function_rho_general}
 e^{i S[\rho, \overline{\rho}]} = e^{- \frac{i}{\nu} \rho\, \Pi_a^{-1} \overline{\rho}} {\cal F} [\nu^{-1} \Pi_a^{-1}\overline{\rho}].
\end{align}
This defines the effective action $S[\rho, \overline{\rho}]$ that allows one to calculate correlation functions of various operators (such as densities or vertex operators) over a given (in general, non-equilibrium) state, encoded by the density matrix $\rho_0$.

In Ref.~\cite{Gutman2010}, the functional ${\cal F}[\overline{V}]$ was determined for the case of free fermions, $\nu=1$. It was shown that, for any given (single-particle) fermionic distribution function $f(\epsilon)$, the function ${\cal F}[\overline{V}]$ is given by the following single-particle Fredholm determinant:
\begin{align} 
 {\cal F} [\overline{V}] &=
   \text{Det} \big[ 1 + (e^{-i\delta_{\overline{V}}(t)} - 1) f (\epsilon) \big ]
 \equiv \Delta [\delta_{\overline{V}}(t)],
   \label{eq:F-V-bar-det}
    \end{align}
where 
\begin{equation}
\delta_{\overline{V}}(t) = \sqrt{2}  \int_{-\infty}^{\infty} dt' \,\overline{V} [\eta v (t + t'), t']
\end{equation}
is the phase acquired by a particle moving in the potential $\overline{V}$ along a light-cone trajectory labeled by the time $t$. In the final equality of Eq.~\eqref{eq:F-V-bar-det}, we introduced the short-hand notation $\Delta$ for the determinant.

The determinant~\eqref{eq:F-V-bar-det} is similar to those appearing in studies of FCS of fermions, which is not surprising in view of the connection to FCS emphasized above. In this determinant, the time $t$ and the energy $\epsilon$ can be understood as canonically conjugate variables, i.e., $[\epsilon, t] = i$, which makes the calculation of the determinant, in general, highly non-trivial.
Of particular interest is the case of a double-step distribution function,
\begin{subequations}
\begin{align}
\label{eq:double_step_fermions}
    &f(\epsilon) = \mathcal{T}f_{0}(\epsilon-e V) + (1-\mathcal{T})f_{0}(\epsilon),\\
    &f_0(\epsilon) = \Theta(-\epsilon),\label{eq:zero_fermi}
\end{align}
\end{subequations}
where $\Theta(\epsilon)$ is the step function. It is assumed here that $eV\gg T$, where $T$ is the temperature, so that the use of the zero-temperature Fermi distribution $f_0(\epsilon)$ is appropriate. The distribution function \eqref{eq:double_step_fermions} describes an electronic system that is driven out of equilibrium by injection (tunneling)  of electrons (charge $e$) with probability $\mathcal{T}$ from a system at voltage $V$.
In the present work, our main focus will be on an extension of this situation to the FQH setting: a FQH edge driven out of equilibrium by tunneling of quasiparticles (that can be either anyons or electrons) from another FQH edge, characterized by voltage $V$ and tunneling probability $\mathcal{T}$. We will assume a dilute limit, $\mathcal{T} \ll 1$, in which case this picture is under control and describes a rare injection of well-defined quasiparticles.

We thus consider now a Laughlin edge (filling $\nu = 1/m$), driven out of equilibrium by tunneling of elementary quasiparticles (with charge $e^* = \nu e$), see Fig.~\ref{fig:Lauglin_setup}.  The non-equilibrium action is given by Eq.~\eqref{eq:partition_function_rho_general}, where we should now specify the functional ${\cal F}[\overline{V}]$. To this end, we note that the FCS of a dilute beam of anyons is similar to that for electrons but with several essential differences: First, the electronic particle number of an anyon is $\nu$ (instead of unity for electrons). Second, the anyon charge that couples to the voltage is $e^* = \nu e$ (instead of $e$ for electrons). Finally, the injected current on the Laughlin edge is $\mathcal{T} V \nu e^2/(2\pi) = \mathcal{T} V \nu^{-1} (e^*)^2/(2\pi)$ instead of $\mathcal{T} V e^2/(2\pi)$ for electrons, with the factor $\nu^{-1}$ being a manifestation of the anyonic statistics of charge-$e^*$ quasiparticles, in the sense that each electronic state effectively translates into $1/\nu$ anyonic quasiparticles. These three considerations suggest the following form of the functional ${\cal F}[\overline{V}]$ entering 
the generating function~\eqref{eq:Partition_function_general} and the action~\eqref{eq:partition_function_rho_general}
for the Laughlin edge:
\begin{align} 
 {\cal F} [\overline{V}] &=
 \left\{  \text{Det} \big[ 1 + (e^{-i\delta_{\overline{V}}(t)} - 1) f (\epsilon) \big ] \right\}^{\frac{1}{\nu}}
\nonumber   \\
   &\equiv  \left\{ \Delta [\delta_{\overline{V}}(t)]
\right\}^{\frac{1}{\nu}},
\label{eq:F-V-bar-det-Laughlin}
    \end{align}
where the phase is now 
\begin{equation}
\delta_{\overline{V}}(t) = \nu \sqrt{2}   \int_{-\infty}^{\infty} dt' \,\overline{V} [\eta v (t' + t), t']
\label{delta-V-t-Laughlin}
\end{equation}
and $f(\epsilon)$ is a distribution function of {\it fermions} with charge $e^* = \nu e$,
\begin{equation}
f(\epsilon) = \mathcal{T} \Theta(-\epsilon+e^* V) + (1-\mathcal{T})\Theta(-\epsilon).
\label{n-epsilon-noneq-laughlin}
\end{equation}
We emphasize that $\mathcal{T}$  in Eq.~\eqref{n-epsilon-noneq-laughlin} should be understood as the actual tunneling probability, i.e., the one that includes the Luttinger-liquid renormalization from the ultraviolet scale up to the characteristic energy scale set by the voltage $V$, i.e., $\mathcal{T} \sim \mathcal{T}_0 |e^*V/\Lambda|^{2\zeta-2}$, where $\mathcal{T}_0$ is the bare tunneling strength, $\Lambda$ is a high-energy cut-off and $\zeta$ the scaling dimension of the involved tunneling operator (see, e.g., Eq.~\eqref{eq:constantvoltagecurrent} below for an explicit example). At this point, we make an approximation by considering $\mathcal{T}$ in Eq.~\eqref{n-epsilon-noneq-laughlin} as energy independent. We will see below that this approximation captures the physics we are interested in, allowing us to demonstrate it in a transparent way.

The anyonic functional~\eqref{eq:F-V-bar-det-Laughlin} differs from 
its fermionic counterpart~\eqref{eq:F-V-bar-det} by factors of $\nu$ (or, respectively, $\nu^{-1}$) entering in three different places, in  correspondence to the three differences discussed above. 
Strictly speaking, Eq.~\eqref{eq:F-V-bar-det-Laughlin} should be viewed as a conjecture at this stage, but we will support it below by comparing its predictions to the result for the FCS of anyons obtained in a simple way. 
Furthermore, we will show that Eq.~\eqref{eq:F-V-bar-det-Laughlin} gives an exact result for the action (and thus for any observable) in the case of the equilibrium state, in which case $f(\epsilon)$ is the fermionic equilibrium (Fermi-Dirac) distribution. Finally, we will extend Eq.~\eqref{eq:F-V-bar-det-Laughlin} to the case when the Laughlin edge is driven out of equilibrium by tunneling of excitations with charge $n\nu e$, with $n>1$.

\subsection{Non-equilibrium Laughlin edge: Full counting statistics of charge}
\label{sec:noneq-Laughlin-FCS}

As our first benchmark calculation, we use the action formulated in the previous section to calculate the FCS of charge for the non-equilibrium Laughlin edge. We show that, for $\mathcal{T}\ll 1$, the result is a Poissonian statistics of charge-$e^*$ quasiparticles as expected. For concreteness, we  choose here the chirality $\eta=+1$.  

The generating function of the FCS  of charge passing from left to right across a point $x_0$ within a time interval $\tau>0$ is defined as
\begin{align} 
   \kappa(\lambda,x_0,\tau) \equiv \langle e^{i\lambda Q(x_0,\tau)/e}e^{-i\lambda Q(x_0,0)/e} \rangle,
   \label{eq:FCS-kappa-definition}
\end{align}
where $Q(x_0,t)\equiv e\int_{-\infty}^{x_0} dx \rho(x,t)$ is 
the operator of the charge located to the left of $x_0$ at time $t$ and $\lambda>0$ is a dimensionless counting parameter. The cumulants of charge fluctuations are obtained by differentiating the cumulant generating function,  $\ln \kappa(\lambda,x_0,\tau)$, near $\lambda=0$,
\begin{align}
\label{eq:FCS_cumulants}
    \langle\langle(\delta Q)^k \rangle\rangle &= (ie\partial_\lambda)^k \ln\kappa(\lambda,x_0,\tau)\mid_{\lambda \to 0}.
\end{align}
In particular, the average charge and noise propagating across the point $x_0$ during the time interval $\tau$ are obtained for $k=1$ and $k=2$, respectively. When presented as an integral along the Keldysh contour, the generating function~\eqref{eq:FCS-kappa-definition} takes the form
\begin{align} 
\label{eq:FCS_Keldysh_def}
  \kappa(\lambda,x_0,\tau) &=   \int \mathcal{D}\rho\mathcal{D}\overline{\rho} \, e^{iS[\rho,\overline{\rho}]} \notag \\
  &\times e^{\frac{i\lambda}{\sqrt{2}} \int dx \Theta(x_0-x)[ \rho(x,\tau) -  \rho(x,0) -\overline{\rho}(x,\tau) -\overline{\rho}(x,0)]}, 
\end{align}
where the action $S[\rho,\overline{\rho}]$ given in Eq.~\eqref{eq:partition_function_rho_general}.

The functional integral in Eq.~\eqref{eq:FCS_Keldysh_def} can be evaluated by the method developed for the fermionic case in Ref.~\cite{Gutman2010}: First, the integral over the classical density component $\rho$ is taken, by using the fact that it enters only linearly in the action. This integration results in a delta function that gives a constraint equation for the quantum component $\overline{\rho}$, cf. Eq.~\eqref{eq:GFgreater2} below. Plugging the solution of this equation back into the functional integral, it is expressed as a functional determinant that depends on the distribution function $f(\epsilon)$ and on a counting phase function $\delta_\tau(t)$ governed by the above solution for $\overline{\rho}$. We omit details of this calculation here, as it is fully analogous to the calculation 
of the Green's functions that is presented in detail below in Sec.~\ref{sec:GF_Laughlin}.
We thus only present the final result here, which reads
\begin{equation}
\label{eq:FCS_laughlin_simple}
\kappa(\lambda,x_0,\tau) = \left\{\overline{\Delta}[\delta_{\tau}(t)]\right\}^{\frac{1}{\nu}}.
\end{equation}
In this expression, $\overline{\Delta}[\delta_{\tau}(t)]$ is a normalized version (see Eq.~\eqref{eq:Determinant_normalization} below for details on normalization) of the determinant [cf. Eq.~\eqref{eq:F-V-bar-det}]
\begin{align} 
\label{eq:Fredholmdet}
    &\Delta[\delta_{\tau}(t)] = \text{Det} \big[ 1 + (e^{-i\delta_{\tau}(t)} - 1) f (\epsilon) \big ].
    \end{align}
In this determinant, the distribution function is given in Eq.~\eqref{n-epsilon-noneq-laughlin} (with $e^* = \nu e$) and the counting phase $\delta_{\tau}(t)$ has the form
\begin{equation}
\label{eq:counting_phase_Laughlin}
\delta_{\tau}(t) = \lambda \nu w_\tau(t,0),
\end{equation}
in which $w_\tau(t_1,t_2)$ is the window function,
  \begin{align}   \label{eq:window_function}
        w_\tau(t_1,t_2) \equiv \Theta(t_2-t_1)-\Theta(t_2-t_1-\tau).
    \end{align}
With Eqs.~\eqref{eq:FCS_laughlin_simple} and~\eqref{eq:Fredholmdet}, we now compute $\langle\langle(\delta Q)^k \rangle \rangle$ in the large-$\tau$ limit. Since cumulants are obtained by expanding around $\lambda \approx 0 $, we can apply the Szeg\H{o} approximation (see the Supplemental Material~\cite{Supplemental_Material} for details) to evaluate the leading asymptotics of the determinant.
In the dilute limit $\mathcal{T}\ll 1$, we find that the generating function evaluates to
\begin{align}    
\label{eq:FCS_double_step}
&\kappa(\lambda,x_0,\tau)\simeq \exp\left\{- \frac{  e V\mathcal{T}\tau}{2\pi } (1- e^{- i \nu  \lambda}) \right\},
\end{align}
which by Eq.~\eqref{eq:FCS_cumulants} yields the average charge
\begin{align}
\label{eq:Laughlin-noneq-charge}
    &\langle Q\rangle = \frac{\nu e^2 V\mathcal{T}\tau}{2\pi}
\end{align}    
and the remaining cumulants
\begin{align}
\label{eq:Schottky_formula}
    &\langle\langle(\delta Q)^k \rangle\rangle = 
    (e \nu)^{k-1}  \langle Q\rangle\,.
\end{align}
This is precisely the FCS of a Poissonian process with fractional charges, see, e.g., Ref.~\cite{Levitov2004Sep}. 
Note that Eq.~\eqref{eq:Schottky_formula} for $k=2$ manifests the well-known Schottky relation between the noise and the average current.

We have thus verified that the action
\eqref{eq:partition_function_rho_general} with 
the functional~\eqref{eq:F-V-bar-det-Laughlin}
and the distribution function~\eqref{n-epsilon-noneq-laughlin} indeed yields the expected Poissonian FCS of particles with charge $e^* = \nu e$.

At this point, it is worth briefly discussing corrections to the Poisson FCS. For free fermions, the distribution \eqref{n-epsilon-noneq-laughlin} leads to binomial statistics, which has the Poisson form to leading order in the small $\mathcal{T}$, but includes also corrections originating from higher-order terms. It has been argued in a number of works that the binomial distribution is relevant also for anyons, see in particular Refs.~\cite{Trauzettel2004effect,Feldman2017Mar,Lee2023May}. 
When terms of higher orders in the small $\mathcal{T}$ are retained, the determinant \eqref{eq:Fredholmdet} generates the binomial distribution. 
The (very interesting) question about the status of the higher-order terms (and, in particular, of the applicability of the binomial FCS to anyons) deserves further studies. In this work, we restrict ourselves to the leading-order terms in $\mathcal{T}$, yielding Poisson statistics.

\subsection{Laughlin edge in equilibrium: Action, full counting statistics, and Green's functions}

\label{sec:Laughlin-equilibrium}
While we are interested in describing a non-equilibrium edge, it is instructive to also inspect what our conjecture for the non-equilibrium action, Eqs.~\eqref{eq:partition_function_rho_general} and~\eqref{eq:F-V-bar-det-Laughlin}, yields in equilibrium, i.e., when we replace the distribution function~\eqref{n-epsilon-noneq-laughlin} by the equilibrium Fermi-Dirac distribution with a temperature $T$,
\begin{align}
    f_{T}(\epsilon)=\frac{1}{e^{\epsilon/T}+1}.
    \label{eq:n_T}
\end{align}
It is known~\cite{Gutman2010} that in this case, an expansion of the logarithm of the determinant in Eq.~\eqref{eq:F-V-bar-det-Laughlin} in powers of the phase functional in the exponential (or, equivalently, in powers of $\overline{V}$) yields the only term that is quadratic in $\overline{V}$. We note that the factor $\nu^{-1}$ in the argument of ${\cal F}$ in Eq.~\eqref{eq:partition_function_rho_general} then cancels the factor $\nu$ in the exponent of Eq.~\eqref{eq:F-V-bar-det-Laughlin}. Thus, the noise part of the action will be the same as for fermions ($\nu=1$), up to a multiplicative factor $1/\nu$ coming from the power of the determinant in 
Eq.~\eqref{eq:F-V-bar-det-Laughlin}.
In equilibrium, the total action in Eq.~\eqref{eq:partition_function_rho_general} thus becomes
\begin{equation}
S[\rho, \overline{\rho}] = - \frac{1}{\nu} \rho\, \Pi_a^{-1} \overline{\rho} - \frac{1}{2\nu}  
\overline{\rho}\, [\Pi^{-1}]_K\:\overline{\rho},
\label{eq:action-equilibrium}
\end{equation}
where $ [\Pi^{-1}]_K$ is the Keldysh component of the inverse polarization operator,
\begin{eqnarray}
[\Pi^{-1}]_K &=& - \Pi_r^{-1} \Pi_K \Pi_a^{-1} \,, \\ 
\Pi_K &=& (\Pi_r - \Pi_a) \coth \frac{\omega}{2T},
\end{eqnarray}
with the advanced component of the polarization operator $\Pi_a(q,\omega)$ given in Eq.~\eqref{eq:advanced_pol} and the retarded component $\Pi_r(q,\omega) = \Pi_a^*(q,\omega)$. Equation~\eqref{eq:action-equilibrium} is the correct, Gaussian, Keldysh action of the finite-$T$ equilibrium Laughlin edge. Correspondingly, the theory defined by Eqs.~\eqref{eq:partition_function_rho_general}, \eqref{eq:F-V-bar-det-Laughlin}, and \eqref{eq:n_T} yields exact results for all observables, including in particular the FCS and the Green's functions (see Sec.~\ref{sec:GF_Laughlin}). Specifically, for the FCS we find, in the large-$\tau$ limit, the generating function
\begin{align}
\label{eq:FCS_equilibrium}
    &\kappa(\lambda,x_0,\tau)  \simeq 
     \exp\left\{-\frac{\nu  T \tau   \lambda ^2}{4\pi}\right\},
\end{align}
and correspondingly the cumulants
\begin{align}
\label{eq:FCS_cumulants_equilibrium}
    &\langle \langle(\delta Q)^k \rangle \rangle =  \begin{cases}
		\displaystyle	\frac{\nu e^2 T \tau}{2\pi}, & \text{for $k=2$},\\[0.2cm]
            0, & \text{otherwise}.
		 \end{cases}
\end{align}

Equations \eqref{eq:FCS_equilibrium}
and
\eqref{eq:FCS_cumulants_equilibrium}
are the correct 
results for the generating function and the cumulants for the Laughlin edge in equilibrium.
These results can be obtained either by evaluating the generating function~\eqref{eq:FCS_Keldysh_def}
with the Gaussian action
\eqref{eq:action-equilibrium} or, equivalently, by expanding the logarithm of the determinant \eqref{eq:Fredholmdet}
in $\delta_\tau(t)$ [where only the quadratic term becomes non-zero for the equilibrium distribution \eqref{eq:n_T}] and substituting the result into 
Eq.~\eqref{eq:FCS_laughlin_simple}.

The following comment is in order here. We note that Eq.~\eqref{eq:FCS_equilibrium} for the FCS generating function is in fact its branch around $\lambda=0$ and holds for $\nu|\lambda| \le \pi$. At $\nu|\lambda| = \pi$, the generating function experiences singularities and is continued periodically, $\kappa(\lambda,x_0,\tau)=\kappa(\lambda+2\pi / \nu,x_0,\tau)$, in correspondence with Eq.~\eqref{eq:FCS-kappa-definition} and with charge quantization in units of $\nu e$. 
When one works with the Gaussian equilibrium action and with the density  fields \eqref{eq:density_def}, one effectively considers the statistics of the smeared charge, Eq.~\eqref{eq:FCS_equilibrium}, which does not reflect charge quantization (i.e., does not exhibit the corresponding periodicity). 
The periodicity can be recovered by an accurate analytical treatment of Toeplitz determinants.
We refer the reader to related discussions for the case of free fermions ($\nu=1$) in Refs.~\cite{Gutman2010,Ivanov2013counting} and for interacting fermions (the conventional Luttinger liquid) in Ref.~\cite{Gutman2010full}. We emphasize that expressions for the FCS generating functions at small $|\lambda|$ are sufficient to obtain all cumulants, as will be done below throughout the paper.
 
In the same way, we can check that our formalism gives exact results for the Green's functions of excitations (see their formal definition in the beginning of Sec.~\ref{sec:GF_Laughlin}) on the equilibrium Laughlin edge:
 \begin{align} 
\label{eq:GF-Laughlin-equilibrium}
\mathcal{G}^{\gtrless}(\tau)&=  \frac{\mp i}{2\pi a} \left(\frac{a}{a\pm iv\tau}\right)^\nu  \Big (\frac{\pi T \tau}{\sinh (\pi T \tau)}\Big)^\nu \,.
\end{align}
In analogy with the FCS, there are two equivalent ways to see this. One possibility is to first reduce the action to the equilibrium form~\eqref{eq:action-equilibrium} and then use it to evaluate the Green's functions. Since the vertex operators~\eqref{eq:vertex_operators} creating and annihilating excitations are exponentiated expressions linear in the density fields, all involved integrals are Gaussian. Further, since $\nu$ enters simply as an overall factor $1/\nu$ in the action~\eqref{eq:action-equilibrium}, the result of the Gaussian integration will contain $\nu$ only as an overall exponent, which is manifest in the Green's function~\eqref{eq:GF-Laughlin-equilibrium}. An alternative path is to first derive a general expression for the Green's functions in terms of a functional determinant and then to reduce it to 
the equilibrium form~\eqref{eq:GF-Laughlin-equilibrium}, see Sec.~\ref{sec:GF_Laughlin}.

\subsection{More general non-equilibrium states of the Laughlin edge}
\label{sec:Laughlin-general-noneq}
We now extend our formalism to a more general class of non-equilibrium states on the Laughlin edge ($\nu = 1/m$). Specifically, we consider the situation where the Laughlin edge is driven out of equilibrium by tunneling of quasiparticles with charge $e^*= n \nu e$. While in the preceding consideration we had $n=1$, we now allow $n$ to be any integer. A special case is $n=m$, for which the tunneling quasiparticles are electrons. 

To deduce the form of the functional ${\cal F}[\overline{V}]$ entering 
the generating function \eqref{eq:Partition_function_general} and the action~\eqref{eq:partition_function_rho_general}
for this more general case of a non-equilibrium Laughlin edge, we generalize the argumentation used in 
Sec.~\ref{sec:formalism} for $n=1$: The electronic particle number of an excitation is now $n\nu$, and the corresponding charge is $e^*=n\nu e$. Furthermore, the statistical exchange angle $\theta$ characterizing the anyonic nature of such excitations is $\theta=\pi n^2 \nu$ [see Eq.~\eqref{eq:Laughlin_exchange_angle}]. These features lead to the following generalization of Eq.~\eqref{eq:F-V-bar-det-Laughlin}:
\begin{align} 
 {\cal F} [\overline{V}] &=
 \left\{  \text{Det} \big[ 1 + (e^{-i\delta_{\overline{V}}(t)} - 1) f (\epsilon) \big ] \right\}^{\frac{1}{\nu n^2}}
\nonumber   \\
   &\equiv  \left\{ \Delta [\delta_{\overline{V}}(t)]
\right\}^{\frac{1}{\nu n^2}},
\label{eq:F-V-bar-det-Laughlin-n}
    \end{align}
where 
\begin{equation}
\delta_{\overline{V}}(t) = \nu n\sqrt{2}   \int_{-\infty}^{\infty} dt' \,\overline{V} [\eta v (t' + t), t']
\label{delta-V-t-n}
\end{equation}
and $f(\epsilon)$ has the same form
\eqref{n-epsilon-noneq-laughlin} as before (for tunneling of excitations driven by the voltage bias $V$), but now
with $e^* = n \nu e$.

We can now perform the benchmarking as above. For the FCS of a non-equilibrium edge (in the dilute limit $\mathcal{T}\ll 1$), we obtain the following generalization of Eqs.~\eqref{eq:FCS_double_step}, 
\eqref{eq:Laughlin-noneq-charge}, and
\eqref{eq:Schottky_formula}:
\begin{align}    
\label{eq:FCS_double_step-n}
&\kappa(\lambda,x_0,\tau)\simeq \exp\left\{- \frac{n \nu  e V\mathcal{T}\tau}{2\pi n^2 \nu} (1- e^{- i n \nu  \lambda}) \right \},
\end{align}
\begin{align}
\label{eq:Laughlin-noneq-charge-n}
    &\langle Q\rangle = \frac{\nu e^2 V\mathcal{T}\tau}{2\pi} \,,
\end{align}    
and 
\begin{align}
\label{eq:Schottky_formula-n}
    &\langle\langle(\delta Q)^k \rangle\rangle = 
    (n e \nu)^{k-1}  \langle Q\rangle\,.
\end{align}
This is the correct FCS of a Poissonian process of quasiparticles with charge $e^* = n\nu e$. 

Next, we consider the equilibrium case, with $f(\epsilon)$ given by   Eq.~\eqref{eq:n_T}. The determinant in  Eq.~\eqref{eq:F-V-bar-det-Laughlin-n} is then a Gaussian functional of the expression in the exponential, and thus its logarithm is proportional to $n^2$. It follows that the factors $n$ totally cancel in the resulting expression for   ${\cal F} [\overline{V}]$, leading to the same equilibrium action \eqref{eq:action-equilibrium} as obtained above for $n=1$. This is the correct result: the equilibrium state of a given Laughlin edge is uniquely defined and should thus be independent of $n$. 

We have thus established that, for any $n$, the action
\eqref{eq:partition_function_rho_general} with 
the functional \eqref{eq:F-V-bar-det-Laughlin-n} yields correctly all benchmark results, including the FCS of a non-equilibrium Laughlin edge as well as its equilibrium properties. 
We can thus proceed with employing this theory for computing the non-equilibrium Green's functions of the Laughlin edge, which is the subject of Sec.~\ref{sec:GF_Laughlin}. 
Subsequently, in  Sec.~\ref{sec:tunneling}, we will use these Green's functions to compute observables for tunneling experiments with quantum point contacts. In Sec.~\ref{sec:multiple-mode-edges}, we will generalize the theory to  edges with more than one mode. 

\section{Non-equilibrium Green's functions of excitations on the Laughlin edge}
\label{sec:GF_Laughlin}

\subsection{General formalism} \label{sec:GF_Laughlin_formalism}
Here, we compute non-equilibrium Green's functions (GFs) of various excitations on the Laughlin edge ($\nu=1/m$). 
The greater and lesser GFs for excitations with charge $n\nu e$ are defined as
\begin{eqnarray}&&
\label{eq_green_function}
\mathcal{G}^>(x_1,t_1;x_2,t_2)\equiv-i\langle
\psi_n(x_1,t_1)\psi_n^\dagger(x_2,t_2)\rangle\,,
\nonumber \\&&
\mathcal{G}^<(x_1,t_1;x_2,t_2)\equiv i\langle
\psi^\dagger_n(x_2,t_2)\psi_n(x_1,t_1)\rangle,
\end{eqnarray}
with the vertex operators
\eqref{eq:vertex_operators}.
We consider first the case $n=1$ corresponding to ``minimal'' anyonic quasiparticle excitations and discuss the generalization to an arbitrary $n$ below.  The  GFs are defined on the Keldysh contour as
\begin{subequations}
    \begin{align}
    \label{eq:greaterGF}
        &\mathcal{G}^>(x_1,t_1;x_2,t_2)\equiv \frac{-i}{2\pi a}\langle T_c e^{i\phi_-(0,\tau)}e^{-i\phi_+(0,0)}\rangle,\\
        \label{eq:lesserGF}
        &\mathcal{G}^<(x_1,t_1;x_2,t_2)\equiv \frac{+i}{2\pi a}\langle T_c e^{-i\phi_-(0,0)}e^{i\phi_+(0,\tau)}\rangle.
    \end{align}
\end{subequations}
Here, we used the fact that, due to Galilean invariance, the GFs depend only on the light-cone variable
\begin{align}
    \tau \equiv t_1-t_2-\eta(x_1-x_2)/v \,,
\end{align}
 so that we can set $x_1=x_2=vt_2=0$ and $t_1 =\tau$ without loss of generality. 
 
 We present now in detail the evaluation of the greater GF~\eqref{eq:greaterGF}; the calculation for the lesser GF proceeds analogously. In the rotated Keldysh basis~\eqref{eq:Keldysh_rotation},  Eq.~\eqref{eq:greaterGF} takes the functional-integral form
\begin{align}
\label{eq:GFgreater1}
   \mathcal{G}^>(\tau) = &\frac{-i}{2\pi a} \int \mathcal{D}\rho\mathcal{D}\overline{\rho} e^{iS[\rho,\overline{\rho}]}\notag \\
    &\times e^{\frac{i}{\sqrt{2}}\left[\phi(0,\tau)-\phi(0,0)-\overline{\phi}(0,\tau)-\overline{\phi}(0,0)\right]},
\end{align}
with the Keldysh action $S[\rho,\overline{\rho}]$ from Eq.~\eqref{eq:partition_function_rho_general}. Due to the linear dependence of the exponent in Eq.~\eqref{eq:GFgreater1} on the classical density component $\rho$, we can integrate it out exactly, which yields
\begin{align}
\label{eq:GFgreater2}
   \mathcal{G}^>(\tau)= &\frac{-i}{2\pi a} \int \mathcal{D}\overline{\rho}\, 
     \left\{ \Delta [\delta_{\overline{V}}(t)]
\right\}^{\frac{1}{\nu}}\Big|_{\overline{V} = \nu^{-1} \Pi_a^{-1}\overline{\rho}}    
    \notag \\
    &\times e^{-\frac{i}{\sqrt{2}}\left(\overline{\phi}(0,\tau)+\overline{\phi}(0,0)\right)}\delta \left(\overline {\rho} - \overline{\rho}^*\right),
\end{align}
where $\Delta [\delta_{\overline{V}}(t)]$
is the Fredholm determinant as defined in Eqs.~\eqref{eq:F-V-bar-det} and~\eqref{eq:F-V-bar-det-Laughlin},   
with $\delta_{\overline{V}}(t)$ given in Eq.~\eqref{delta-V-t-Laughlin}. Furthermore, $\overline{\rho}^*$ in Eq.~\eqref{eq:GFgreater2}  is the  advanced solution to the differential equation
\begin{align}
\label{eq:constraint_Laughlin}
    \eta\partial_t \overline{\rho}(x,t) + \partial_x v\overline{\rho}(x,t) = -\eta \nu j(x,t)
\end{align}
with the source term
\begin{align}
\label{eq:source_j}
    j(x,t) \equiv  \frac{1}{\sqrt{2}}\delta(x)\left(\delta(t-\tau)-\delta(t)\right).
\end{align}
 The derivation of Eq.~\eqref{eq:constraint_Laughlin}  uses the form \eqref{eq:advanced_pol} of the polarization operator in the action \eqref{eq:partition_function_rho_general}, which implies 
\begin{align}
\label{eq:pol_convolution}
    & \partial_x \int dt' dx' \Pi_{a}^{-1}(x,t;x',t')\overline{\rho}(x',t') \notag \\&=
    2\pi \left[v \partial_x\overline{\rho}(x,t) + \eta  \partial_t \overline{\rho}(x,t) \right]\,.
\end{align}
This produces the left-hand-side of Eq.~\eqref{eq:constraint_Laughlin} (up to a factor $2\pi$).
The advanced solution to Eq.~\eqref{eq:constraint_Laughlin} is
\begin{align}
\label{eq:rho_sol_Laughlin}
    &\overline{\rho}^*(x,t) \notag \\
    &=
    \frac{\nu}{\sqrt{2}} \Theta(-\eta x)\big[ \delta(x+ \eta v(\tau - t)) - \delta(x - \eta v t) \big]\,.
\end{align}
The determinant entering Eq.~\eqref{eq:GFgreater2} and governing the dependence of the GFs on the distribution function becomes 
$\Delta[\delta_\tau(t)]$, Eq.~\eqref{eq:Fredholmdet}, with the phase function $\delta_\tau(t)$ given by
\begin{align}
\label{eq:scatteringPhase}
      & \delta_{\tau}(t) = 2\theta w_\tau(t,0) \,.
\end{align}
Here, we identified the Laughlin anyon braiding angle $2\theta=2\pi \nu$ (see Eq.~\eqref{eq:Laughlin_exchange_angle} for $n=1$) and used the notation for the window function, Eq.~\eqref{eq:window_function}.
The result Eq.~\eqref{eq:scatteringPhase} for the phase is obtained by substituting $\overline{V}=\Pi_a^{-1}\overline{\rho}/\nu$ into 
Eq.~\eqref{delta-V-t-Laughlin}, 
resulting in~\cite{Gutman2010}
\begin{align}
\label{eq:gen_phase_formula}
    \delta_\tau(t) =  2\pi\sqrt{2}\eta \lim_{t'\to-\infty}\int^{\eta v(t'+t)}_0dx'\,\overline{\rho}(x',t').
\end{align}
Inserting the density solution~\eqref{eq:rho_sol_Laughlin} into Eq.~\eqref{eq:gen_phase_formula}
and using the identity 
\begin{equation}
\label{eq:useful_phase_identity}
    \eta\lim_{t'\to-\infty}\int^{\eta v(t'+t)}_0\delta(t'-\eta x'/v+p)dx' = -v\Theta(p-t),
    \end{equation}
    where $p$ is a generic variable with dimension time and $\Theta(t)$ is the step function,
we arrive at Eq.~\eqref{eq:scatteringPhase}.

Importantly, the determinant~\eqref{eq:Fredholmdet} entering Eq.~\eqref{eq:GFgreater2} exhibits an ultraviolet divergence, thus requiring an ultraviolet regularization (length scale $a$ in our notation). It is convenient to normalize the determinant to its value for the zero-temperature equilibrium distribution function~\cite{Supplemental_Material}. We thus define
    \begin{align}   \label{eq:Determinant_normalization}
    \overline{\Delta}[\delta_{\tau}(t)] \equiv \frac{\Delta[\delta_{\tau}(t)]}{\Delta[\delta_{\tau}(t)]_{T=0}}.
    \end{align}
Since the singular behavior at $a\to 0$ is independent of the distribution function, the normalized determinant 
 $\overline{\Delta}[\delta_{\tau}(t)]$ is free of ultraviolet singularities. Gathering the above results, we arrive at the final result for the quasiparticle GFs:
 \begin{align} 
 \label{eq:GFgreater4}
\mathcal{G}^{\gtrless}(\tau)&=  \frac{\mp i}{2\pi a} \left(\frac{a}{a\pm iv\tau}\right)^\nu
 \left\{\overline{\Delta} [\delta_{\tau} (t)]\right\}^{\frac{1}{\nu}}.
\end{align}
Here, we have included also the result for the lesser GF, which is obtained by an analogous calculation. For equilibrium at $T=0$, the determinant in  Eq.~\eqref{eq:GFgreater4} is by definition equal to unity, and the GFs reduce to the known bosonization result~\cite{Wen_CLL_edge_1990}. 

As was already pointed out in Sec.~\ref{sec:Laughlin-equilibrium}, our formalism yields exact results for GFs in equilibrium. Let us discuss how this follows from Eq.~\eqref{eq:GFgreater4}. For a finite-temperature equilibrium distribution function, the normalized determinant evaluates to~\cite{Supplemental_Material}
\begin{equation}
\overline{\Delta}[\delta_{\tau} (t)]= \Big (\frac{\pi T \tau}{\sinh (\pi T \tau)}\Big)^{\nu^2},
\end{equation}
which, after substitution in Eq.~\eqref{eq:GFgreater4},
correctly reproduces the equilibrium result, Eq.~\eqref{eq:GF-Laughlin-equilibrium}.

We proceed now by generalizing the above derivation that yielded Eq.~\eqref{eq:GFgreater4} for GFs of excitations at a non-equilibrium Laughlin edge. Specifically, we consider its generalization in two respects. First, the non-equilibrium state can be produced by tunneling of quasiparticles with  $e_1^*\equiv n_1 \nu e$, see
Sec.~\ref{sec:Laughlin-general-noneq}.
Second, we can study, in this non-equilibrium state, the GFs of excitations with charge $e_2^*\equiv n_2 \nu e$.
In general, $n_1$ and $n_2$ can be arbitrary integers; the result \eqref{eq:GFgreater4} was derived for $n_1 = n_2 =1$. The action of our theory is now given by Eqs.~\eqref{eq:partition_function_rho_general},
\eqref{eq:F-V-bar-det-Laughlin-n},
and
\eqref{delta-V-t-n}. Extending the above analysis to this case,
 we find that the GFs~\eqref{eq:GFgreater4} generalize to 
 \begin{align} \label{eq:GFgreater_gen_n}
\mathcal{G}^{\gtrless}(\tau)&=  \frac{\mp i}{2\pi a} \left(\frac{a}{a\pm iv\tau}\right)^{n_2^2\nu}\overline{\Delta} [\delta_{\tau} (t)]^{\frac{1}{n_1^2\nu}}.
\end{align}
The scattering phase is now
\begin{align}
    \label{eq:scatteringPhase_n1n2}
      & \delta_{\tau}(t) = \delta_0 w_\tau(t,0), 
\end{align}
with the amplitude
\begin{align} 
\label{eq:amplitude}
    \delta_0 =2\theta_{12},
\end{align}
given in terms of the mutual braiding angle $2\theta_{12}$ in Eq~\eqref{eq:theta12_def}. The non-equilibrium distribution function has the same form
\eqref{n-epsilon-noneq-laughlin} as before but now with $e^* \mapsto e^*_1 = n_1 \nu e$. 

At finite-temperature equilibrium, Eq.~\eqref{eq:GFgreater_gen_n} becomes independent of $n_1$ as expected and reduces to the correct 
equilibrium result given by
Eq.~\eqref{eq:GF-Laughlin-equilibrium}
with the substitution $\nu \mapsto n_2^2\nu$.

\subsection{Long-time asymptotic behavior}
\label{sec:GF-asymptotic}

Having established the general form 
\eqref{eq:GFgreater_gen_n}
of the quasiparticle GFs on the Laughlin edge, we next investigate their long-time asymptotic behavior for the non-equilibrium, double-step distribution function \eqref{n-epsilon-noneq-laughlin}. 
When considered in the time representation, the operator in Eq.~\eqref{eq:Fredholmdet} entering the 
determinant $\Delta[\delta_{\tau}(t)]$ has the structure of a Toeplitz matrix, i.e., it's entries depends only on time differences. The limit of long time $\tau$ corresponds to a large size of this Toeplitz matrix. Its large-size asymptotic behavior can conveniently be explored with the Szeg\H{o} approximation and, more accurately, by using the (extended) Fisher-Hartwig conjecture.
We now proceed by analyzing the GFs within the Szeg\H{o} approximation in Sec.~\ref{sec:szego}. After discussing limitations of this approximation, we will go beyond it and explore the GF asymptotics within the extended Fisher-Hartwig conjecture
in Sec.~\ref{sec:fisher-hartwig}.

\subsubsection{Szeg\H{o} approximation}
\label{sec:szego}

We first consider the case of $n_1 = n_2 = 1$, when the non-equilibrium state is produced by the minimal quasiparticle excitations and the GFs for the same type of quasiparticle excitations are considered, $e_1^*=e^*_2=e^*=\nu e$. In this case, the leading long-time (i.e., $|\tau| \gg 1/|e^*V|$) asymptotics can be captured within the Szeg\H{o} approximation.
This approximation is obtained by writing the determinant in Eq.~\eqref{eq:Fredholmdet} as $\exp({\rm tr} \ln \ldots)$ 
and then treating the trace in the quasiclassical approximation (i.e., replacing it by an integral over the energy-time phase space). The result reads
\begin{align} \label{eq:logarithmbranch}
   \Delta[\delta_{\tau}(t)] \sim \exp \Big[\frac{|\tau|}{2\pi} \int_{-\Lambda}^{\Lambda} d\epsilon & \Big\{\ln \big[1+ (e^{-i \delta_0 \text{sgn}(\tau)}-1) f(\epsilon)\big] \nonumber \\ &- \frac{i \epsilon\delta_0 \text{sgn}(\tau)}{2\Lambda} \Big \}\Big]\,,
\end{align}
where $\Lambda \equiv v/a$ is the ultraviolet energy cutoff and $\delta_0$ is the amplitude of the scattering phase~\eqref{eq:amplitude}, with $n_1=n_2=1$. The second line in Eq.~\eqref{eq:logarithmbranch} appears upon proper regularization of the determinant~\cite{Supplemental_Material}.
Using Eq.~\eqref{n-epsilon-noneq-laughlin} for the distribution function $f(\epsilon)$ and performing the energy integration in the dilute limit, $\mathcal{T} \ll 1$, we find that the large-$\tau$ asymptotics of the normalized determinant in Eq.~\eqref{eq:GFgreater4} is given by 
\begin{align}
\label{eq:det_Poisson}
    \overline{\Delta} [\delta_{\tau} (t)] \simeq \exp \left[-\frac{
    \mathcal{T}|e^*V \tau|}{2\pi}(1-e^{-i2\pi \nu \text{sgn}(e^*V\tau)}) \right].
\end{align}
For the derivation of Eq.~\eqref{eq:det_Poisson}, see the Supplemental Material~\cite{Supplemental_Material}.
Plugging the determinant~\eqref{eq:det_Poisson} into the GFs~\eqref{eq:GFgreater4}, we obtain 
 \begin{align} 
 \label{eq:GF_Poisson}
 \mathcal{G}^{\gtrless}(\tau)&\simeq  \frac{\mp i}{2\pi a} \left(\frac{a}{a\pm iv\tau}\right)^\nu \nonumber \\
 & \times
 \exp\left[-\frac{\mathcal{T}|eV \tau|}{2\pi}(1-e^{-i2\pi \nu\text{sgn}(e V\tau)}) \right].
\end{align} It is worth noting that the product $V \mathcal{T}$ that enters Eq.~\eqref{eq:GF_Poisson} is proportional to the average edge current~\cite{Levkivskyi2016Apr,Rosenow2016Apr,Supplemental_Material}
\begin{align}
\label{eq:average_I}
    \langle I_0\rangle = \frac{\nu e^2 V \mathcal{T}}{2\pi},
\end{align}
cf. Eq.~\eqref{eq:Laughlin-noneq-charge-n}. Once more, we emphasize that for the FQH edge, the transmission $\mathcal{T}$ that enters~\eqref{eq:average_I} must be understood as a re-normalized quantity with a generally non-linear dependence on the voltage $V$. With this in mind, identifications analogous to~\eqref{eq:average_I} can be done for all products of the  type $\mathcal{T}V$ in subsequent formulas, so that the corresponding results can be expressed in terms of experimentally measurable average edge currents. We see that Eq.~\eqref{eq:GF_Poisson} differs from the equilibrium GF by an exponential factor. This factor can be split into a product of an oscillatory exponential $\exp (-i\omega_V \tau)$, that can be viewed as an effective shift of the chemical potential~\cite{Safi2020Jul,zhang2025effectivelinearresponsenonequilibrium}, 
with the frequency
\begin{align}
\label{eq:omega_V}
    \omega_V \equiv \frac{eV\mathcal{T} \sin(2\theta)}{2\pi},
\end{align}
where $2\theta=2\pi \nu$ is the braiding phase of the minimal Laughlin quasiparticles, and an exponentially decaying factor
$\exp(-|\tau|\gamma_\phi)$ with the dephasing rate 
\begin{equation}
\label{eq:gamma_deph}
\gamma_\phi = \frac{|eV|\mathcal{T} \big(1- \cos (2\theta)\big)}{2\pi} \,.
\end{equation}
Thus, the GFs of anyons ($\nu = 1/m < 1$) necessarily exhibit non-equilibrium dephasing. This is in stark contrast with the free-fermion case, $\nu=1$, where such a dephasing is absent (see, e.g,  Eq.~\eqref{eq:GFs_free_fermions} below). The non-equilibrium dephasing
of the anyon GFs in  Eq.~\eqref{eq:GF_Poisson}
can thus be attributed to the strongly correlated nature of the Laughlin edge. 

The Szeg\H{o}-approximation result \eqref{eq:GF_Poisson} for the non-equilibrium GFs is identical to that obtained in Refs.~\cite{Levkivskyi2016Apr,Rosenow2016Apr} where the authors assumed Poissonian statistics for the injected charge distribution. Here, we have found the same result 
by using the non-equilibrium Keldysh action with the 
double-step distribution function~\eqref{n-epsilon-noneq-laughlin} for the injected anyons. 
Since we used the Poissonian FCS as a benchmark of our formalism, 
see Sec.~\ref{sec:noneq-Laughlin-FCS},
the agreement is not surprising. 

 Generalizing the derivation
of Eq.~\eqref{eq:GF_Poisson} to 
generic integers $n_1$ and $n_2$, we find for the GFs within the Szeg\H{o} approximation 
\begin{align} 
 \label{eq:GF_Poisson_gen}
 \mathcal{G}^{\gtrless}(\tau)&\simeq  \frac{\mp i}{2\pi a} \left(\frac{a}{a\pm iv\tau}\right)^{n_2^2\nu} \nonumber \\
 & \times
 \exp\left[-
 \frac{\mathcal{T}|e V \tau|}{2\pi n_1}(1-e^{-i \delta_0 \text{sgn}(eV\tau)})\right],
\end{align}
where $\delta_0=2\theta_{12}=2\pi \nu n_1 n_2$ is the mutual braiding angle~\eqref{eq:amplitude} of type-$n_1$ and type-$n_2$ quasiparticles. Thus, the needed modification to the oscillation frequency~\eqref{eq:omega_V} and the dephasing rate~\eqref{eq:gamma_deph} is obtained by substituting $2\theta\to 2\theta_{12}$ and $V\to V/n_1$ in these expressions.

We now discuss limitations of the approximation \eqref{eq:GF_Poisson_gen}. They become particularly manifest if we consider the case of $\delta_0 = 2\pi j$ with an integer $j$. Equation~\eqref{eq:GF_Poisson_gen} reduces then to the equilibrium GFs, i.e., it misses completely the non-equilibrium physics, see
Refs.~\cite{Rosenow2016Apr,Schiller2023anyon, Thamm2024effect,Iyer2024finite,Varada2025May} for related discussions. A particular case is that of free fermions, $\nu=1$ and $n_1=n_2=1$, which yields $\delta_0 = 2\pi$. Another realization of such a situation is the Laughlin edge, $\nu=1/m$, with $n_1 =1$ and $n_2 = m$ (or, alternatively, $n_1 = m$ and $n_2=1$), which also yields $\delta_0 = 2\pi$ (for discussions on experimental realizations of these cases, see Sec.~\ref{sec:Andreev_setups} below). Thus, the approximation \eqref{eq:GF_Poisson_gen} does not fully capture the non-equilibrium nature of the edge, which calls for a more accurate treatment. 
As we will demonstrate in Sec.~\ref{sec:fisher-hartwig},
the required essential improvement is provided by the generalized Fisher-Hartwig conjecture, which cures the above deficiency of the Szeg\H{o} approximation. This improvement thereby yields the complete large-$\tau$ asymptotics for GFs of anyons on a non-equilibrium FQH edge. \newline

 \subsubsection{Generalized Fisher-Hartwig conjecture}
\label{sec:fisher-hartwig}

The Fisher-Hartwig conjecture applies to Toeplitz determinants with singularities. In our case, these are the singularities of the distribution function $f(\epsilon)$. The non-equilibrium situation of our main interest, with $f(\epsilon)$ 
 having discontinuities at two ``Fermi edges'', Eq.~\eqref{n-epsilon-noneq-laughlin}, is exactly of this type. 

For generality, we consider here the case 
of generic $n_1$ and $n_2$. The non-equilibrium state is thus produced by injection of quasiparticles with charge $e_1^* = n_1  \nu e$ and in this state, we study correlations of quasiparticles with charge $e^*_2=n_2\nu e$. The scattering phase amplitude entering in Eq.~\eqref{eq:GFgreater_gen_n} is thus $\delta_0 = 2\theta_{12}=2\pi \nu n_1 n_2$, the mutual braiding phase for these two types of quasiparticles. The application of the generalized Fisher-Hartwig conjecture \cite{Gutman2011non-equilibrium} leads then to the following long-time asymptotics of the normalized determinant $\overline{\Delta}[\delta_{\tau}(t)]$~\cite{Supplemental_Material}: 
\begin{widetext}
\begin{align} \label{eq:HartwigFisher2longfullmaintext}
     \overline{\Delta} [\delta_{\tau}(t)] \simeq  \frac{e^{i e_1^* V \beta_1 \tau}}{G( 1+\frac{\delta_0}{2 \pi })
    G( 1-\frac{\delta_0}{2 \pi})} & \sum_{k = -\infty}^{\infty }  \Big [ \left(\frac{1}{|e_1^* V \tau|} \right)^{- 2 (\beta_0 + k) (\beta_1 -k)}  e^{-i k e_1^* V \tau}
    \nonumber \\ & 
    \times G(1 + \beta_0 + k) G(1-\beta_0-k) G(1 + \beta_1 - k) G(1-\beta_1 + k) \Big]\,. 
\end{align}
\end{widetext}
Here, the $\tau$-dependent parameters $\beta_0 $ and 
$\beta_1 $ are generated by the Fermi-edge singularities and are given by
\begin{subequations} \label{eq:beta1full}
\begin{align}
    \beta_1 &= - \frac{i}{2\pi} \text{sgn}(e_1^*V\tau) \ln [ 1+ \mathcal{T} (e^{-i \delta_0 \text{sgn}(e_1^*V\tau)} -1 ) ]\,,
    \label{eq:beta1}
    \\
    \beta_0  &= -\frac{\delta_0}{2\pi} - \beta_1 \,,
\end{align}
\end{subequations}
with $G(z)$ being the Barnes $G$-function.
The relation $\overline{\Delta}[\delta_{\tau} (t)] = (\overline{\Delta}[\delta_{-\tau} (t)])^*$ follows from the following property of the Barnes $G$-function, $[G(z)]^* = G(z^*)$. 

The overall exponential factor $e^{i e_1^* V \beta_1 \tau}$ in Eq.~\eqref{eq:HartwigFisher2longfullmaintext} is, for $\mathcal{T} \ll 1$, the same as the exponential factor in the Szeg\H{o} approximation 
 \eqref{eq:GF_Poisson_gen}. Crucially, in the more general result~\eqref{eq:HartwigFisher2longfullmaintext}, this exponential factor is multiplied by a sum over terms labeled by an index $k$ denoting the branches of the complex logarithm in 
 Eq.~\eqref{eq:beta1}. For each branch, there is an oscillatory exponential factor $e^{-i k e_1^* V \tau}$ as well as a power-law factor
 $\propto (1/|\tau|)^{-2(\beta_0+k)(\beta_1-k)}$, multiplied with a number of Barnes $G$-functions. The original form of the Fisher-Hartwig conjecture contained only the term with the dominant  exponent (i.e., the leading large-$|\tau|$ asymptotics in our case); its generalized form, derived in Ref.~\cite{Gutman2011non-equilibrium}, yields a more accurate result by including a sum over all branches $k$.  For a broader mathematical exposition of the Fisher-Hartwig conjecture with references to earlier works, e.g, Ref.~\cite{deift2011asymptotics}. We also mention that the generalized Fisher-Hartwig conjecture was applied to various non-equilibrium 1D many-body problems
in Refs.~\cite{Gutman2011non-equilibrium,Protopopov2011many-particle,Protopopov2012Luttinger,Protopopov2013correlations,NgoDinh2013analytically,Ivanov2010phase,Abanov2011quantum,Ivanov2013counting,Ivanov2013characterizing,Ivanov2013Fisher,Ivanov2014Fisher}.

The different branches $k$ entering in Eq.~\eqref{eq:HartwigFisher2longfullmaintext} are thus characterized by different power-law exponents 
$-2(\beta_0+k)(\beta_1-k)$, and the branches with smaller exponents dominate at long times $\tau$. In the dilute limit, $\mathcal{T} \ll 1$, we have $|\beta_1| \ll 1$, so that the dominant branches become determined by the phase amplitude $\delta_0$, see Eq.~\eqref{eq:beta1full}. In particular, when $\delta_0$ lies in the range $(0,4\pi)$ (and not too close to $0$ and $4\pi$), the two dominant branches are $k=0$ and $k=1$. Focusing on $\delta_0 \in (0,4\pi)$ and keeping the contributions of these two branches only, the combination of Eq.~\eqref{eq:GFgreater_gen_n} and Eq.~\eqref{eq:HartwigFisher2longfullmaintext} produces the following GFs in the long-time limit, $|\tau| \gg 1/|e_1^*V|$,
\begin{widetext}
\begin{align} \label{eq:GreenfunFullHartwig}
\mathcal{G}^{\gtrless}(\tau) &\simeq   \frac{\mp i}{2\pi a} \frac{a^{n_2^2\nu}}{ (a \pm i v\tau)^{n_2^2\nu}}  \exp [- \frac{\mathcal{T} |e_1^* V \tau| }{2\pi n_1^2}  (1- e^{- i\delta_0 \text{sgn}(e_1^* V \tau)}) ] \Big(\frac{1}{
|e_1^*V \tau|} \Big)^{-\frac{2n_2}{\pi n_1} \mathcal{T} \sin \left(\frac{\delta_0}{2}\right) \exp [i \frac{\delta_0}{2} \text{sgn}(e_1^* V \tau)]}
 \nonumber \\ 
 & \times 
  \Big [1+ \frac{\mathcal{T}}{n_1^2 \nu} e^{-i \delta_0 \text{sgn}(e_1^* V \tau)/2} \Big \{\frac{\delta_0}{2\pi^2} \sin \big(\frac{\delta_0}{2} \big) \Big(2-\psi (1- \frac{\delta_0}{2\pi}) - \psi (1+\frac{\delta_0}{2\pi}) \Big) - e^{-i e_1^*V \tau}  \frac{1}{\big(\Gamma(\frac{\delta_0}{2\pi})\big)^2} \Big( \frac{1}{|e_1^*V\tau|} \Big)^{2(1- \frac{\delta_0}{2\pi})}\Big \} \Big]\,.
\end{align}
\end{widetext}
Here, $\Gamma(z)$ is the gamma function and $\psi(z)=\Gamma'(z)/\Gamma(z)$ is the digamma function. Equation~\eqref{eq:GreenfunFullHartwig}, which improves the Szeg\H{o} approximation result~\eqref{eq:GF_Poisson_gen}, is one of the key results of this work.

It is instructive to benchmark the generalized Fisher-Hartwig conjecture by applying it to the free-fermion case, i.e., by taking $\delta_0 \rightarrow 2\pi$, $e_1^* = e$, $n_1= n_2 = \nu = 1$. In this limit, the GFs~\eqref{eq:GreenfunFullHartwig} reduce to~\cite{Supplemental_Material} 
\begin{align}
\label{eq:GFs_free_fermions}
    \mathcal{G}^{\gtrless}(\tau) &=   \frac{\mp i}{2\pi a} \frac{a}{ (a \pm i v\tau)} (1-\mathcal{T} + \mathcal{T} e^{-i e V \tau})\,.
\end{align}
This is the exact result for GFs of free fermions with a double-step electronic distribution function. 
While the generalized Fisher-Hartwig formula \eqref{eq:HartwigFisher2longfullmaintext} is in general of asymptotic (long-time) nature, in the case of $\delta_0=2\pi$ it becomes exact and is not limited to long times $\tau$. 
In this case, the contributions from all branches other than $k=0$ and $k=1$ are strictly zero. We also note that Eq.~\eqref{eq:GFs_free_fermions} is valid for any $\mathcal{T}$.

The free-fermion setup ($\nu=1$ edge) is not the only one for which the phase amplitude $\delta_0$ is equal to $2\pi$ and for which the Szeg\H{o} approximation misses completely the non-equilibrium physics, as discussed below Eq.~\eqref{eq:GF_Poisson_gen}.
Indeed, the $\delta_0=2\pi$ situation is realized also for Laughlin $\nu=1/m$ edges in two important cases: (i) $n_1=1$, $n_2=m$ and (ii) $n_1=m$, $n_2=1$. (This is a manifestation of the well-known fact that an electron and an anyon have trivial mutual braiding statistics.)
As pointed out above, the Fisher-Hartwig formula for the determinant becomes exact for $\delta_0=2\pi$, yielding 
\begin{align}
\label{eq:G_a_inj_e_tunn}
    \mathcal{G}^{\gtrless}(\tau)&= \frac{\mp i}{2\pi a} \left(\frac{a}{a\pm iv\tau}\right)^{m} \left[1-\mathcal{T} + \mathcal{T} e^{-i \frac{e V \tau}{m}}\right]^m
\end{align}
for $n_1=1$, $n_2=m$, and
\begin{align} 
    \label{eq:G_e_inj_a_tunn}
    \mathcal{G}^{\gtrless}(\tau)&=\frac{\mp i}{2\pi a} \left(\frac{a}{a\pm iv\tau}\right)^{\frac{1}{m}} \left[1-\mathcal{T} + \mathcal{T} e^{-ie V \tau}\right]^{\frac{1}{m}},
\end{align}
for $n_1=m$, $n_2=1$. For $m=1$, the expressions~\eqref{eq:G_a_inj_e_tunn}-\eqref{eq:G_e_inj_a_tunn} reduce to the free fermion result~\eqref{eq:GFs_free_fermions} (valid for any $\mathcal{T}$). For Laughlin edges, we have derived these formulas in the dilute limit $\mathcal{T}\ll 1$ assumed throughout the paper. It remains to be understood whether these formulas are applicable also beyond the dilute limit.

The results for GFs obtained in
Sec.~\ref{sec:GF-asymptotic} will be used in
Sec.~\ref{sec:tunneling} where we will study the current and noise for tunneling between two Laughlin edges, with one or both of them being out of equilibrium. 

\section{Tunneling between Laughlin edges}
\label{sec:tunneling}
\begin{figure}[t!]
    \centering
\includegraphics[width=0.9\columnwidth]{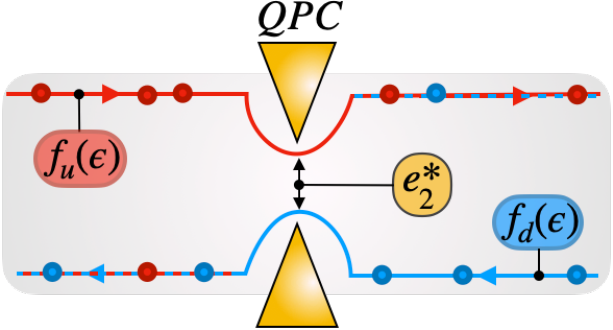}  
    \caption{
    Schematic representation of a setup for studying tunneling current and noise between non-equilibrium Laughlin edges, see Sec.~\ref{sec:tunneling}. Two $\nu = 1/m$ Laughlin edges, characterized by distributions $f_u(\epsilon)$ and $f_d(\epsilon)$ of quasiparticles of charge $e_1^*=n_1 \nu e$  
    [see Eq.~\eqref{n-epsilon-noneq-laughlin-ud}],
    are connected by a central QPC where quasiparticles of charge $e_2^* = n_2 \nu e$ can tunnel. }
    \label{fig:QPC}
\end{figure}

In this Section, we will use our formalism  to compute the tunneling current and noise across a quantum point contact (QPC) connecting two Laughlin edges, see  Fig.~\ref{fig:QPC}.

\subsection{General expressions for the tunneling current and noise}

We consider two $\nu=1/m$ Laughlin edges, labeled $u$ and $d$, with a single QPC connecting them. Non-equilibrium states of both edges are prepared by tunneling of quasiparticles of charge $e_1^* = n_1\nu e$, as discussed in Sec.~\ref{sec:GF_Laughlin}. They impinge then on the central QPC connecting the two edges $u$ and $d$, which is located at $x=x_{\rm qpc}$. The tunneling Hamiltonian $H_T$ and the tunneling current operator $I_T$ at this central QPC are given by
\begin{align} \label{eq:tunnelingdef}
    H_T = A^\dagger + A\,,\; I_T =  -i e_2^* (A-A^\dagger)\,,  \; A = \xi \psi_d^{\dagger} \psi_u^{}.
\end{align}
Here, the tunneling operators are of the form $\psi_{u,d}\sim \exp[in_2\phi_{u,d}]$ (where we leave the spatial argument $x=x_{\rm qpc}$ implicit for brevity), the tunneling charge is $e_2^*= n_2 \nu e$, and the tunneling amplitude is denoted by $\xi$. In the weak tunneling limit, i.e., keeping terms up to second order in $\xi$ (first-order terms vanish due to charge conservation), one can derive the expectation value of the tunneling current as
\begin{align} \label{eq:tunnelingcurrent}
    \langle I_T \rangle =  e_2^* \int_{-\infty}^{\infty} d\tau \langle [A^\dagger (0), A (\tau)] \rangle\,.
\end{align}
Importantly, this formula is not limited to equilibrium, i.e.,
the expectation value $\langle...\rangle$ here is generically taken over non-equilibrium states of the edges (characterized by non-equilibrium distribution functions). The average tunneling current~\eqref{eq:tunnelingcurrent} can further be expressed in terms of the generic GFs~\eqref{eq:GFgreater_gen_n} as
\begin{align} \label{eq:tunneling_current}
    \langle I_T \rangle &= e_2^*|\xi|^2 
    \int_{-\infty}^{\infty} d\tau \left[ \mathcal{G}_u^<(-\tau)\mathcal{G}_d^>(\tau) - \mathcal{G}_u^>(\tau)\mathcal{G}_d^<(-\tau) \right], 
\end{align}
where the indices $u$ and $d$ refer to the GFs for their respective edges. We also consider the symmetrized, zero-frequency tunneling noise $S_T$, defined as
\begin{align}
\label{eq:ST_def}
    &S_T \equiv \int_{-\infty}^{\infty} dt \langle \delta I_T(t)\delta I_T(0)+\delta I_T(0)\delta I_T(t)\rangle, \notag \\ &\delta I_T(t) \equiv I_T(t)-\langle I_T(t) \rangle,
\end{align}
with the tunneling current operator $I_T$ given in Eq.~\eqref{eq:tunnelingdef}. Evaluating~\eqref{eq:ST_def} to second order in $\xi$, one finds that it can be written in terms of the GFs as
\begin{align} \label{eq:tunnelin_noise}
   S_T&= 2(e_2^*)^2|\xi|^2 
    \int_{-\infty}^{\infty} d\tau \left[ \mathcal{G}_u^<(-\tau)\mathcal{G}_d^>(\tau) + \mathcal{G}_u^>(\tau)\mathcal{G}_d^<(-\tau) \right]. 
\end{align}
In the following subsections, we will use the results for GFs obtained within the  non-equilibrium bosonization theory
in Sec.~\ref{sec:GF_Laughlin}
to evaluate the tunneling current~\eqref{eq:tunneling_current} and noise~\eqref{eq:tunnelin_noise} for various non-equilibrium states of Laughlin edges injected towards the QPC.  

\subsection{Voltage bias}
\label{sec:tunneling_constant_bias}

As a first benchmark, we consider here the standard setup of a constant bias voltage $V_0\equiv V_u-V_d$ applied across the QPC. The normalized determinants in Eq.~\eqref{eq:GFgreater_gen_n} then evaluate exactly~\cite{Supplemental_Material} to
\begin{align}
\label{eq:det_constant_bias}
  \left\{\overline{\Delta}[\delta_{u,d}(t)]\right\}^\frac{1}{n_1^2\nu} = 
    \left[e^{-in_1\nu e (\delta_0/2\pi) V_{u,d}\tau}\right]^\frac{1}{n_1^2\nu}
    =
    e^{-ie_2^* V_{u,d} \tau},
\end{align}
where we used $\delta_0 = 2\pi n_1n_2 \nu$.
Using this result for the GFs in Eq.~\eqref{eq:tunneling_current} and integrating, we obtain
\begin{align} \label{eq:constantvoltagecurrent}
    \langle I_T \rangle = \frac{|e_2^*||\xi|^2 a^{2\zeta} \text{sgn} (V_0)}{2 \pi v^{2\zeta}  a^2} \frac{|e_2^*V_0|^{2 \zeta - 1}}{\Gamma (2 \zeta)}, 
\end{align}
where we have introduced the notation $\zeta=\nu n_2^2$, which is twice the scaling dimension of the operators $\psi_{u,d}\sim \exp[in_2\phi_{u,d}]$.  Likewise, using Eq.~\eqref{eq:det_constant_bias} in Eq.~\eqref{eq:tunnelin_noise} and integrating,  we find 
\begin{align} \label{eq:constantvoltagenoise}
   S_T =\frac{2 (e_2^*)^2|\xi|^2 a^{2\zeta} }{2 \pi v^{2\zeta} a^2} \frac{|e_2^*V_0|^{2 \zeta - 1}}{\Gamma (2 \zeta)}\,.
\end{align}
By comparing the tunneling current \eqref{eq:constantvoltagecurrent} and the tunneling noise \eqref{eq:constantvoltagenoise}, we further obtain the well-known zero-temperature, Poissonian shot-noise formula,
\begin{align}
\label{eq:Gen_Fano_app}
 S_T = 2 |e_2^*| \langle I_T \rangle  \text{sgn} (V_0),  
\end{align}
which can be used to extract the tunneling charge $e_2^*=n_2 \nu e$ via the Fano factor,
\begin{align}
\label{eq:Fano_def}
    F \equiv \frac{S_T}{2|e|\langle I_T\rangle}.
\end{align}
Equations~\eqref{eq:constantvoltagecurrent},~\eqref{eq:constantvoltagenoise}, and~\eqref{eq:Gen_Fano_app} thus show that our non-equilibrium formalism correctly reproduces basic results (see e.g., Ref.~\cite{Wen_Topological_1995}) for tunneling between Laughlin edges.

\subsection{Temperature bias}
As a second benchmark, we consider injecting two thermal equilibrium distributions, taken at temperatures $T_u$ and $T_d$, respectively. We choose also the voltage biases $V_u=V_d=0$. The determinants in Eq.~\eqref{eq:GFgreater_gen_n} then evaluate exactly~\cite{Supplemental_Material} to
\begin{align}
\label{eq:det_thermal_bias}
    \left\{\overline{\Delta}[\delta_{u,d}(t)]\right\}^\frac{1}{n_1^2\nu} =  \left(\frac{\pi T_{u,d} \tau}{\sinh(\pi T_{u,d} \tau)}\right)^{\zeta}.
\end{align}
In this case, we find zero tunneling current, $\langle I_T \rangle =0$, but finite tunneling noise
\begin{align} \label{eq:delta_T_noise}
   S_T&=\frac{4(e_2^*)^2|\xi|^2}{(2\pi a)^2} 
    \int_{-\infty}^{\infty} d\tau \left(\frac{a}{a+ iv\tau}\right)^{2\zeta} \notag \\
    &\times\left(\frac{\pi T_{u} \tau}{\sinh(\pi T_{u} \tau)}\right)^{\zeta}\left(\frac{\pi T_{d} \tau}{\sinh(\pi T_{d} \tau)}\right)^{\zeta},
\end{align}
which is the correct expression~\cite{Rech2020Aug} for so-called ``delta-$T$'' noise, i.e, non-equilibrium charge noise generated by a pure temperature bias, in the FQH regime. Analytical evaluation of the integral in Eq.~\eqref{eq:delta_T_noise} is generally not possible. However, it can be evaluated numerically, or analytically in certain regimes, e.g., the strong bias regime $T_u/T_d\ll 1$ or the weak bias regime $T_{u,d}\approx \overline{T}\pm \Delta T/2$, where $\overline{T}\equiv (T_u+T_d)/2$, $\Delta T\equiv (T_u-T_d)/2$, with $\Delta T\ll \overline{T}$. For further details on calculations in these regimes, we refer to Refs.~\cite{Rech2020Aug,Zhang2022May, Acciai2025Feb}.

\subsection{Tunneling between non-equilibrium edges}
\label{sec:tun-noneq-Laughlin-edges}

We now move on to the case when both edges  $u$ and $d$ (or at least one of them) are characterized by non-equilibrium distributions of the type \eqref{n-epsilon-noneq-laughlin} injected towards the QPC connecting the two edges:
\begin{eqnarray}
f_u(\epsilon) &=& \mathcal{T}_u \Theta(-\epsilon+e_1^* V_0 + e_1^* V_u) + (1-\mathcal{T}_u)\Theta(-\epsilon + e_1^* V_0), \nonumber \\
f_d(\epsilon) &=& \mathcal{T}_d \Theta(-\epsilon+e_1^* V_d) + (1-\mathcal{T}_d)\Theta(-\epsilon).
\label{n-epsilon-noneq-laughlin-ud}
\end{eqnarray}
Here, we assumed that the non-equilibrium states of the edges $u$ and $d$ are created by tunneling of quasiparticles with the same charge $e_1^* = n_1 \nu e$. It is straightforward  to generalize the analysis below to the case of different $n_{1,u}$ and $n_{1,d}$ for the two edges. For generality, we allowed in Eq.~\eqref{n-epsilon-noneq-laughlin-ud} for different parameters 
$(V_u, \mathcal{T}_u)$ and $(V_d, \mathcal{T}_d)$ characterizing the non-equilibrium states of the edges. We have also included a possible constant voltage $V_0$ between the edges. 

As above, we assume that the tunneling current between the edges $u$ and $d$ is carried by quasiparticles of charge $e_2^*= e\nu n_2$. We also assume dilute injections, $\mathcal{T}_{u,d}\ll 1$, corresponding to dilute beams of quasiparticles impinging on the QPC. For $n_1 =n_2 =1$ and $V_0=0$, this setup is known as the anyon collider geometry~\cite{Rosenow2016Apr,Bartolomei2020Apr}.

In the 
Szeg\H{o} approximation, the GFs of charge-$e_2^*$ quasiparticles on both edges are given by
\begin{align} 
 \label{eq:GF_Poisson_collider}
 \mathcal{G}_{u,d}^{\gtrless}(\tau)&\simeq  \frac{\mp i}{2\pi a} \left(\frac{a}{a\pm iv\tau}\right)^{\zeta} e^{-\gamma_{u,d}|\tau|}e^{-i\omega_{u,d}\tau},
\end{align}
with the scaling dimension $\zeta = n_2^2 \nu $, the dephasing rates 
\begin{equation}
\gamma_{u,d} = \frac{|e V_{u,d}|\mathcal{T}_{u,d}\left(1-\cos(2\theta_{12})\right)}{2\pi n_1},
\end{equation}
and the oscillation frequencies 
\begin{eqnarray}
\omega_{u} &=& \frac{e V_{u}\mathcal{T}_{u}\sin(2\theta_{12} )}{2\pi n_1} + e_2^* V_0, \nonumber \\
\omega_{d} &=& \frac{e V_{d}\mathcal{T}_{d}\sin(2\theta_{12} )}{2\pi n_1}.
\label{omega-u-d}
\end{eqnarray}
We recall that $2\theta_{12} = 2\pi n_1 n_2 \nu$ is the braiding angle of type-$n_1$ and type-$n_2$ Laughlin quasiparticles.

To calculate the tunneling current and the noise, we should substitute formulas for the GFs into
Eqs.~\eqref{eq:tunneling_current} and 
\eqref{eq:tunnelin_noise}. However, depending on the values of the scaling exponent $\zeta = n_2^2\nu$ and the exchange phase $2\theta_{12} = 2\pi n_1 n_2 \nu$, we must distinguish between several cases:

\begin{itemize}
\item[(i)] Substituting Eq.~\eqref{eq:GF_Poisson_collider} into
Eqs.~\eqref{eq:tunneling_current} and 
\eqref{eq:tunnelin_noise}, we observe that,
if the scaling dimension $\zeta$ satisfies \textcolor{red}{$\zeta < 1$}, the integral over $\tau$ is determined by long times, see Sec.~\ref{sec:transport-zeta-less-12} and the Supplemental Material \cite{Supplemental_Material} for more details. 
This means that the Szeg\H{o}  
approximation, which is of infrared (long-time) nature, is relevant. 
For a generic value of the braiding angle $2\theta_{12}$ (different from an integer multiple of $2\pi$), the Szeg\H{o}  approximation will give a parametrically dominant contribution to the tunneling current and noise for $\mathcal{T}_{u}, \mathcal{T}_{d} \ll 1$. 
This is because the $k=0$ branch of the Fisher-Hartwig formula \eqref{eq:HartwigFisher2longfullmaintext} [which is well approximated by the Szeg\H{o} formula \eqref{eq:GF_Poisson_collider}]
yields for the tunneling current and noise the results that increase when $\mathcal{T}_{u}, \mathcal{T}_{d} \to 0$, whereas contributions of other branches are proportional to $\mathcal{T}_{u}$ and $ \mathcal{T}_{d}$. 

\item[(ii)] If $\zeta > 1$, the time integral is governed by short times. For generic values of the phase $2\theta_{12}$, 
the Szeg\H{o} and Fisher-Hartwig approximations, which are of infrared character, are not applicable in this time range. Calculation of the tunneling current and of the noise thus requires an analysis of Toeplitz determinants entering the formulas for GFs at sufficiently short times. We will not address this problem in the present paper, leaving it for future studies. One can, however, extend the validity of the Szeg\H{o} and Fisher-Hartwig approximation by considering differential conductance and noise with respect to the constant voltage $V_0$. Indeed, taking a derivative of Eq.~\eqref{eq:GF_Poisson_collider} with respect to $V_0$ produces an additional factor of $\tau$, so that the 
integrals for the differential conductance and noise are of infrared character in the extended range
$\zeta < 3/2$.

\item[(iii)] For $2\theta_{12}$ equal to $2\pi$, we encounter a special situation. In this case, 
the Szeg\H{o} approximation misses completely the non-equilibrium character of the state, see Sec.~\ref{sec:szego}. Furthermore,
the Fisher-Hartwig formula contains then contributions of two branches only ($k=0$ and $k=1$) and is exact, as discussed in the end of Sec.~\ref{sec:fisher-hartwig}.  Thus, the current and noise can be calculated by using the Fisher-Hartwig formula for the GFs entering Eqs.~\eqref{eq:tunneling_current} and 
\eqref{eq:tunnelin_noise}, independently of the value of $\zeta$. This calculation can be extended to the case when $2\theta_{12} = 2 \pi j$  with $j=1, 2, \ldots$. The Fisher-Hartwig formula is exact also in this situation, with $j+1$ logarithm branches contributing.   
\end{itemize}
Below, we analyze all these three cases.

\subsection{Tunneling between non-equilibrium edges for $\zeta < 1$}
\label{sec:transport-zeta-less-12}

We proceed now by considering the case (i) from the classification in Sec.~\ref{sec:tun-noneq-Laughlin-edges}: $\zeta < 1$ and $2\theta_{12}$ different from an integer multiple of $2\pi$. 
Using the GFs~\eqref{eq:GF_Poisson_collider} in the tunneling current expression~\eqref{eq:tunneling_current}, we obtain the integral
\begin{align}
\label{eq:IT_integral}
    \langle I_T \rangle &= \frac{2ie_2^*|\xi|^2}{(2\pi a)^2} 
    \int_{-\infty}^{\infty} d\tau  \left(\frac{a}{a+ iv\tau}\right)^{2\zeta} \notag \\ \times &e^{-(\gamma_u+\gamma_d)|\tau|}\sin(\tau(\omega_u-\omega_d)).
\end{align}
Using now the assumption $\zeta<1$, we can safely take the limit $a\to 0$ in the integrand. We then find (see the Supplemental Material \cite{Supplemental_Material})
\begin{align}
\label{eq:current_dilute_app}
   \langle I_T\rangle & =  \frac{2ie_2^*|\xi|^2}{(2\pi a)^2} 
    \int_{-\infty}^{\infty} d\tau  \left(\frac{a}{a+ iv\tau}\right)^{2\zeta} \notag \\&\times e^{-(\gamma_u+\gamma_d)|\tau|}\sin(\tau(\omega_u-\omega_d))
   \notag \\ & \simeq \frac{e_2^* a^{2\zeta} |\xi|^2}{2(\pi a)^2 v^{2\zeta}} \frac{\pi}{\Gamma(2 \zeta) \cos(\pi\zeta)} \notag \\ & \times \text{Im}\big[(-i (\omega_u-\omega_d) + (\gamma_u+\gamma_d))^{2\zeta - 1} \big]  \notag \\ &=\frac{e_2^* a^{2\zeta} |\xi|^2}{2(\pi a)^2 v^{2\zeta}} \frac{\pi}{\Gamma(2 \zeta) \cos(\pi \zeta)} \notag \\& \times \frac{\sin \big[(1-2\zeta)\tan^{-1} (p)    \big]}{(|\omega_u-\omega_d|^2 + (\gamma_u+\gamma_d)^2)^{1/2-\zeta}}\,,
\end{align}
where we introduced the dimensionless bias parameter
\begin{equation}
\label{eq:bias_parameter}
p =\frac{\omega_u - \omega_d}{\gamma_u + \gamma_d} \,.
\end{equation}

Likewise, the tunneling noise~\eqref{eq:tunnelin_noise} becomes
\begin{align}
\label{eq:noise_dilute_app}
    S_T &= \frac{4(e_2^*)^2|\xi|^2}{(2\pi a)^2}
    \int_{-\infty}^{\infty} d\tau  \left(\frac{a}{a+ iv\tau}\right)^{2\zeta} \notag \\ &\times e^{-(\gamma_u+\gamma_d)|\tau|}\cos(\tau(\omega_u-\omega_d))\notag \\ 
    & \simeq \frac{(e_2^*)^2 |\xi|^2}{(\pi a)^2v^{2\zeta}}\frac{\pi  a^{2 \zeta}}{\Gamma(2 \zeta) \sin(\pi  \zeta)} \notag \\ &\times\text{Re}\big[(-i (\omega_u-\omega_d) + (\gamma_u+\gamma_d))^{2\zeta - 1} \big] \notag \\
    &= \frac{(e_2^*)^2 |\xi|^2}{(\pi a)^2v^{2\zeta}} \frac{\pi  a^{2\zeta}}{\Gamma(2 \zeta) \sin(\pi \zeta)} \notag \\ &\times \frac{\cos \big[(1-2\zeta) \tan^{-1} (p) \big]}{(|\omega_u-\omega_d|^2 + (\gamma_u+\gamma_d)^2)^{1/2-\zeta}}.
\end{align}

Using Eqs.~\eqref{eq:current_dilute_app} and \eqref{eq:noise_dilute_app}, we find that the Fano factor~\eqref{eq:Fano_def} evaluates to 
\begin{equation}
\label{eq:gen_fano}
F = \frac{e_2^*}{|e|}\cot (\pi\zeta) \cot \big[(1-2\zeta) \tan^{-1} (p) \big].
\end{equation}

The Fano factor~\eqref{eq:gen_fano} thus depends on two dimensionless parameters: the scaling dimension $\zeta=n_2^2\nu$ and the bias parameter $p$, Eq.~\eqref{eq:bias_parameter}, of which the latter in particular contains information about
the mutual braiding angle $2\theta_{12}=2\pi \nu n_1 n_2$ of type-$n_1$ and type-$n_2$ quasiparticles. 

We now analyze the Fano factor in several limiting regimes. First,  we consider a sufficiently strong constant bias, $|V_0|\gg \mathcal{T}_u |V_u|,\: \mathcal{T}_d |V_d|$. Then $\tan^{-1}(p)\to  \text{sgn}(e^*_2V_0) \pi /2$, which yields for the Fano factor~\eqref{eq:gen_fano} the limiting value $F\to \nu n_2 \text{sgn}(V_0)$. This corresponds  to tunneling of charge $e_2^*=n_2\nu e$ quasiparticles under a constant bias $V_0$. We thus recover the result from Sec.~\ref{sec:tunneling_constant_bias}. 

Second, we consider the case of $V_0 = 0$. Then, the parameter $p$ in Eq.~\eqref{eq:bias_parameter} reduces to
\begin{equation}
p = \frac{e V_{u}\mathcal{T}_{u} - e V_{d}\mathcal{T}_{d}}{|e V_{u}\mathcal{T}_{u}| + |e V_{d}\mathcal{T}_{d}|} \cot (\theta_{12}) = \frac{\langle I_u \rangle-\langle I_d \rangle}{|\langle I_u \rangle| + |\langle I_d \rangle|} \cot (\theta_{12}),
\end{equation}
where in the last equality, we identified the injected average currents $\langle I_{u,d} \rangle=\nu e^2\mathcal{T}_{u,d}V_{u,d}/(2\pi)$, cf. Eq.~\eqref{eq:average_I}.
A further simplification happens if (in addition to $V_0=0$) one of the voltages $V_u$, $V_d$ is zero or if they have opposite signs, $\text{sgn}(eV_u) = - \text{sgn}(eV_d)$. The former case, in which one of voltages vanishes, was experimentally realized in Ref.~\cite{Lee2023May}.
We get then $p = \text{sgn}(eV_u) \cot (\theta_{12})$, so that the Fano factor~\eqref{eq:gen_fano} simplifies to
\begin{align}
\label{eq:Fano_single_bias}
    F &= \nu n_2\cot(\pi \zeta)\cot\big[(1-2\zeta)(\frac{\pi}{2}-\tilde{\theta}_{12})]\text{sgn}(V_u),\notag \\
    \tilde{\theta}_{12} &=\theta_{12} \text{ mod } \pi \,,   \qquad 0 <  \tilde{\theta}_{12} < \pi \,.
\end{align}
For the case $V_d=0$ and $n_1=n_2=1$, this result was reported in Ref.~\cite{Lee2023May}. 

Finally, we note that for $\omega_u=\omega_d$, the bias parameter $p=0$ and the Fano factor \eqref{eq:gen_fano} diverges.  This is due to the fact that the tunneling current 
$\langle I_T\rangle$, Eq.~\eqref{eq:current_dilute_app},
vanishes for $\omega_u=\omega_d$ (i.e., $p=0$) while the tunneling noise \eqref{eq:noise_dilute_app} remains finite.
The condition $\omega_u=\omega_d$ can be straightforwardly resolved for varying $V_0$ (for other parameters fixed) by using 
Eqs.~\eqref{omega-u-d}.
 
We return now to the conditions of applicability of Eqs.~\eqref{eq:current_dilute_app}, \eqref{eq:noise_dilute_app}, 
\eqref{eq:gen_fano}, and \eqref{eq:Fano_single_bias}.
The results for the integrals presented in 
Eqs.~\eqref{eq:current_dilute_app} and \eqref{eq:noise_dilute_app} are leading terms that scale with small injected currents $\langle I_{u}\rangle$ and $\langle I_{d}\rangle$ as $\langle I_T\rangle, S_T \propto I_{\rm max}^{2\zeta-1}$, where $I_{\rm max} = \text{max}\{|\langle I_{u}\rangle|, |\langle I_{d}\rangle| \}$ and we set $V_0 = 0$. The leading correction to this result scales as a first power of $I_{\rm max}$, see the Supplemental Material~\cite{Supplemental_Material}, so that its ratio to the leading term is $\sim (a I_{\rm max} / e_1^* v)^{2-2\zeta}$. This shows that the condition of applicability with respect to the exponent $\zeta$ is $\zeta < 1$. This condition can be deduced also by inspection of the integrals in Eqs.~\eqref{eq:current_dilute_app} and \eqref{eq:noise_dilute_app}. Indeed, the results as given in these formulas are governed by long times, $\tau \sim e_1^*/I_{\rm max}$. For times $\tau$ much shorter than this scale, the sine function in Eq.~\eqref{eq:current_dilute_app} can be expanded in $\tau$, so that the integral  is of the type $\int d\tau / \tau^{2\zeta-1}$ and is thus governed by long times for $\zeta < 1$.   In Eq.~\eqref{eq:noise_dilute_app}, 
the factor $e^{-(\gamma_u+\gamma_d)|\tau|}\cos(\tau(\omega_u-\omega_d))$ is equal to unity to leading order in $\tau$ at short times, which yields the equilibrium noise (equal to zero at zero temperature). The leading correction (which reflects the non-equilibrium conditions) is of first order in $\tau$, resulting again in the condition $\zeta < 1$.

It is worth recalling that, in addition to the conditions $\zeta < 1$ and $2\theta_{12} \ne 2\pi j$, we also used here an assumption of dilute beams, $\mathcal{T}_{u,d}\ll 1$. This latter assumption has allowed us to neglect corrections to the Szeg\H{o} approximation for the GFs, which are represented by the 
last factor in the first line and by the factor in the second line of the Fisher-Hartwig formula~\eqref{eq:GreenfunFullHartwig}. These factors lead to contributions to the tunneling current and noise that have relative smallness $ \sim \mathcal{T}_{u,d}^{2-2\zeta}\ll 1$ as compared to the leading expressions \eqref{eq:current_dilute_app}  and \eqref{eq:noise_dilute_app}.
We see that the two conditions, $\mathcal{T}_{u,d}\ll 1$ and $\zeta < 1$, indeed allow us to neglect the corrections to the Szeg\H{o} approximations. 
 
Thus,
the conditions of applicability of the results for tunneling current and noise obtained within the Szeg\H{o} approximation in this section are (in addition to $\mathcal{T}_{u,d}\ll 1$)
$\zeta < 1$ and $2\theta_{12} \ne 2\pi j$.
 It is worth mentioning that these conditions are at variance with those stated in Ref.~\cite{Thamm2024effect} for the dominance of long-time contributions to the transport integrals. Specifically, the 
authors of Ref.~\cite{Thamm2024effect} argued that the conditions are $\zeta < 1/2$ and $\theta_{12} < \pi/2$. Our condition on $\zeta$ is weaker, and for $\theta_{12}$ we don't find any restriction apart from it not being an integer multiple of $\pi$. We also note that our condition for $\zeta$ is consistent with Ref.~\cite{Schiller2023anyon} where transport at elevated temperatures was analyzed.

\subsection{Tunneling between non-equilibrium edges for $\zeta < 3/2$: \ Differential conductance and noise}
\label{sec:diff_FF}

We recall that the calculations in Sec.~\ref{sec:transport-zeta-less-12} were restricted to $\zeta <1$. When $\zeta>1$, the time integrals for the tunneling current and the noise are dominated by short-time range that is not captured by the Szeg\H{o} and Fisher-Hartwig approximations. To extend our analysis to the broader regime $\zeta < 3/2$, we therefore want to consider transport quantities that are insensitive to the ultraviolet features in this regime. For this purpose, we introduce a differential Fano factor, defined as \footnote{We note that a quantity related to (but different from) the differential Fano factor~\eqref{eq:Fano_diff_def} was introduced in Ref.~\cite{Zhang2025Mar}, under the name of ``witness function'' and with different motivations, for the Laughlin-edge tunneling setup with $n_1=n_2=1$.} 
\begin{align}
\label{eq:Fano_diff_def}
    F_d \equiv \frac{\partial_{V_0}S_T}{2|e|\partial_{V_0}\langle I_T \rangle}.
\end{align}
Performing the derivatives of the tunneling current and noise, we find 
\begin{align}
\label{eq:current_dilute_app_deriv}
   \partial_{V_0}\langle I_T\rangle & =  \frac{2i(e_2^*)^2|\xi|^2}{(2\pi a)^2} 
    \int_{-\infty}^{\infty} d\tau  \left(\frac{a}{a+ iv\tau}\right)^{2\zeta}\tau \notag \\&\times e^{-(\gamma_u+\gamma_d)|\tau|}\cos(\tau(\omega_u-\omega_d))
   \notag \\ & \simeq \frac{(e_2^*)^2|\xi|^2}{2\pi^2 a v} 
    \int_{-\infty}^{\infty} d\tau  \left(\frac{a}{a+ iv\tau}\right)^{2\zeta-1} \notag \\&\times e^{-(\gamma_u+\gamma_d)|\tau|}\cos(\tau(\omega_u-\omega_d))
     \notag \\ &= - \frac{(e_2^*)^2 a^{2\zeta} |\xi|^2}{2(\pi a )^2 v^{2\zeta}} \frac{\pi}{\Gamma(2 \zeta-1) \cos(\pi \zeta)} \notag \\& \times \frac{\cos \big[(2-2\zeta)\tan^{-1} (p)    \big]}{(|\omega_u-\omega_d|^2 + (\gamma_u+\gamma_d)^2)^{1-\zeta}}\,,
\end{align}
and
\begin{align}
\label{eq:noise_dilute_app_deriv}
     \partial_{V_0}S_T &= -\frac{4(e_2^*)^3|\xi|^2}{(2\pi a)^2}
    \int_{-\infty}^{\infty} d\tau  \left(\frac{a}{a+ iv\tau}\right)^{2\zeta}\tau \notag \\ &\times e^{-(\gamma_u+\gamma_d)|\tau|}\sin(\tau(\omega_u-\omega_d))\notag \\ 
   &\simeq \frac{i(e_2^*)^3|\xi|^2}{\pi^2 a v} 
    \int_{-\infty}^{\infty} d\tau  \left(\frac{a}{a+ iv\tau}\right)^{2\zeta-1} \notag \\&\times e^{-(\gamma_u+\gamma_d)|\tau|}\sin(\tau(\omega_u-\omega_d))
     \notag \\ &= \frac{(e_2^*)^3 a^{2\zeta} |\xi|^2}{(\pi a)^2 v^{2\zeta}} \frac{\pi}{\Gamma(2 \zeta-1) \sin(\pi \zeta)} \notag \\& \times \frac{\sin \big[(2-2\zeta)\tan^{-1} (p)    \big]}{(|\omega_u-\omega_d|^2 + (\gamma_u+\gamma_d)^2)^{1-\zeta}}\,.
\end{align}
Importantly, the derivatives with respect to $V_0$ produce additional factors $\tau$ in the integrands. As a result, the exponents of the equilibrium contributions change from $2\zeta$ to $2\zeta - 1$, rendering the integrals 
dominated by the long-time behavior
for $\zeta <3/2$. Substituting then Eq.~\eqref{eq:current_dilute_app_deriv} and~\eqref{eq:noise_dilute_app_deriv} into Eq.~\eqref{eq:Fano_diff_def}, the differential Fano factor becomes
\begin{align}
    \label{eq:Fano_diff_eval}
    F_d = \frac{e^*_2}{|e|} \cot (\pi \zeta) \tan \big[(2-2\zeta) \tan^{-1} (p) \big]\,. 
\end{align}
We now first analyze Eq.~\eqref{eq:Fano_diff_eval} in the case $|V_0|\gg \mathcal{T}_u |V_u|, \mathcal{T}_d |V_d|$. We find
\begin{align}
\label{eq:Fano_diff_largevoltage}
    F_d = \nu n_2\, \text{sgn}(V_0), 
\end{align}
which is the same result one obtains for the Fano factor for the constant bias setup, Sec.~\ref{sec:tunneling_constant_bias}. 
Moving on to the case $V_0 = 0$ and $\text{sgn}(eV_u) = - \text{sgn}(eV_d)$, we find 
\begin{align}
\label{eq:Fano_diff_balance}
    F_d &=  \nu n_2\cot(\pi \zeta)\tan\big[(2-2\zeta)(\frac{\pi}{2}-\tilde{\theta}_{12})]\text{sgn}(V_u), \notag \\
    \tilde{\theta}_{12} &=\theta_{12} \text{ mod } \pi\,,   \qquad 0 <  \tilde{\theta}_{12} < \pi \,,
\end{align}
which is the differential analog  of Eq.~\eqref{eq:Fano_single_bias}. 

\subsection{Tunneling between non-equilibrium edges for $2\theta_{12}=2\pi$}
\label{sec:Andreev_setups}
We consider now the tunneling between two non-equilibrium Laughlin edges, or between a non-equilibrium and an equilibrium edge, for the case where the mutual braiding angle $2\theta_{12}$ equal to $2\pi$. 
As discussed in the
end of Sec.~\ref{sec:fisher-hartwig}, the Szeg\H{o} approximation is then not sufficient, as it misses completely the non-equilibrium physics. This property explains difficulties with treating this case via approaches of Refs.~\cite{Levkivskyi2016Apr, Rosenow2016Apr,Schiller2023anyon,Varada2025May}. At the same time, our generic formalism remains applicable.

Since the Szeg\H{o} approximation is insufficient in the case $2\theta_{12}$ equals $2\pi$, we have to use the Fisher-Hartwig formula, independently of the value of the scaling dimension $\zeta$. This situation is the case (iii) from the classification in Sec.~\ref{sec:tun-noneq-Laughlin-edges}.
 Importantly, as was also emphasized in Sec.~\ref{sec:fisher-hartwig}, the Fisher-Hartwig formula for the Toeplitz determinant becomes exact for the phase $\delta_0 = 2\theta_{12} = 2\pi$, with only two logarithm branches ($k=0$ and $k=1$) contributing. 

When $m = \nu^{-1}$ is a prime number (which is the case for the experimentally relevant Laughlin edges at fillings $\nu=1/3$, 1/5, and 1/7), there are only two possibilities to have $2\theta_{12} \equiv  2\pi n_1 n_2 \nu = 2\pi$: either $n_1 = 1$ and $n_2 = m$, or, alternatively, $n_1 = m$ and $n_2 = 1$. We consider now both these settings. As in Sec.~\ref{sec:tun-noneq-Laughlin-edges}, both edges $u$ and $d$ can be in general in non-equilibrium states characterized by the voltages $V_{u,d}$ and the dilution parameters  $\mathcal{T}_{u,d}\ll 1$. 

We begin by considering the case $n_1 = 1$ and $n_2 = m$, i.e., a setup where dilute beams of charge-$e/m$ anyons on $\nu=1/m$ Laughlin edges are sent towards a central QPC set in the strong back-scattering regime, favoring electron tunneling. The GFs on the edges are then of the type given in Eq.~\eqref{eq:G_a_inj_e_tunn} and the the scaling dimension is $\zeta=m^2\nu=m$. Let us assume first that the edge $u$ is out of equilibrium, while the edge $d$ is in equilibrium, i.e., $V_d=0$ (see Ref.~\cite{Glidic2023Jan} for a recent experimental implementation of this setup). Then, using $\mathcal{T}_u\ll 1$, we have 
\begin{align} \label{eq:GFgreater_Andreev_u}
 \mathcal{G}_u^{\gtrless}(\tau)&\approx \frac{\mp i}{2\pi a} \left(\frac{a}{a\pm iv\tau}\right)^{m} \left[1+m\mathcal{T}_u(e^{-i\frac{eV_u\tau}{m}}-1)\right],\\
 \label{eq:GFgreater_Andreev_d}
 \mathcal{G}_d^{\gtrless}(\tau)&=  \frac{\mp i}{2\pi a} \left(\frac{a}{a\pm iv\tau}\right)^{m}.
\end{align}
Inserting these GFs into the equations for the tunneling current~\eqref{eq:tunneling_current} and noise~\eqref{eq:tunnelin_noise}, we find
\begin{align} \label{eq:constantvoltagecurrent_Andreev}
    &\langle I_T \rangle = \frac{|e| m\mathcal{T}_u|\xi|^2 a^{2m} \text{sgn} (V_u)}{2 \pi v^{2m} a^2} \frac{|eV_u/m|^{2 m - 1}}{\Gamma (2 m)},\\
    \label{eq:constantvoltagenoise_Andreev}
   &S_T =\frac{2 e^2 m \mathcal{T}_u|\xi|^2 a^{2m} }{2 \pi v^{2m}  a^2} \frac{|eV_u/m|^{2 m - 1}}{\Gamma (2 m)}.
\end{align}
Generalization to a non-zero $V_d$ is straightforward: this will give additive contributions in the current and noise that are obtained from  Eqs.~\eqref{eq:constantvoltagecurrent_Andreev} and \eqref{eq:constantvoltagenoise_Andreev}
by substitution $V_u \mapsto - V_d$ and 
$\mathcal{T}_u \mapsto \mathcal{T}_d$. 
It is also straightforward to include a constant voltage $V_0$ as in Eq.~\eqref{n-epsilon-noneq-laughlin-ud}: it yields additive contributions to the current and noise as calculated in Sec.~\ref{sec:tunneling_constant_bias}.
Since $V_0$ does not yield here any additional physics, we set $V_0=0$ as follows. 

Using Eqs.~\eqref{eq:constantvoltagecurrent_Andreev} and \eqref{eq:constantvoltagenoise_Andreev}, we find the Fano factor~\eqref{eq:Fano_def}  for the case $V_d=0$, 
\begin{align}
    F =\frac{S_T}{2|e|\langle I_T\rangle} =\text{sgn}(V_u)\,.
    \label{eq:Fano-n1-1-n2-m}
\end{align}
This is the same value as obtained for charge-$e$ (electron) tunneling between equilibrium edges with a constant bias voltage, cf. Sec.~\ref{sec:tunneling_constant_bias}. Thus, the non-equilibrium character of the Laughlin edge does not affect the Fano factor for the case of $2\theta_{12} = 2\pi$, i.e., due to the trivial braiding between electrons and minimal anyons.  We also find that $|F|=1$ holds also for non-zero $V_d$ if $\text{sgn}{V_d} = - \text{sgn}{V_u}$.

It is worth recalling that $\mathcal{T}_u$ in Eqs.~\eqref{eq:constantvoltagecurrent_Andreev} and \eqref{eq:constantvoltagenoise_Andreev}
should be understood as the actual tunneling probability, i.e., the one that includes the Luttinger-liquid renormalization from the ultraviolet energy scale $\Lambda=v/a$ to the scale set by the voltage $V_u$, see the comment below
Eq.~\eqref{n-epsilon-noneq-laughlin}. In the present case, this renormalization has the form $\mathcal{T}_u \sim \mathcal{T}^{(0)}_u |eV_u/\Lambda|^{2/m-2}$, where $\mathcal{T}^{(0)}_u$ is the bare (ultraviolet) value. We emphasize that this does not affect the Fano factor~\eqref{eq:Fano-n1-1-n2-m}.

We next investigate the ``dual'' setup, 
with $n_1=m$ and $n_2=1$, producing the scaling dimension $\zeta=1/m$. In this case, dilute beams of electrons on $\nu=1/m$ Laughlin edges impinge on a QPC set in the weak back-scattering regime, favoring anyon (charge-$e/m$) tunneling.  The local GFs at the QPC are then of the form~\eqref{eq:G_e_inj_a_tunn}. As before, let us first take the $u$ edge to be out of equilibrium while keeping the $d$ edge in equilibrium.  With $\mathcal{T}_u\ll1$, we then have
\begin{align}
   \mathcal{G}_u^{\gtrless}(\tau)&\approx  \frac{\mp i}{2\pi a} \left(\frac{a}{a\pm iv\tau}\right)^{\frac{1}{m}} \left[1+\frac{\mathcal{T}_u}{m}(e^{-ieV_u\tau}-1)\right],\\
   \mathcal{G}_d^{\gtrless}(\tau)&=  \frac{\mp i}{2\pi a} \left(\frac{a}{a\pm iv\tau}\right)^{\frac{1}{m}}.
\end{align}
These GFs produce the tunneling current~\eqref{eq:tunneling_current} and noise~\eqref{eq:tunnelin_noise}
\begin{align}
\label{eq:tunneling_current_anyons}
    &\langle I_T \rangle = \frac{|e|\mathcal{T}_u|\xi|^2 a^{\frac{2}{m}} \text{sgn} (V_u)}{2 \pi m^2 v^{\frac{2}{m}} a^2} \frac{|eV_u|^{\frac{2}{m}  - 1}}{\Gamma (2/m)},\\ 
    \label{eq:tunneling_noise_anyons}
   &S_T =\frac{2 e^2 \mathcal{T}_u|\xi|^2 a^{\frac{2}{m}} }{2  \pi m^3v^{\frac{2}{m}}  a^2} \frac{|eV_u|^{\frac{2}{m} - 1}}{\Gamma (2/m)},
   \end{align}
 and thus the Fano factor  
\begin{align}
   \label{eq:Fano_anyons}
    &F = \frac{S_T}{2|e|\langle I_T\rangle} =\frac{1}{m} \text{sgn}(V_u).
\end{align}
As before, $\mathcal{T}_u$ in 
Eqs.~\eqref{eq:tunneling_current_anyons}-\eqref{eq:tunneling_noise_anyons}
is the actual (renormalized tunneling probability), which in this case is related to its bare (ultraviolet) value via
 $\mathcal{T}_u \sim \mathcal{T}^{(0)}_u |eV_u/\Lambda|^{2m-2}$.  The Fano factor 
 \eqref{eq:Fano_anyons} has the same value as for charge-$e/m$ anyon tunneling between equilibrium edges with a constant bias voltage, cf. Sec.~\ref{sec:tunneling_constant_bias}. We see again that the non-equilibrium character of the Laughlin edge does not affect the Fano factor for the trivial braiding  $2\theta_{12} = 2\pi$. An extension to the situation with non-zero $V_d$ is again straightforward, see the comment below Eqs.~\eqref{eq:constantvoltagecurrent_Andreev} and \eqref{eq:constantvoltagenoise_Andreev}. The result~\eqref{eq:Fano_anyons} holds also for non-zero $V_d$ if $\text{sgn}(eV_d) = - \text{sgn}(eV_u)$.

\begin{figure}[t!]
    \centering
    \subfloat[][]{
\includegraphics[width=0.9\columnwidth]{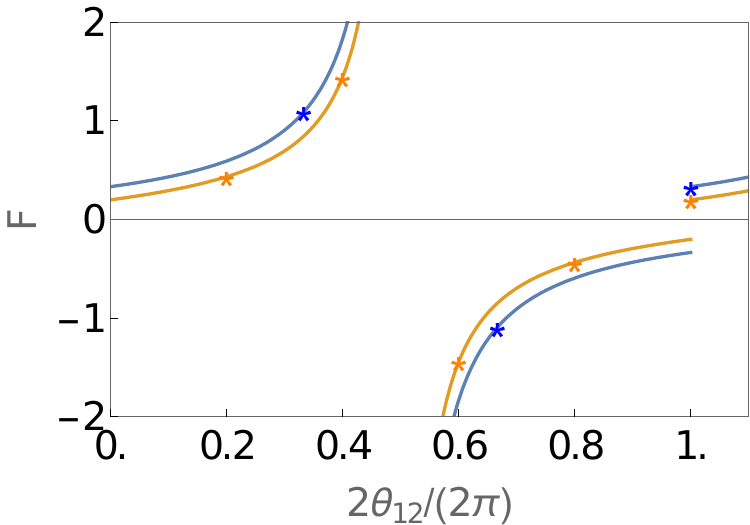} 
\put(-220,140){\large\textbf{(a)}}
}
\par\vspace{-3mm}
\subfloat[][]{
  \includegraphics[width=0.9\columnwidth]{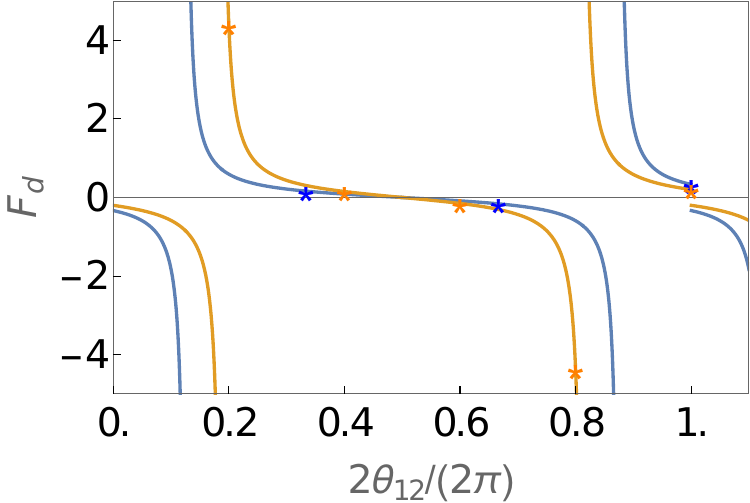} 
  \put(-220,140){\large\textbf{(b)}}
}
    \caption{{\bf (a) }Fano factor $F$ and {\bf (b) }differential Fano factor $F_d$ for QPC tunneling transport between two Laughlin edges, see Fig.~\ref{fig:QPC}, with $\nu=1/3$ (blue) and $\nu=1/5$ (orange),  as functions of the mutual braiding phase $2\theta_{12}$ (normalized by $2\pi$). The edge $u$ is driven out of equilibrium by quasiparticle injection with charge $e_1^*=n_1\nu e$ under a bias $V_u>0$, while the edge $d$ is at zero-temperature equilibrium. The offset voltage $V_0$ is set to zero, $V_0 = 0$. The tunneling quasiparticle at the central QPC is taken to be elementary ($n_2=1$), so that the scaling dimension $\zeta=\nu n_2^2<1/2$. The solid lines represent Eq.~\eqref{eq:Fano_single_bias} in panel {\bf (a)}  and Eq.~\eqref{eq:Fano_diff_balance} in panel {\bf (b)}, with
physical values (integer $n_1$) marked by stars. For the special point $2\theta_{12} = 2\pi$, the stars represent Eq.~\eqref{eq:Fano_anyons} in both panels. }
\label{fig:Fanovsbraidingangles}
\end{figure}
\subsection{Fano factors vs the braiding phase $2\theta_{12}$}

We now highlight important features of the Fano factor $F$ and the differential Fano factor $F_d$ for Laughlin edges. Figure~\ref{fig:Fanovsbraidingangles} illustrates the dependence of $F$ and $F_d$ in a tunneling QPC setup, see Fig.~\ref{fig:QPC}, on the mutual braiding phase 
$2\theta_{12}$ between injected and tunneling quasiparticles, at fillings 
$\nu=1/3$ and $\nu = 1/5$. The shown data correspond to a setup with the edge $u$ driven out of equilibrium by dilute quasiparticle injection and the edge $d$ being in equilibrium ($V_u>0$, $V_d=0$, and $V_0=0$). The tunneling at the central QPC is assumed to involve the ``minimal'' quasiparticles ($n_2=1$), so that the scaling dimension  satisfies $\zeta<1/2$, justifying our analysis of both $F$ and $F_d$.  The lines represent Eq.~\eqref{eq:Fano_single_bias} for $F$ and Eq.~\eqref{eq:Fano_diff_balance} for $F_d$, and the star symbols mark the physically relevant braiding angles corresponding to integer $n_1$. For the special point $2\theta_{12} = 2\pi$, the stars correspond to Eq.~\eqref{eq:Fano_anyons}.

Let us discuss key features of the behavior of $F$  as shown in Fig.~\ref{fig:Fanovsbraidingangles}\textcolor{blue}{(a)}. First, $F$ is $2\pi$-periodic in the braiding phase $2\theta_{12}$. Second, it diverges at $2\theta_{12} = \pi$, where the dimensionless parameter $p$, Eq.~\eqref{eq:bias_parameter}, and the tunneling current, Eq.~\eqref{eq:current_dilute_app}, vanish. Across this point, $F$ changes sign: it is positive for $0<2\theta_{12}<\pi$ and negative for $\pi<2\theta_{12}<2\pi$. The negative values of $F$ (and the tunneling current) can be traced back to negative values of $\omega_u$, Eq.~\eqref{omega-u-d}.
  
Furthermore, $F$ exhibits another discontinuity at the special point
 $2\theta_{12} = 2\pi$ (corresponding to $n_1 =\nu^{-1}=m$).
 Interestingly, the true value of the Fano factor at this point, $F = 1/m$, see Eq.~\eqref{eq:Fano_anyons},
matches the limit $2\theta_{12} \rightarrow 2\pi +0^+$ of Eq.~\eqref{eq:Fano_single_bias}. 

The differential Fano factor $F_d$, Fig.~\ref{fig:Fanovsbraidingangles}\textcolor{blue}{(b)}, also exhibits a strong (and $2\pi$-periodic) dependence on the braiding phase $2\theta_{12}$. 
It is worth recalling that $F$ and $F_d$, as shown in the figure, are calculated (apart from the special point $2\theta_{12} = 2\pi$) in the 
Szeg\H{o} approximation, with corrections being relatively small due to the dilution parameter $\mathcal{T}_u\ll 1$.

This completes our analysis of non-equilibrium Laughlin-edge setups. In Sec.~\ref{sec:multiple-mode-edges}, we will generalize the theory to include non-equilibrium FQH edges with two (or more) modes and the transport between such edges.
 
\section{Non-equilibrium bosonization of complex fractional quantum Hall edges}
\label{sec:multiple-mode-edges}

\subsection{K-matrix theory and setup }
\label{sec:K-matrx-setup}
We now extend the non-equilibrium theory of the single-mode (Laughlin) edge developed above to more complex edge structures with more than two modes. For simplicity, we will focus on edges with two modes; a generalization to generic Abelian FQH edges is conceptually straightforward. We will distinguish two classes of two-mode edges: those with co-propagating and those with counter-propagating edge modes. 

A two-mode FQH edge (which can be intrinsic or engineered) can be described by the pair~\cite{Wen_Topological_1995}
\begin{align}
\label{eq:KMat2x2}
    K = \begin{pmatrix}
    \nu_1^{-1} & 0\\
    0 & \nu_2^{-1}
    \end{pmatrix}, \quad \vec{q}=\begin{pmatrix}
        1\\1
    \end{pmatrix},
\end{align}
called the $K$-matrix and the charge vector, respectively. In the $K$-matrix, $\nu_{1,2}^{-1}$ are integers describing the filling factor discontinuities corresponding to each of the modes.  The charge vector $\vec{q}$ sets the fundamental charge in the theory. The two mode chiralities are given as $\eta_{1,2}=\text{sgn}(\nu_{1,2})\in \left\{+1,-1\right\}$. Equal (opposite) chiralities correspond to co-propagating (counter-propagating) modes. The particular choice of basis in Eq.~\eqref{eq:KMat2x2} is not unique, but all physical observables are independent of this choice, see, e.g, Ref.~\cite{Wen_Topological_1995}.

\begin{figure}[t!]
    \centering
\includegraphics[width=\columnwidth]{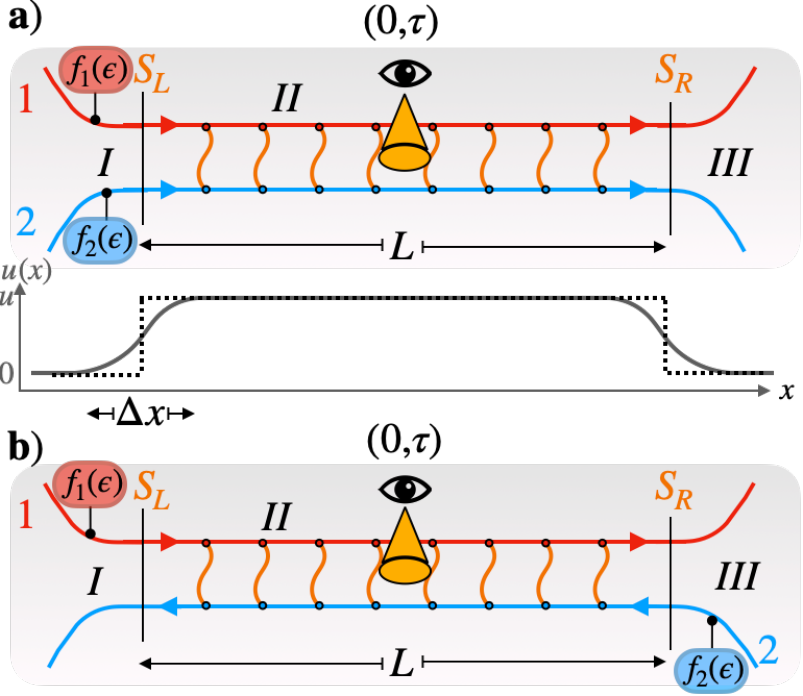}  
    \caption{Setups with non-equilibrium complex FQH edges hosting two modes, either co-propagating [panel (a)] or counter-propagating [panel (b)]. In the central region II of length $L$ where the local GFs are studied (depicted by an eye symbol), there is an inter-mode interaction of strength $u$. In the regions I and III, the interaction is assumed to be zero. The interaction profile $u(x)$ is schematically shown in the bottom of panel (a) for smooth (solid line) and sharp (dotted line) interfaces I - II and II - III. The characteristic length scale on which the interactions spatially change is denoted $\Delta x$. The interfaces are described by the plasmon scattering matrices $S_{L/R}$. For the mode 1, a distribution $f_1(\epsilon)$ is created in region I. For the mode 2, a distribution $f_2(\epsilon)$ is created in the region I [panel (a)] or the region III [panel (b)]. At least one of these distributions propagating towards the central region II is non-equilibrium. }
    \label{fig:complex_setup}
\end{figure}
The setup that we consider includes an interacting edge segment (denoted region II) of length $L$ connected to non-interacting regions (``leads'') I and III, see Fig.~\ref{fig:complex_setup}.   Setups of this type have been considered to study transport properties of conventional Luttinger liquids~\cite{Safi_Transport_1995,Maslov_Landauer_conductance_1995,Ponomarenko_Renormalization_1995,Trauzettel2004appearance,Gutman2010,Gutman2008nonequilibrium,Berg2009fractional,Kuhne2015measuring},
a multi-mode FQH edge or an interface between two FQH states (i.e., a ``line junction'' of FQH edges)~\cite{Sen2008line,
Protopopov_transport_2_3_2017, Nosiglia2018,Spanslatt2021,Spanslatt_Noise_2022, Yutushui2024, Park2024}. It is also a fundamental building block for models of more complex geometries such as interferometers~\cite{Gutman2010,Kovrizhin2009exactly, Chirolli2013Jul,Park2015Dec,Iyoda2025Apr,Kovrizhin2009exactly,Levkivskyi2009noise-induced,NgoDinh2013analytically}. We assume that the edge is clean so that we neglect tunneling between the modes. 

We are interested in the GFs of excitations in the interacting segment II. These GFs  determine the tunneling current and the noise for two such complex edges coupled via a QPC in the region II. We assume that distribution functions $f_1(\epsilon)$ and $f_2(\epsilon)$ are created in the non-interacting regions and propagate towards region II, see Fig.~\ref{fig:complex_setup}. This setup can be directly implemented in experiments with engineered (line-junction) edges. In other cases, the assumption that non-equilibrium states are created in non-interacting segments may be a simplification of an actual setup. In any case, the analysis of our setup below demonstrates how the inter-mode interaction  affects non-equilibrium properties of a multi-mode FQH edge and the novel features that it brings in comparison to the single mode Laughlin edges. 

The two interfaces I-II and II-III where the interaction strength changes cause scattering of density waves (``plasmons'')~
\cite{Safi_Transport_1995,Maslov_Landauer_conductance_1995,Ponomarenko_Renormalization_1995,Safi1999Dec}. 
This scattering is described by the matrices
\begin{align}
\label{eq:scattering_matrices}
    S_L \equiv \begin{pmatrix}
        t_L &-r_L \\
        r_L & t_L
    \end{pmatrix}, \quad S_R \equiv \begin{pmatrix}
        t_R &r_R \\
        -r_R & t_R
    \end{pmatrix},
\end{align}
in which the transmission $t_j$ ($j=L,R$) and reflection $r_j$ amplitudes satisfy $t_j^2+r_j^2=1$. The values  of these amplitudes depend not only on the filling factor discontinuities $\nu_a$ ($a=1,2$) and the interaction strength $u$ in region II but also on the ``sharpness'' of the switching-on of the interaction at the interfaces in comparison with the characteristic plasmon wavelength. We will discuss the explicit form of these amplitudes below. 

\subsection{Co-propagating edge modes: Non-equilibrium action}
\label{sec:Co_prop}
As a prototypical example of an edge with co-propagating modes, we consider the $\nu=4/3$ state, which consists of two edge modes associated with the filling factor discontinuities $\nu_1= 1$ and $\nu_2 = 1/3$.  The corresponding $K$-matrix and the charge vector $\vec{q}$ edge read 
\begin{align} \label{eq:Kmatrix43}
    K = \begin{pmatrix}
    1 & 0\\
    0 & 3
    \end{pmatrix}\,, \quad \vec{q} = \begin{pmatrix}
    1 \\ 1
    \end{pmatrix}\,. 
\end{align}
The Hamiltonian for the interacting $\nu=4/3$ edge reads
\begin{align}
\label{eq:H25}
    H &=\pi \int dx \left( v_1\rho_{1}^2+3v_{2}\rho_{2}^2+2u(x)\rho_{1}\rho_{2}\right),
\end{align}
where the short-range repulsive interaction $u(x)=u>0$ for $|x|<L/2$ (region II), and zero otherwise, see Fig.~\ref{fig:complex_setup}. The density operators, denoted $\rho_1 = \partial_x \phi_1/(2\pi)$ and $\rho_{2} = \partial_x \phi_2/(2\pi)$, obey the commutation relations
\begin{subequations}
\begin{align}
&[\rho_{1}(x),\rho_{1}(y)]=-\frac{i}{2\pi}\partial_x\delta(x-y),\\
    &[\rho_{2}(x),\rho_{2}(y)]=-\frac{1}{3}\frac{i}{2\pi}\partial_x\delta(x-y)\,,
\end{align}
\end{subequations}
where $\phi_1$ and $\phi_2$ are chiral bosonic fields that describe the edge mode for $\nu_1=1$ and $\nu_2=1/3$, respectively.
We choose both filling factors with positive sign, which corresponds to mode chiralities $\eta_1=\eta_{2}=+1$ (right-movers).

The Hamiltonian~\eqref{eq:H25} is diagonalized by the matrix
\begin{align}
\label{eq:Lambda_25}
   \Lambda=\begin{pmatrix}
        \cos \gamma & -\sin \gamma\\
        \frac{\sin \gamma}{\sqrt{3}} & \frac{\cos \gamma}{\sqrt{3}} 
    \end{pmatrix}, \; \tan (2\gamma) = \frac{2}{\sqrt{3}}\frac{u}{v_{1}-v_{2}},
\end{align}
which scales the $K$-matrix~\eqref{eq:Kmatrix43} and also preserves its matrix signature: $\Lambda^T K \Lambda = \text{diag}(1,1)$. The Hamiltonian then becomes
\begin{align}
\label{eq:H_diag}
    H=\pi \int dx \left(v_+ \rho^2_++v_- \rho^2_-\right),
\end{align}
with eigenmodes
\begin{align} \label{eq:eigenmode43}
 \begin{pmatrix}
        \rho_+\\
        \rho_-
    \end{pmatrix} =\Lambda^{-1}\begin{pmatrix}
        \rho_1\\
        \rho_{2}
    \end{pmatrix},
\end{align}
and their velocities
\begin{align}
    v_\pm = v_{1,2}\cos^2 \gamma+v_{2,1}\sin^2\gamma\pm \frac{u\sin(2\gamma)}{\sqrt{3}}.
\end{align}
Note that even in the absence of interactions, i.e., when $\gamma=0$, the basis transformation~\eqref{eq:Lambda_25} rescales the density field of the 1/3 mode. To ensure a stable Hamiltonian bounded from below, we assume that $u^2 \leq 3 v_1 v_2$ such that $v_\pm>0$.
On the Keldysh contour, the interaction term in the Hamiltonian~\eqref{eq:H25} is represented by classical and quantum components in the action 
\begin{align}
    S_{\rm int} = -2\pi\int dt dx\,  u(x) \left( \overline{\rho}_1\rho_{2}+ \overline{\rho}_2\rho_{1}  \right).
\end{align}

We now consider a non-equilibrium state, as illustrated in Fig.~\ref{fig:complex_setup}, where excitations are injected separately into individual edge modes (1 and 1/3).
Specifically, in analogy with the above analysis of a non-equilibrium Laughlin edge, we assume that mode 1 is driven out of equilibrium by injection of excitations with a charge $n_{1,1}e$ and is characterized by the distribution function $f_1(\epsilon)$, while mode 1/3 is driven out of equilibrium by injection of excitations with a charge $n_{1,2}e/3$ and is characterized by the distribution function $f_2(\epsilon)$. The corresponding excitations are described by the vertex operators
\begin{align} \label{eq:excitations43}
 \psi_{n_{1,1}}^{\dagger} \sim e^{-i n_{1,1} \phi_1} \,, \qquad \psi_{n_{1,2}}^{\dagger} \sim e^{-i n_{1,2} \phi_2}\,.
\end{align}
The first index 1 in the notations
 $n_{1,1}$ and $n_{1,2}$ serves to distinguish the excitations~\eqref{eq:excitations43}, injected to drive the edge out of equilibrium, from excitations described by $(n_{2,1}, n_{2,2})$, Eq.~\eqref{eq:vertex_43}, for which GFs and tunneling transport in the generated non-equilibrium state will be studied in Secs.~\ref{sec:GFs_co} and~\ref{sec:transport_co}.
 Applying the Keldysh formulation introduced for the Laughlin states, 
see Eqs.~\eqref{eq:partition_function_rho_general} and \eqref{eq:F-V-bar-det-Laughlin-n}, we find the full non-equilibrium action for the $\nu=4/3$ edge:
\begin{align}
\label{eq:action_43}
    e^{iS[\rho,\overline{\rho}]} &= e^{- i \rho_1\, \Pi_{{1,a}}^{-1} \overline{\rho}_1} \Delta [\Pi_{1,a}^{-1}\overline{\rho}_1]^{1/(n_{1,1})^2}\notag \\
    &\times e^{- i 3 \rho_{2}\, \Pi_{2,a}^{-1} \overline{\rho}_{2}} \Delta [3\Pi_{2,a}^{-1}\overline{\rho}_{2}]^{3/(n_{1,2})^2} \notag \\
    &\times e^{-i\int dt dx 2\pi u(x)\left( \overline{\rho}_1\rho_{2} + \overline{\rho}_{2}\rho_1  \right)}.
\end{align}
Using the action~\eqref{eq:action_43}, we will now compute, in Sec.~\ref{sec:FCS_co} and Sec.~\ref{sec:GFs_co}, the  FCS and the quasiparticle GF's for the $\nu=4/3$ edge in the setup of Fig.~\ref{fig:complex_setup}\textcolor{blue}{(a)}. 

\subsection{Co-propagating edge modes: Full counting statistics}
\label{sec:FCS_co}
The FCS generating function was defined in Eq.~\eqref{eq:FCS-kappa-definition} and represented as a Keldysh functional integral in Eq.~\eqref{eq:FCS_Keldysh_def}. For the $\nu=4/3$ edge, the total charge
 to the left of the point $x_0$ at time $t$
is given by
\begin{align}
\label{eq:Q_43}
    Q(x_0,t) =e \int_{-\infty}^{x_0}dx\left[\rho_1(x,t) + \rho_{2}(x,t) \right].
\end{align}
After plugging the charge~\eqref{eq:Q_43} into the integral representation~\eqref{eq:FCS_Keldysh_def} and using the action~\eqref{eq:action_43}, the calculations are carried out in analogy to those for the Laughlin edge. As a result, the generating function is expressed in terms of two normalized determinants according to
\begin{align}
\label{eq:FCS_25}
    \kappa(\lambda,x_0,\tau) =\left\{\overline{\Delta}[\delta_{1, \tau}(t)]\right\}^{1/(n_{1,1})^2}\left\{\overline{\Delta}[\delta_{2,\tau}(t)]\right\}^{3/(n_{1,2})^2}.
\end{align}
What remains to be computed is the explicit form of the counting phases entering these determinants. We obtain these phases by integrating the quantum components of the incoming density fields according to Eq.~\eqref{eq:gen_phase_formula}. To find the quantum components, we integrate out the classical components in the action and obtain the following coupled equations 
\begin{align}
\label{eq:eom25_FCS}
&K \partial_t  \begin{pmatrix}
        \overline{\rho}_1 \\
        \overline{\rho}_{2} 
    \end{pmatrix} + \partial_x \begin{pmatrix}
        v_1 & u\\
        u & 3v_{2}
    \end{pmatrix}\begin{pmatrix}
        \overline{\rho}_1 \\
        \overline{\rho}_{2} 
    \end{pmatrix} = -\frac{\lambda }{2\pi}j(x-x_0,t) \vec{q}\,. 
\end{align}
Here, the source term $j(x,t)$ is given in Eq.~\eqref{eq:source_j} and we used $\eta_1=\eta_{2}=+1$. The vector structure on the right-hand-side is determined by the bare-mode charge vector in Eq.~\eqref{eq:Kmatrix43}

We now proceed with solving Eq.~\eqref{eq:eom25_FCS} for the three-region segment in Fig.~\ref{fig:complex_setup}\textcolor{blue}{(a)}. To solve the equations, we first diagonalize Eq.~\eqref{eq:eom25_FCS} with the matrix $\Lambda$ in Eq.~\eqref{eq:Lambda_25}. Then, Fourier-transforming to frequency space, Eq.~\eqref{eq:eom25_FCS} becomes
\begin{align}
\label{eq:eoms_25_diag}
    &-i\omega  \begin{pmatrix}
        \overline{\rho}_+ \\
        \overline{\rho}_- 
    \end{pmatrix} + \partial_x \begin{pmatrix}
         v_+ && 0\\
        0 && v_-
    \end{pmatrix} \begin{pmatrix}
        \overline{\rho}_+ \\
        \overline{\rho}_- 
    \end{pmatrix}   \notag \\&
    =-\frac{\lambda }{2\pi}j(x-x_0,\omega)\begin{pmatrix}
      q_+ \\ q_-
    \end{pmatrix}.
\end{align}
Here, the right-hand-side contains two ``eigenmode charges'' given by 
\begin{align}
\label{eq:eigencharges_25}
    \begin{pmatrix}
        q_+\\
        q_-
    \end{pmatrix} \equiv \Lambda^T \vec{q} = \begin{pmatrix}
       \cos \gamma+\frac{\sin \gamma}{\sqrt{3}} \\[0.2cm] \frac{\cos \gamma}{\sqrt{3}}-\sin\gamma
    \end{pmatrix},
\end{align}
which thus depend on the interaction strength. Further, $j(x,\omega)$ is the frequency representation of the source~\eqref{eq:source_j}, which reads
\begin{align}
\label{eq:j_x_omega}
     j(x,\omega) = \frac{1}{\sqrt{2}}\delta(x)\left(e^{i\omega \tau}-1\right).
\end{align}

We now choose the charge-measurement point $x_0$ to lie within the interacting
region~II [see Fig.~\ref{fig:complex_setup}\textcolor{blue}{(a)}], i.e., $|x_0| < L/2$. Probing the FCS and the associated fractionalization in this region is particularly insightful from the point of view of connections to properties of GFs and the tunneling transport observables studied below. Calculations of FCS in the regions I and III can be performed in a similar way. To proceed, we solve the coupled equations~\eqref{eq:eom25_FCS} separately in each region.
In the non-interacting regions~I and~III, the equations reduce to the homogeneous
form with $u(x)=j(x,\omega)=0$, whereas in region~II we solve Eq.~\eqref{eq:eoms_25_diag} in the presence of the source term $j(x,\omega)$.
In all regions, the general solution is a superposition of two co-propagating
modes. 

According to the causal structure of the action~\eqref{eq:action_43}, we need the \textit{advanced} solutions that vanish at times larger than those in the source $j(x-x_0,t)$. This constraint requires that only incoming modes
are present. Since region~III contains only outgoing modes and the source acts in region~II, the advanced
condition enforces $\overline{\rho}_{1,2}=0$ in region~III. With these conditions, we obtain the solutions in each region as follows:
\begin{subequations}
\label{eq:planewave25}
\begin{align}
\bar\rho_{1,2}(x,\omega)
&=
\begin{cases}
A^{\text{I}}_{1,2}\,e^{ i\omega x / v_{1,2}}, & \qquad  x<-\tfrac{L}{2},\\
0, & \qquad x>\tfrac{L}{2},
\end{cases}
\label{eq:planewave25a}
\\[4pt]
\bar\rho_{\pm}(x,\omega)
&=
e^{ i\omega x / v_{\pm}}
\big(
A^{\text{II}}_{\pm}
- \mathcal{J}_{\pm} \big),
\quad |x|<\tfrac{L}{2}\,,
\label{eq:planewave25b}
\end{align}
\end{subequations}
where $A_{1,2}^{\text{I}}$, $A_{\pm}^{\text{II}}$ are constants and 
\begin{align}
\label{eq:extended_source}
    \mathcal{J}_{\pm} = \Theta(x-x_0)\,
\frac{\lambda q_{\pm}}{2\pi\sqrt{2}\,v_{\pm}}\,
e^{- i\omega \eta_{\pm} x_0 / v_{\pm}}\,
\big(e^{i\omega \tau}-1\big)\,,
\end{align}
where the co-propagating eigenmodes in region II have chiralities $\eta_{\pm} = +1$. The four constants,
$A^{\mathrm{I}}_{1,2}$ and $A^{\mathrm{II}}_{\pm}$,
are fixed by the boundary conditions at the interfaces between the
non-interacting and interacting regions, which require the continuity of
the energy current densities across each boundary.
These conditions are conveniently implemented using the unitary scattering
matrices~\eqref{eq:scattering_matrices}, which relate the densities on the two
sides of the interfaces as follows:
\begin{align} 
\begin{pmatrix}
v_+\overline{\rho}_+(\omega)\\
v_-\overline{\rho}_-(\omega)
\end{pmatrix}_{\rm II}
&=
S_L
\begin{pmatrix}
v_1\overline{\rho}_1(\omega)\\
\sqrt{3}\,v_2\overline{\rho}_2(\omega)
\end{pmatrix}_{\rm I},
\label{eq:SL_bc}
\end{align}
for the I - II boundary and
\begin{align}
\begin{pmatrix}
v_1\overline{\rho}_1(\omega)\\
\sqrt{3}\,v_2\overline{\rho}_2(\omega)
\end{pmatrix}_{\rm III}
&=
S_R
\begin{pmatrix}
v_+\overline{\rho}_+(\omega)\\
v_-\overline{\rho}_-(\omega)
\end{pmatrix}_{\rm II}
\label{eq:SR_bc}
\end{align}
for the II - III boundary. By solving 
Eqs.~\eqref{eq:planewave25},
\eqref{eq:SL_bc}, and \eqref{eq:SR_bc}
for the constants $A_{1,2}^\text{I}$, we obtain the advanced solutions in region I as 
\begin{subequations}
\label{eq:advancedsolution43}
\begin{align}
    \overline{\rho}_1(x,\omega) &= \frac{\lambda}{2\pi \sqrt{2} v_1} [r_L q_- e^{-\frac{i\omega}{v_-} (x_0+\frac{L}{2})} + t_L q_+ e^{-\frac{i\omega}{v_+} (x_0+\frac{L}{2})}] \nonumber \\
    & \times 
    e^{\frac{i\omega}{v_1} \big(x+ \frac{L}{2} \big)} (e^{i \omega \tau } -1)\,, \quad x<-\frac{L}{2}\,,
\end{align}
\begin{align}
     \overline{\rho}_2(x,\omega) &= \frac{\lambda}{2\pi \sqrt{6} v_2} [-r_L q_+ e^{-\frac{i\omega}{v_+} (x_0+\frac{L}{2})} + t_L q_- e^{-\frac{i\omega}{v_-} (x_0+\frac{L}{2})}] \nonumber \\
    & \times 
    e^{\frac{i\omega}{v_2} \big(x+ \frac{L}{2} \big)} (e^{i \omega \tau } -1)\,, \quad x<-\frac{L}{2}\,. 
\end{align}
\end{subequations}
The densities~\eqref{eq:advancedsolution43} in the frequency domain can be Fourier transformed into the time domain as 
\begin{subequations}
\label{eq:density43_timedomain}
\begin{align}
   \overline{\rho}_1(x,t) &= -\frac{\lambda}{2\pi \sqrt{2} v_1} \Big[r_L q_- \Upsilon_{\tau} (t,\frac{x-x_0}{v_1} + t_{1,-}) \nonumber \\
   & \qquad \qquad + t_L q_{+} \Upsilon_{\tau}(t,\frac{x-x_0}{v_1} + t_{1,+}) \Big], \\
   \overline{\rho}_2(x,t) &= -\frac{\lambda}{2\pi \sqrt{6} v_2} \Big[-r_L q_+ \Upsilon_{\tau} (t,\frac{x-x_0}{v_2} + t_{2,+}) \nonumber \\
   & \qquad \qquad + t_L q_{-} \Upsilon_{\tau}(t,\frac{x-x_0}{v_2} + t_{2,-}) \Big]\,,
\end{align}
\end{subequations}
where the function $\Upsilon_{\tau}(t_1,t_2)$ is defined as 
\begin{align}
\label{eq:delta_spikes}
    \Upsilon_{\tau} (t_1, t_2) \equiv \delta (t_1 - t_2) - \delta (t_1- t_2 -\tau)\,,
\end{align}
and the times $t_{1,\pm}$ and $t_{2,\pm}$ are given by 
\begin{subequations}
    \begin{align}
    t_{1,\pm} = (x_0 + \frac{L}{2}) \Big(\frac{1}{v_1} - \frac{1}{v_\pm } \Big)\,, \\
     t_{2,\pm} = (x_0 + \frac{L}{2}) \Big(\frac{1}{v_2} - \frac{1}{v_\pm } \Big)\,. 
\end{align} \label{eq:timeshifts43}
\end{subequations}

Finally, the counting phases $\delta_{1,\tau}$ and $\delta_{2,\tau}$ are obtained by integrating the density solutions in Eq.~\eqref{eq:density43_timedomain} [cf. Eq.~\eqref{eq:gen_phase_formula}],
\begin{subequations}
\label{eq:phases_43}
\begin{align}
    \delta_{1,\tau}(t) &= 2\pi\sqrt{2}\, n_{1,1}
    \lim_{t'\to -\infty}\int^{v_1(t'+t)}_{x_0} dx'\,\overline{\rho}_{1}(x',t')\,, \\
    \delta_{2,\tau}(t) &= 2\pi\sqrt{2}\, n_{1,2}
    \lim_{t'\to -\infty}\int^{v_2(t'+t)}_{x_0} dx'\,\overline{\rho}_{2}(x',t')\,.
\end{align}
\end{subequations}
We recall that the integers $n_{1,1}$ and $n_{1,2}$ characterize the charge of the excitations (in units of the charge of the corresponding elementary excitations) injected into the edge modes with filling factors $\nu_1=1$ and $\nu_2=1/3$, respectively, as described by the vertex operators \eqref{eq:excitations43}. Evaluating the integrals~\eqref{eq:phases_43} with aid of Eq.~\eqref{eq:useful_phase_identity}, we find that each counting phase entering the generating function~\eqref{eq:FCS_25} can be expressed as a superposition of two rectangular time pulses,
\begin{subequations}
\label{eq:phases_43_FCS}
\begin{align}
    \delta_{1,\tau}(t) &= \delta_{1,+}\, w_{\tau}(t,-t_{1,+})
    + \delta_{1,-}\, w_{\tau}(t,-t_{1,-})\,,\\
    \delta_{2,\tau}(t) &= \delta_{2,+}\, w_{\tau}(t,-t_{2,+})
    + \delta_{2,-}\, w_{\tau}(t,-t_{2,-})\,,
\end{align}
\end{subequations}
where $w_{\tau}(t_1, t_2)$ is the window function defined in Eq.~\eqref{eq:window_function} and 
the time shifts $t_{1,\pm}$ and $t_{2,\pm}$ are given by Eq.~\eqref{eq:timeshifts43}.
The amplitudes of the individual pulses are
\begin{subequations}
\label{eq:phases_43_amplitudes}
\begin{align}
    \delta_{1,+} =  \lambda t_L q_+ n_{1,1}\,, 
    &\quad \delta_{1,-} =  \lambda r_L q_- n_{1,1}\,, \\
    \delta_{2,+} = -\frac{ \lambda r_L q_+ n_{1,2}}{\sqrt{3}}\,, 
    &\quad \delta_{2,-} = \frac{ \lambda t_L q_- n_{1,2}}{\sqrt{3}}\,.
\end{align}
\end{subequations}

Comparing with the result for the Laughlin edge, Eq.~\eqref{eq:counting_phase_Laughlin}, we see that the counting phases, Eq.~\eqref{eq:phases_43_amplitudes}, exhibit fractionalization. Let us discuss in more detail how this comes about. For this purpose, we define the counting pulses for all $x$ and $t$ according to 
\begin{subequations} \label{eq:totalcharge_xandt}
\begin{align}
    q_{1} (x,t) &= e\frac{2\pi  \sqrt{2} }{\lambda} \int_{-\infty}^{x} dx' \overline{\rho}_1(x',t)\,, \\
    q_{2} (x,t) &= e\frac{2\pi  \sqrt{2} }{\lambda} \int_{-\infty}^{x} dx' \overline{\rho}_2(x',t)\,,
\end{align}
\end{subequations}
where $\overline{\rho}_1$ and $\overline{\rho}_2$ denote the advanced solutions of the coupled equations~\eqref{eq:eom25_FCS}. It is useful to consider them as functions of $t$ for different $x$. For $x=x_0 - 0$, these are rectangular pulses of duration $\tau$ with the amplitudes $e$ and $e/3$, respectively, see Fig.~\ref{fig:pulsedynamics}\textcolor{blue}{(a)}. The counting pulses are non-zero for all $x < x_0$, since we consider the advanced solution. Moving to the left in the coordinate $x$ corresponds to propagation backward in time. Within the interacting region II, these pulses fractionalize into two eigenmode pulses 
(labeled by $+$ and $-$) moving with the velocities $v_+$ and $v_-$, respectively. Upon crossing to the non-interacting region I, i.e., for $x < -L/2$, each of these eigenmode pulses separates into pulses in the modes 1 and 1/3 that move with the velocities $v_1$ and $v_2$, respectively. The corresponding amplitudes in the mode 1 resulting from both eigenmodes are $q^{\text{(p)}}_{1,+}$ and $q^{\text{(p)}}_{1,-}$. Similarly, in the mode 1/3 one gets two pulses with amplitudes $q^{\text{(p)}}_{2,+}$ and $q^{\text{(p)}}_{2,-}$,
\begin{subequations}
\label{eq:accumulatedcharges}
\begin{align}
    q_{1} (x,t) = q_{1,+}^{\text{(p)}} w_{\tau}(t,-t_{1,+}) + q_{1,-}^{\text{(p)}}w_{\tau}(t,-t_{1,-})\,, \\ q_{2} (x,t) = q_{2,+}^{\text{(p)}} w_{\tau}(t,-t_{2,+}) + q_{2,-}^{\text{(p)}}w_{\tau}(t,-t_{2,-})\,,
\end{align}
\end{subequations}
This process of fractionalization of counting pulses is illustrated in Fig.~\ref{fig:pulsedynamics}\textcolor{blue}{(a)}.
The counting phases $\delta_{1,\pm}$ and $\delta_{2,\pm}$ are determined by the fractionalized pulse amplitudes
$q_{1,\pm}^{\text{(p)}}$ and $q_{2,\pm}^{\text{(p)}}$ in the non-interacting region I:
\begin{align} \label{eq:FCScharges_phases_correspondence}
  e\delta_{1,\pm} =  n_{1,1}\lambda\, q_{1,\pm}^{\text{(p)}}\,, \qquad 
   e\delta_{2,\pm} =  n_{1,2}\lambda\, q_{2,\pm}^{\text{(p)}}
  \,. 
\end{align}
As we show below by explicitly calculating the FCS, the fractionalization of counting phases \eqref{eq:phases_43_amplitudes} implies a corresponding fractionalization of charges of excitations due to inter-mode interaction. We will also discuss there the physics underlying this charge fractionalization. The fractionalization of anyonic excitations that plays a central role in this paper, including both fractionalization of the fractional charge and the fractional braiding phases, can be viewed as a counterpart of the electron fractionalization in conventional Luttinger liquids that has been discussed in the context of various observables in previous works~\cite{Gutman2010,Gutman2010full,Levkivskyi2009noise-induced,Safi_Transport_1995,LeHur2005dephasing,LeHur2006electron,LeHur2008charge,Steinberg2008charge}. 

The amplitudes of the scattering matrices in Eq.~\eqref{eq:scattering_matrices} depend sensitively on how abruptly the interaction is switched on at the interfaces. When the interaction varies on a length scale much shorter than the relevant plasmon wavelength $\sim v_{\pm}/\omega$ (so-called "sharp" interfaces), the transmission and reflection amplitudes are determined by the interaction parameter $\gamma$ in Eq.~\eqref{eq:Lambda_25}, and take the form
\begin{align} \label{eq:interfacetransmissionamplitudes}
t_L = t_R = \cos \gamma\,, \quad r_L = r_R = -\sin \gamma\,.
\end{align}
 This result is obtained by integrating the equation of motion \eqref{eq:eom25_FCS} over a small spatial interval around the interface, using the rotation matrix~\eqref{eq:Lambda_25} on the side II of the interface, and comparing the result with the definition of the scattering matrix, Eqs.~\eqref{eq:scattering_matrices} and \eqref{eq:SL_bc}-\eqref{eq:SR_bc}.

By contrast, when the interaction changes smoothly on a length scale much longer than the plasmon wavelength (so-called "adiabatic" interfaces), the interfaces become perfectly transmitting,
\begin{align} \label{eq:interfacetransmissionamplitudes_adaibatic}
t_L = t_R = 1\,, \quad r_L = r_R = 0\,.
\end{align}
In the following, we analyze the scattering phases entering Eq.~\eqref{eq:phases_43_FCS}, and the resulting form of the FCS, for both cases of sharp and adiabatic interfaces.

\subsubsection{Sharp interfaces}
\label{sec:43_FCS_sharp}

For sharp interfaces, two distinct regimes can be identified: the long-time (or equivalently short-length) limit $(x_0 + L/2) \ll v_{\pm} |\tau|$ and the short-time (or long-length) limit $(x_0 + L/2) \gg v_{\pm} |\tau|$. Here, $(x_0 + L/2)$ is the distance between the interface position, $x=-L/2$ and the observation point $x = x_0$. In the short-length regime, the two square pulses generated by the fractionalization of the counting pulse do not separate when reaching the interface. As a result, the scattering phases $\delta_{1,\tau}(t)$  and $\delta_{1,\tau}(t)$ can be well approximated a single pulse of duration $\tau$. Summing up the amplitudes of the two pulses, Eqs.~\eqref{eq:phases_43_amplitudes}, for each of the modes 1 and 1/3, and using the sharp-interface conditions \eqref{eq:interfacetransmissionamplitudes}, the counting phases \eqref{eq:phases_43_FCS} become
\begin{subequations}\label{eq:FCS_43_long_time_int_region}
\begin{align}
\delta_{1,\tau}(t) &\simeq n_{1,1} \lambda w_\tau(t,0)\,, \\
\delta_{2,\tau}(t) &\simeq n_{1,2} \frac{\lambda}{3} w_\tau(t,0)\,.
\end{align}
\end{subequations}
This is exactly the result that we would have in the absence of interaction. Consequently, in the short-length (long-time) limit, the FCS becomes insensitive to the inter-channel interaction. This behavior of FCS is an extension of the known result that the dc conductance of an interacting 1D system is governed by properties of the non-interacting regions (``leads''), which holds for conventional Luttinger liquids~\cite{Maslov_Landauer_conductance_1995,Safi_Transport_1995,Ponomarenko_Renormalization_1995} and for FQH edges or linear junctions~\cite{Protopopov_transport_2_3_2017}. 

By contrast, in the long-length regime with $(x_0 + L/2) \gg v_{\pm} |\tau|$, the fractionalized counting pulses become well separated in time, and the coherence between them becomes negligible. As a result, each determinant in Eq.~\eqref{eq:FCS_25} factorizes into two independent contributions~\cite{Gutman2010,Protopopov2013correlations},
\begin{subequations}\label{eq:countingphases_43_short_time_int_region}
\begin{align}
\overline{\Delta}[\delta_{1, \tau}(t)]&\simeq \overline{\Delta}[\delta_{1,+}w_\tau(t,0)] \overline{\Delta}[\delta_{1,-}w_\tau(t,0)], \\
\overline{\Delta}[\delta_{2,\tau}(t)] &\simeq \overline{\Delta}[\delta_{2,+}w_\tau(t,0)] \overline{\Delta}[\delta_{2,-}w_\tau(t,0)],
\end{align}
\end{subequations}
with respective amplitudes~\eqref{eq:phases_43_amplitudes}. Thus, in the long-length limit, the FCS resolves four independent counting pulses with non-universal amplitudes, which reflects, as shown below, the interaction-induced fractionalization of charge of excitations. 

\begin{figure}[tbp]
    \centering
\subfloat[][]{
\includegraphics[width=0.9\columnwidth]{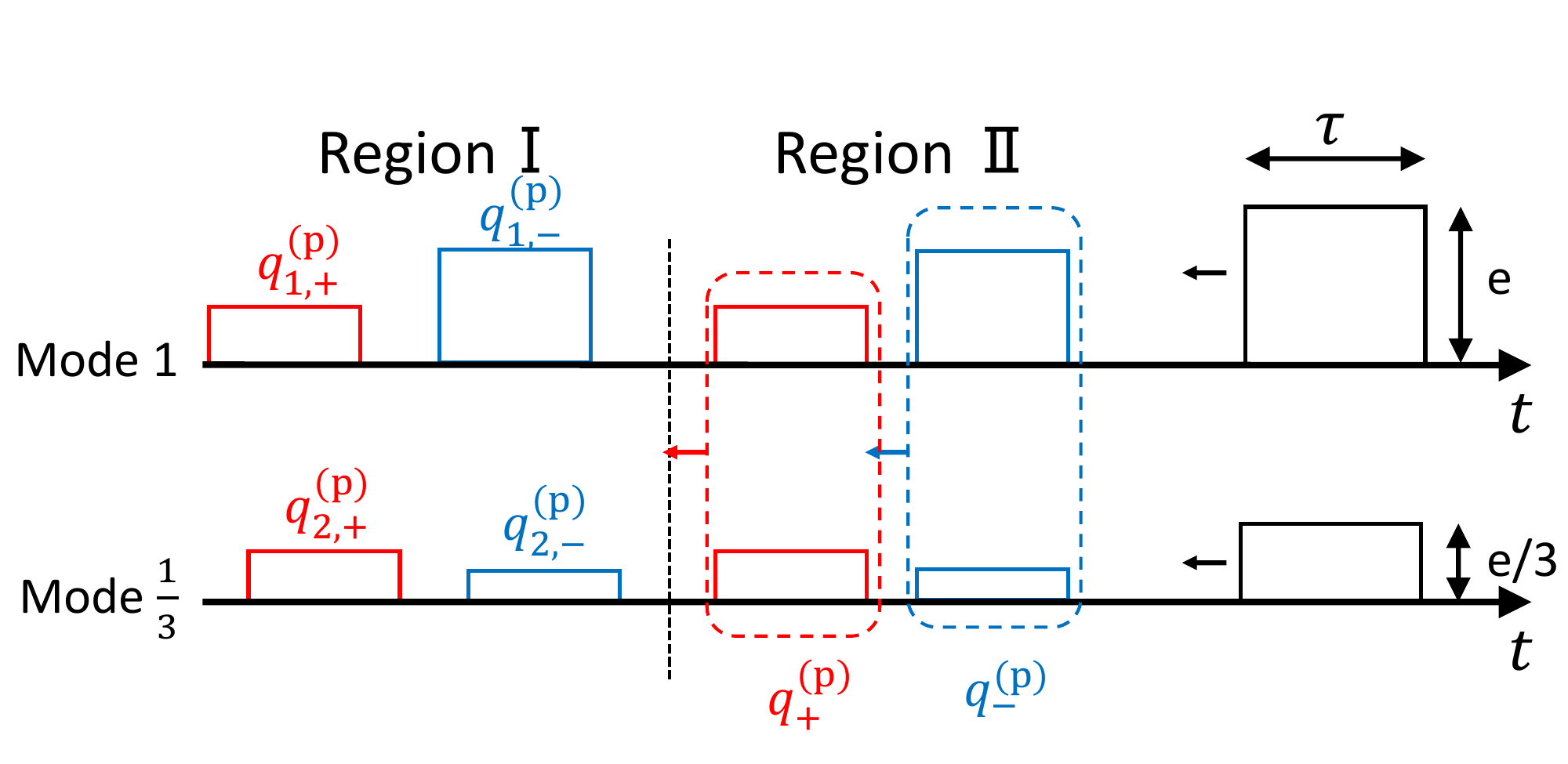}  
\put(-220,85){\large\textbf{(a)}}
}
\par\vspace{-10mm}
\subfloat[][]{
\includegraphics[width=0.9\columnwidth]{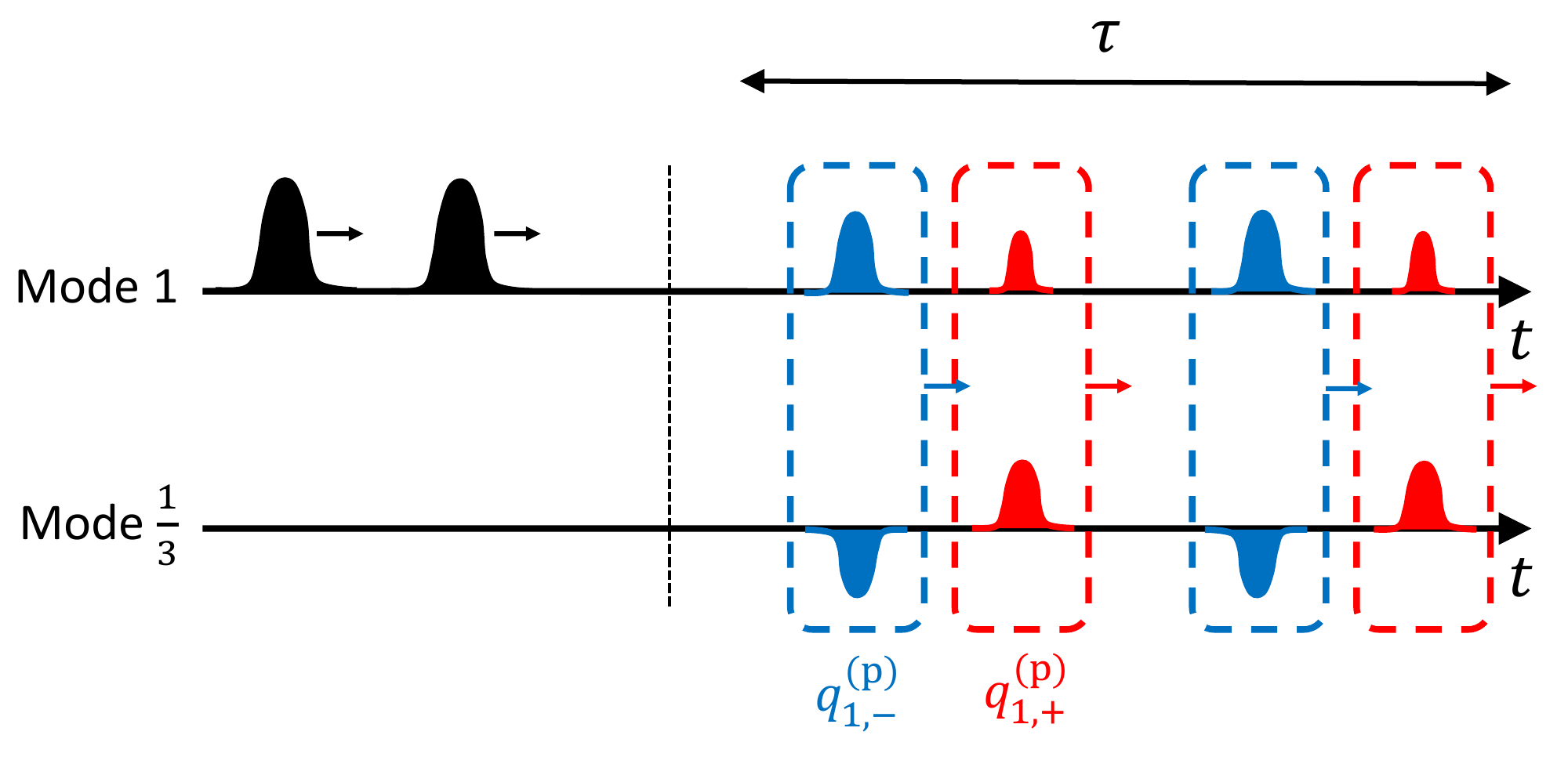}  
 \put(-220,85){\large\textbf{(b)}}
}
\par\vspace{-10mm}
\subfloat[][]{
\includegraphics[width=0.9\columnwidth]{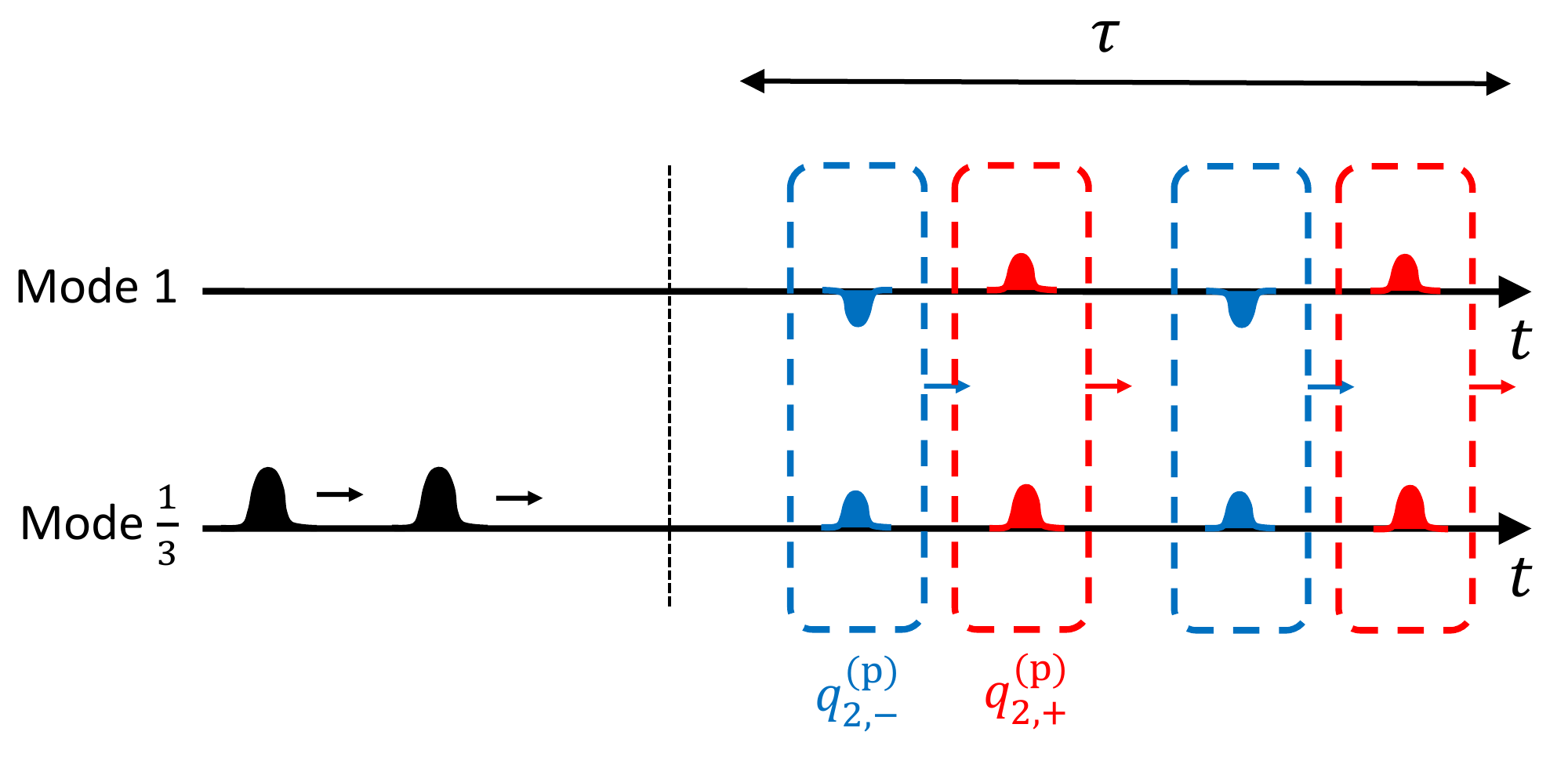}  
 \put(-220,85){\large\textbf{(c)}}
}
\caption{Illustrations of fractionalization due to intermode interactions on the $\nu=4/3$ edge with a sharp interface. (a) Dynamics of the counting pulses~\eqref{eq:totalcharge_xandt} for FCS  of charge passing across a point $x_0$ in the interacting region II during the time $\tau$. The pulse configurations are shown at three spatial locations. At $x=x_0 - 0$ (right), there is a single pulse in each mode, with quantized amplitudes $e$ and $e/3$, respectively
For $-L/2  < x_0$ (middle), the pulses fractionalize into those corresponding to two eigenmodes labeled $+$ and $-$. For $x < -L/2$ (non-interacting region I, left), the eigenmode pulses separate into pulses in modes 1 and 1/3 that move with velocities $v_1$ and $v_2$.
The amplitudes of these fractionalized pulses, $q_{1,\pm}^{\text{(p)}}$ and $q_{2,\pm}^{\text{(p)}}$, govern the counting phases $\delta_{1,\pm}$ and $\delta_{2,\pm}$
[see Eqs.~\eqref{eq:phases_43_amplitudes} and \eqref{eq:FCScharges_phases_correspondence}] and the fractional charges of excitations as found by calculating the FCS in the interacting region, 
Eqs.~\eqref{eq:43chargefluctuations_sharp},\eqref{eq:pulsecharges43}, and \eqref{average-charges-FCS-43}. (b-c) Physical picture of charge fractionalization. (b) In the non-interacting region I, the mode 1 is driven out of equilibrium by dilute injection of excitations with charge $n_{1,1}e$, implying a random Poissonian train of such excitations. Upon entering the interacting region II, each excitation fractionalizes into two excitations that correspond to eigenmodes and are characterized by charges $n_{1,1}q_{1,+}^{\text{(p)}}$ and $n_{1,1}q_{1,-}^{\text{(p)}}$. (c) Similarly, if the 1/3 mode is driven out of equilibrium in region I by injection of excitations with charge $n_{1,2}e/3$, they will fractionalize into excitations with charges $n_{1,2}q_{2,+}^{\text{(p)}}$ and $n_{1,2}q_{2,-}^{\text{(p)}}$ upon entering the interacting region II. The FCS, Eqs.~\eqref{eq:43chargefluctuations_sharp}, \eqref{eq:pulsecharges43}, and \eqref{average-charges-FCS-43}, detects a superposition of Poissonian processes with these fractional charges.}
\label{fig:pulsedynamics}
\end{figure}

With these results for the counting phases, we are in a position to evaluate the FCS generating function~\eqref{eq:FCS_25} for sharp interfaces.
We consider a non-equilibrium situation, with generic excitations~\eqref{eq:excitations43} injected into the modes 1 and 1/3 in region I, with voltages $V_1$ and $V_2$, respectively, and with transmissions $\mathcal{T}_1,\mathcal{T}_2 \ll 1$. The distribution function for each of the modes has then the double-step structure~\eqref{n-epsilon-noneq-laughlin} with quasiparticle charges $e^* \mapsto e_{1,a}^* = n_{1,a}\nu_a e$ for the two modes $\nu_1 = 1$ and $\nu_2 = 1/3$:
\begin{eqnarray}
f_1(\epsilon) &=& \mathcal{T}_1 \Theta(-\epsilon+e_{1,1}^* V_1) + (1-\mathcal{T}_1)\Theta(-\epsilon), \nonumber \\
f_2(\epsilon) &=& \mathcal{T}_2 \Theta(-\epsilon+e_{1,2}^* V_2) + (1-\mathcal{T}_2)\Theta(-\epsilon).
\label{eq:doublestep_43}
\end{eqnarray}
To evaluate the determinants entering the generating function, we use the Szeg\H{o} approximation~\cite{Supplemental_Material}. In the long-length limit for sharp interfaces, the factorized form of the determinants \eqref{eq:countingphases_43_short_time_int_region} leads to a factorization also of the generating function into four independent contributions,
\begin{align}
\label{eq:kappa_43_sharp_1}
    \kappa(\lambda,x_0,\tau)
    &\simeq \prod_{s=\pm}
    \kappa_{1,s}(\lambda,x_0,\tau)\,
    \kappa_{2,s}(\lambda,x_0,\tau)\,,
\end{align}
with
\begin{subequations}\label{eq:kappa_43_sharp_2}
\begin{align}
 \kappa_{1,s}(\lambda,x_0,\tau)
    &=
    \exp\!\left[
    - \frac{e V_1 \mathcal{T}_1\tau}{2\pi n_{1,1}}
    \left(1- e^{- i \delta_{1,s}}\right)
    \right]\,, \\ 
    \kappa_{2,s}(\lambda,x_0,\tau)
    &=
    \exp\!\left[
    - \frac{e V_2 \mathcal{T}_2\tau}{2\pi n_{1,2}}
    \left(1- e^{- i \delta_{2,s}}\right)
    \right]\,. 
\end{align}
\end{subequations}
According to Eqs.~\eqref{eq:kappa_43_sharp_1}-\eqref{eq:kappa_43_sharp_2}
in combination with Eq.~\eqref{eq:phases_43_amplitudes} for the phases amplitudes $\delta_{1,\pm}$ and $\delta_{2,\pm}$, the cumulants of charge fluctuations~\eqref{eq:FCS_cumulants} evaluate to
\begin{align} \label{eq:43chargefluctuations_sharp}
     &\langle\langle(\delta Q)^k \rangle\rangle = 
(n_{1,1}q^{(\text{p})}_{1,+})^{k-1}  \langle Q_{1, +}\rangle + (n_{1,1}q^{(\text{p})}_{1,-})^{k-1}  \langle Q_{1, -}\rangle \nonumber \\
    &\,\,\,+(n_{1,2}q^{(\text{p})}_{2,+})^{k-1}  \langle Q_{2,+}\rangle + (n_{1,2}q^{(\text{p})}_{2,-})^{k-1}  \langle Q_{2,-}\rangle\,, 
\end{align}
where the charges $q^{(\text{p})}_{1,\pm}$ and $q^{(\text{p})}_{2,\pm}$ are given by  [see Eq.~\eqref{eq:FCScharges_phases_correspondence}] 
\begin{subequations} \label{eq:pulsecharges43}
\begin{align}
    q^{(\text{p})}_{1,+} &= q_+ t_L e\,, \qquad \,\,\, q^{(\text{p})}_{1,-} =  q_-r_L e\,, \\ q^{(\text{p})}_{2,+} &= -r_L \frac{q_+}{\sqrt{3}} e\,, \quad  q^{(\text{p})}_{2,-} = \frac{q_-}{\sqrt{3}}t_L e\,,
\end{align}
\end{subequations}
and the average charges passing  across the point $x_0$ during the time interval $\tau$ are
\begin{subequations}
\label{average-charges-FCS-43}
\begin{align}
\langle Q_{1,+}\rangle  &= (q_+ t_L \mathcal{T}_1 e^2 V_1 \tau)/(2\pi)\,, \\ 
    \langle Q_{1,-}\rangle  &= (q_-r_L\mathcal{T}_1 e^2 V_2 \tau)/(2\pi)\,,\\ \langle Q_{2,+}\rangle  &= -(q_+ r_L\mathcal{T}_2 e^2 V_2 \tau)/(2\pi \sqrt{3})\,, \\ 
    \langle Q_{2,-}\rangle  &= (q_- t_L \mathcal{T}_2 e^2 V_2 \tau)/(2\pi \sqrt{3})\,. 
\end{align}
\end{subequations}
Equation~\eqref{eq:43chargefluctuations_sharp} is the FCS of four independent Poissonian processes, corresponding to quasiparticles carrying charges $n_{1,1}q_{1,\pm}^{\text{(p)}}$ and $n_{1,2}q_{2,\pm}^{\text{(p)}}$. 
We recall that $q_\pm$ and $r_{L,R}$ in these formulas are expressed in terms of the interaction parameter $\gamma$ via Eqs.~\eqref{eq:eigencharges_25} and
\eqref{eq:interfacetransmissionamplitudes}, respectively. The four fractional charges $n_{1,1}q_{1,\pm}^{\text{(p)}}$ and $n_{1,2}q_{2,\pm}^{\text{(p)}}$ are thus continuously varying functions of $\gamma$, i.e., they are {\it not} topologically quantized. 

In the above, we have obtained this charge fractionalization by calculating the FCS, as a result of the fractionalization of the counting pulse in the process of its advanced evolution  from the point $x_0$ to the region I, see Fig.~\ref{fig:pulsedynamics}\textcolor{blue}{(a)}. We present now a transparent physical interpretation of the obtained result. If the mode $\nu_1 =1$ is driven out of equilibrium in region I by injection of excitations with a charge $n_{1,1}e$, we have there a dilute random (Poissoinian) train of these excitations. When each of them crosses the (sharp) interface to the region II, it fractionalizes into two excitations corresponding to eigenmodes, with charges $n_{1,1}q_{1,\pm}^{\text{(p)}}$, see Fig.~\ref{fig:pulsedynamics}\textcolor{blue}{(b)}. Similarly, If the mode $\nu_2 =1/3$ is driven out of equilibrium in region I by injection of excitations with a charge $n_{1,2}e/3$,  each of fractionalizes, upon crossing into the interacting region II into two excitations corresponding to eigenmodes, with charges $n_{1,2}q_{2,\pm}^{\text{(p)}}$. This yields exactly the four fractional charges as we have found evaluating the FCS. It is easy to verify that the following equalities hold:
\begin{subequations} \label{eq:totalinjectedcharge_43}
\begin{align}
    n_{1,1}\sum_{s=\pm} q_{1,s}^{\text{(p)}} &= n_{1,1} e\,, \\ 
    n_{1,2}\sum_{s=\pm} q_{2,s}^{\text{(p)}} &= n_{1,2} \frac{e}{3}\,.
\end{align}
\end{subequations}
They are a manifestation of charge conservation in the process of interaction-induced splitting of an injected quasiparticle into two excitations. 

\subsubsection{Adiabatic interfaces}
\label{sec:adiabatic_interfaces_43}

We consider now the case of smooth interfaces, with the characteristic scale $\Delta x$ at which the interaction changes 
(see Fig.~\ref{fig:complex_setup})
satisfying $\Delta x \gg v_{\pm} \tau$. In this case, for characteristic frequencies corresponding to the counting pulse duration, $\omega \sim 1/\tau$, we have $\omega \gg v_{\pm}/\Delta x$, implying the adiabatic limit \eqref{eq:interfacetransmissionamplitudes_adaibatic} for the scattering matrix. However, the counting pulses contain also low-frequency components (since the time integrals of the pulses are non-zero). For these components, the interfaces are effectively sharp, and the scattering amplitudes take the form~\eqref{eq:interfacetransmissionamplitudes}. Although the low-frequency components carry only a small spectral weight, we will see that their inclusion is essential to ensure charge conservation.
\begin{figure}[t]
    \centering
\includegraphics[width=0.9\columnwidth]{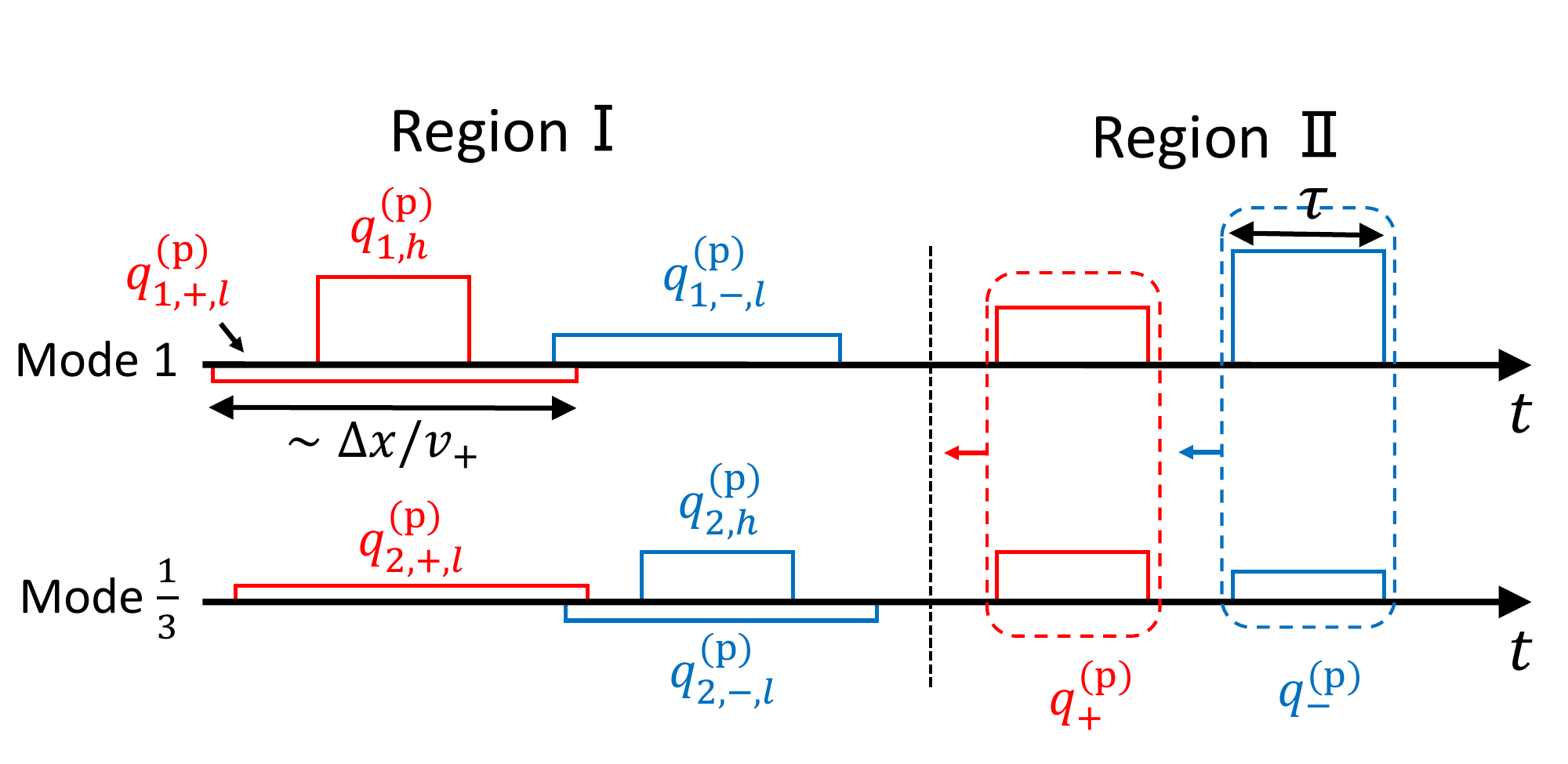}  
\caption{Dynamics of fractionalization of FCS counting pulses 
\eqref{eq:totalcharge_xandt}
for a 4/3 edge with a smooth (``adiabatic'') interface.
The fractionalization within the interacting region II takes place in the same way as in Fig.~\ref{fig:pulsedynamics}\textcolor{blue}{(a)}.
At the I - II interface, the inter-mode interactions vary on a length scale $\Delta x \gg v_{\pm}\tau$.
For characteristic frequencies $\omega \sim 1/\tau$, the $+$ $(-)$ eigenmodes are then adiabatically transmitted into the $\nu_1=1$ (respectively, $\nu_2 = 1/3$) mode, yielding pulses of duration $\tau$ and with amplitudes $q_{1,h}^{\text{(p)}}$ and $q_{2,h}^{\text{(p)}}$, respectively, see
Eq.~\eqref{eq:adiabatic_charges}.
The amplitudes $q_{1,h}^{\text{(p)}}$ and $q_{2,h}^{\text{(p)}}$ determine the fractionalized charges observed in FCS, Eq.~\eqref{eq:43chargefluctuations}. 
In addition, the low-frequency components of the counting pulses in region II, for which the interface is effectively sharp,  generate, upon traversal the interface to region I, pulses with a large temporal width $\sim \Delta x /v_{\pm}$ and a small amplitude suppressed by the factor $v_{\pm} \tau /\Delta x$. These low-frequency components of the counting phases in region I contribute only to the average charge crossing the point $x_0$ in time $\tau$, ensuring charge conservation, Eq.~\eqref{eq:kappa_43_adiabatic_2_low}.}
\label{fig:pulsedynamics_adiabatic}
\end{figure}

To evaluate the determinants in Eq.~\eqref{eq:FCS_25} for the case of adiabatic interfaces, we first separate the counting pulses into high-frequency ($\omega >v_{\pm}/\Delta x$) and the low-frequency components ($\omega <v_{\pm}/\Delta x$). The total determinant for each of the modes 1 and 1/3 can then be expressed, to leading approximation, in terms of high-  and low-frequency sectors (labeled  $h$ and $l$, respectively),
\begin{subequations}
\label{eq:FCS_43_long_time_int_region_adiabatic}
\begin{align}
\overline{\Delta}[\delta_{1,\tau} (t)] \simeq \overline{\Delta}_h[\delta_{1,\tau} (t)] \overline{\Delta}_l[\delta_{1,\tau} (t)]\,, \\ 
\overline{\Delta}[\delta_{2,\tau} (t)] \simeq \overline{\Delta}_h[\delta_{2,\tau} (t)] \overline{\Delta}_l[\delta_{2,\tau} (t)]\,.
\end{align}
\end{subequations}
At high frequencies, the counting pulses are transmitted from region II to region I adiabatically, i.e., according to  Eq.~\eqref{eq:interfacetransmissionamplitudes_adaibatic}. Applying these scattering coefficients to the whole counting pulse, we obtain for each of the modes in the region I a single counting pulse of duration $\tau$, with the amplitudes \eqref{eq:pulsecharges43}  \begin{align} \label{eq:adiabatic_charges}
  q_{1,h}^{\text{(p)}} = q_+ e \,, \quad 
  q_{2,h}^{\text{(p)}} =\frac{q_- }{\sqrt{3}}e\,,
\end{align}
see Fig.~\ref{fig:pulsedynamics_adiabatic}.
This yields for the determinants of scattering phases entering the FCS
\begin{subequations}
\label{eq:HF_43}
\begin{align}
   \overline{\Delta}_h[\delta_{1,\tau} (t)] \simeq  \overline{\Delta} [n_{1,1}q_{1,h}^{\text{(p)}}\lambda w_\tau(t,0)]\,, \\
    \overline{\Delta}_h[\delta_{2,\tau} (t)] \simeq   \overline{\Delta} [n_{1,2}q_{2,h}^{\text{(p)}}\lambda w_\tau(t,0)]\,.
\end{align}
\end{subequations}

As we show below, Eq.~\eqref{eq:adiabatic_charges} yields fractionalized excitation charges in the case of adiabatic interfaces. However, this result for the counting pulses in the region I cannot be the whole truth. Indeed, it is easy to prove, using Eqs.~\eqref{eq:eom25_FCS}
and Eq.~\eqref{eq:totalcharge_xandt} that the time integrals of counting pulses $\int_{-\infty}^\infty dt q_1(x,t) $ and $\int_{-\infty}^\infty dt q_2(x,t) $ are independent of $x$ for $x < x_0$. These integrals are thus given by their values at $x= x_0 - 0$ [determined by the source term in Eqs.~\eqref{eq:eom25_FCS}],
\begin{align}
\label{eq:counting-pulse-integrals}
    \int_{-\infty}^{\infty} dt\,q_1(x,t) = e \tau\,, \quad
     \int_{-\infty}^{\infty} dt\,q_2(x,t) = \frac{e}{3} \tau\,.
\end{align}
While we apply here this conservation law to time integrals of counting pulses, it is in fact nothing but the charge conservation. Clearly, pulses of duration $\tau$ and amplitudes \eqref{eq:pulsecharges43} violate the conservation law \eqref{eq:counting-pulse-integrals} in the presence of intermode interaction ($\gamma\ne 0$). The reason is pointed out above: while we applied the adiabatic transmission to the whole counting pulse, its  low-frequency components in fact experience sharp interfaces. In the time representation, the low-frequency components should necessarily spread over a long time scale $\sim \Delta x / v_\pm$. It follows that there will be an additional contribution to the counting pulses in region I that spread over a scale $\sim \Delta x / v_\pm$ and
can be approximately written as 
\begin{subequations} 
\label{eq:accumulatedcharges_low}
\begin{align}
    q_{1,l} (x,t) = q_{1,+,l}^{\text{(p)}} w_{\Delta x/v_+}(t,-t_{1,+}) + q_{1,-,l}^{\text{(p)}}w_{\Delta x/v_-}(t,-t_{1,-})\,, \\ q_{2,l} (x,t) = q_{2,+,l}^{\text{(p)}} w_{\Delta x/v_+}(t,-t_{2,+}) + q_{2,-,l}^{\text{(p)}} w_{\Delta x/v_-}(t,-t_{2,-})\,,
\end{align}
\end{subequations}
see Fig.~\ref{fig:pulsedynamics_adiabatic}.
The corresponding amplitudes are prescribed by the charge conservation law, which yields 
\begin{subequations}
\label{eq:phases_43_amplitudes_adiabatic}
\begin{align}
    q_{1,+,l}^{\text{(p)}} &= (q_{1,+}^{\text{(p)}}-q_{1,h}^{\text{(p)}}) \Big(\frac{v_{+} \tau}{\Delta x}\Big)\,, \\
    q_{1,-,l}^{\text{(p)}} &=  q_{1,-}^{\text{(p)}} \Big(\frac{v_{-} \tau}{\Delta x} \Big)\,, \\
    q_{2,+,l}^{\text{(p)}} &= q_{2,+}^{\text{(p)}} \Big(\frac{v_{+} \tau}{\Delta x} \Big)\,,  \\
    q_{2,-,l}^{\text{(p)}} &=  (q_{2,-}^{\text{(p)}}-q_{2,h}^{\text{(p)}}) \Big( \frac{v_{-} \tau}{\Delta x} \Big)\,.
\end{align}
\end{subequations}
Here  $q_{1,\pm}^{\text{(p)}}$ and $q_{2,\pm}^{\text{(p)}}$ are amplitudes of counting pulses in each of the eigenmodes $\pm$ in region II, which are given by Eq.~\eqref{eq:pulsecharges43}. 

The low-frequency charge pulses~\eqref{eq:accumulatedcharges_low} contribute to the determinants through
\begin{align}
\label{eq:FCS_43_long_time_int_region_adiabatic_phases_low}
\overline{\Delta}_l[\delta_{1,\tau} (t)] \simeq &\prod_{s=\pm} \overline{\Delta}[n_{1,1}   q^{\text{(p)}}_{1,s,l} \lambda w_{\Delta x/v_{s}}(t,0)]\,, \nonumber \\  \overline{\Delta}_l[\delta_{2,\tau} (t)] \simeq &\prod_{s=\pm}
\overline{\Delta}[n_{1,2}  q^{\text{(p)}}_{2,s,l} \lambda w_{\Delta x/v_{s}}(t,0)]\,. 
\end{align}
We note that the amplitudes \eqref{eq:phases_43_amplitudes_adiabatic} are very small since they are
strongly suppressed by the factors $ v_{\pm} \tau/\Delta x\ll 1$, which
will allow us below to expand  with respect to them. This justifies our approximation by rectangular pulses in
Eq.~\eqref{eq:accumulatedcharges_low}: in fact, the precise temporal shape of the contribution $q_{1,l}(x,t)$ and 
$q_{2,l}(x,t)$ is not essential. Only the corresponding time integrals are important, and they are dictated by the charge conservation \eqref{eq:counting-pulse-integrals}.

We evaluate now the FCS generating function~\eqref{eq:FCS_25} for the non-equilibrium situation described by the distribution functions \eqref{eq:doublestep_43} as already considered for sharp interfaces in Sec.~\ref{sec:43_FCS_sharp}.
Similar to the case of sharp interfaces, we apply the Szeg\H{o} approximation to the determinant \eqref{eq:FCS_43_long_time_int_region_adiabatic}, which
leads to the generating functions 
\begin{align}
\label{eq:kappa_43_adiabatic_1}
    \kappa(\lambda,x_0,\tau)
    &\simeq 
    \kappa_{1, l}(\lambda,x_0,\tau) \kappa_{1, h}(\lambda,x_0,\tau) \nonumber \\  \quad & \times 
    \kappa_{2, l}(\lambda,x_0,\tau)  \kappa_{2, h}(\lambda,x_0,\tau)\,.
\end{align}
Here, in natural notations, the factors labeled by $h$ originate from the counting pulses with the amplitudes 
\eqref{eq:adiabatic_charges}, 
\begin{subequations}\label{eq:kappa_43_adiabatic_2_high}
\begin{align}
 \kappa_{1,h}(\lambda,x_0,\tau)
    &=
    \exp\!\left[
    - \frac{e V_1 \mathcal{T}_1\tau}{2\pi n_{1,1}}
    \left(1- e^{- i \lambda n_{1,1} q_{1,h}^{\text{(p)}}/e}\right)
    \right]\,, \\ 
    \kappa_{2, h}(\lambda,x_0,\tau)
    &=
    \exp\!\left[
    - \frac{e V_2 \mathcal{T}_2\tau}{2\pi n_{1,2}}
    \left(1- e^{- i \lambda n_{1,2} q_{2,h}^{\text{(p)}}/e}\right)
    \right]\,, 
\end{align}
\end{subequations}
while 
the factors labeled by $l$ are produced by the additional low-frequency counting-pulse contribution
\eqref{eq:accumulatedcharges_low},  
\begin{subequations}
\label{eq:kappa_43_adiabatic_2_low}
\begin{align}
 \kappa_{1,l}(\lambda,x_0,\tau)
    &=
    \exp\!\left[-
    \frac{ie^2 V_1 \mathcal{T}_1\lambda\tau}{2\pi}
    (1-q_+) \right]\,, \\ 
    \kappa_{2, l}(\lambda,x_0,\tau)
    &=
   \exp\!\left[-
    \frac{ie^2 V_2\mathcal{T}_2\lambda\tau}{2\pi} \Big(  \frac{1}{3}- \frac{q_-}{\sqrt{3}}\Big)
    \right]\,.
\end{align}
\end{subequations}
When deriving Eq.~\eqref{eq:kappa_43_adiabatic_2_low}, we used the smallness of the amplitudes \eqref{eq:phases_43_amplitudes_adiabatic}
due to the adiabatic condition $v_{\pm}\tau/\Delta x \ll 1$ and kept only the leading order contribution in this parameter in the exponential.   
 
From Eqs.~\eqref{eq:kappa_43_adiabatic_1}-\eqref{eq:kappa_43_adiabatic_2_low}, we obtain all cumulants of charge fluctuations by using Eq.~\eqref{eq:FCS_cumulants}. 
Since the exponents in Eq.~\eqref{eq:kappa_43_adiabatic_2_low} are linear in $\lambda$, these factors only contribute to the average charge, without affecting higher moments. 
The total average charge (i.e., the $k=1$ cumulant] passing across the point $x_0$ during the time interval $\tau$ is found as
\begin{align}
\label{eq:average_charge_43_adiabatic}
\langle Q \rangle &=  \langle  Q_1\rangle + \langle Q_2\rangle  \,, \nonumber \\ \text{with }\langle Q_{1}\rangle  &= \frac{\mathcal{T}_1 e^2 V_1 \tau}{2\pi}\,, \quad 
    \langle Q_{2}\rangle  = \frac{1}{3}\frac{\mathcal{T}_2 e^2 V_2 \tau}{2\pi}\,.
\end{align}
It is independent of interactions and obeys charge conservation (i.e., is equal to the average injected charge during the same time interval) as expected. The higher cumulants ($k>1$) evaluate to
\begin{align} \label{eq:43chargefluctuations}
     \langle\langle(\delta Q)^k \rangle\rangle &= 
    (n_{1,1} q_+e)^{k-1}  \langle Q_{+}\rangle +  (n_{1,2} \frac{q_-}{\sqrt{3}} e)^{k-1}  \langle Q_{-}\rangle \nonumber \\ \text{with }\langle Q_{+}\rangle  &= \frac{q_+\mathcal{T}_1 e^2 V_1 \tau}{2\pi}\,, \quad 
    \langle Q_{-}\rangle  = \frac{q_-\mathcal{T}_2 e^2 V_2 \tau}{2\pi \sqrt{3}}\,.
\end{align}

The results~\eqref{eq:average_charge_43_adiabatic}-\eqref{eq:43chargefluctuations} constitute a superposition of FCS of two independent Poissonian processes with fractionalized quasiparticles charges $e q_+ n_{1,1}$ and $e q_- n_{1,2}/\sqrt{3}$.
[We recall that $q_+$ and $q_-$ here are the eigenmode charges given by Eq.~\eqref{eq:eigencharges_25}.] 
As in the case of sharp interfaces, these fractional charges are continuous
functions of the interaction parameter $\gamma$ and thus not quantized.
The physical interpretation of this result (which is an ``adiabatic counterpart'' of the physical picture for the case of sharp interfaces discussed in the end of
Sec.~\ref{sec:43_FCS_sharp}
and illustrated in
Fig.~\ref{fig:pulsedynamics}) is as follows. In the region I, we have a dilute Poissonian train of excitations with the charge $n_{1,1}e$ in the mode $\nu_1=1$ and of excitations with the charge $n_{1,2}e/3$ in the mode $\nu_2 =1/3$. When crossing the adiabatically smooth interface to region II, an excitation $n_{1,1}e$ from the mode 1 gives rise to an excitation with a charge $e q_+ n_{1,1}$, and an excitation $n_{1,2}e/3$ from the mode 1/3 gives rise to an excitation with a charge $e q_- n_{1,2}/\sqrt{3}$. This result cannot be the whole truth, however, since it would violate charge conservation. The remaining charge in each scattering process spreads over a large distance $\sim \Delta x$. These charges, originating from all scattering processes, combine into an essentially uniform charge background, which is  represented by Eq.~\eqref{eq:kappa_43_adiabatic_2_low}
and is essentially a shift of the chemical potentials of the 1 and 1/3 modes in the interacting region II. 

In the non-interacting limit $\gamma\rightarrow 0$, the fractionalized charges $e q_+ n_{1,1}$ and $e q_- n_{1,2}/\sqrt{3}$ reduce to the injected charges $e n_{1,1}$ and $e n_{1,2}/3$, as they should. At the special value of the interaction, $\gamma=\tan^{-1}(1/\sqrt{3})=\pi/6$, a charge-neutral decoupling occurs: The charges become $eq_+=2e n_{1,1}/\sqrt{3}$ and $eq_-=0$, respectively, with only one of the eigenmodes contributing to the FCS.

This completes our analysis of the FCS 
for an interacting edge with co-propagating modes ($\nu=4/3$). 
Below, in Sec.~\ref{sec:GFs_co} and \ref{sec:transport_co}, we will 
demonstrate how the interaction-induced fractionalization observed in FCS modifies the counting phases entering the GFs and, consequently, alters the tunneling transport properties.  

\subsection{Co-propagating edge modes: Green's functions}
\label{sec:GFs_co}

We now compute the GFs for generic edge excitations in the $\nu = 4/3$ state. These excitations are described by the vertex operators 
\begin{align}
\label{eq:vertex_43}
   \psi^{\dagger}_{\vec{n}_2} \sim e^{-i \vec{n}_2 \cdot \vec{\phi}} = \exp [-i (n_{2,1} \phi_1 + n_{2,2}\phi_2)]\,, 
\end{align}
where the integer-valued vector $\vec{n}_2=(n_{2,1},n_{2,2})^{T}$ specifies the number of elementary excitations in edge modes $\nu_1 = 1$ and $\nu_2 = 1/3$, carrying charges $e$ and $e/3$, respectively. We evaluate the GFs defined in Eq.~\eqref{eq_green_function} at coinciding spatial points $x_1=x_2=0$ within the interacting region II and at the time separation $t_1-t_2\equiv\tau$, see Fig.~\ref{fig:complex_setup}\textcolor{blue}{(a)}. 
In the Keldysh representation, the greater GF  takes the form (the lesser GF is represented analogously)
\begin{align}
\label{eq:GF_43}
    \mathcal{G}^>(\tau) &\equiv \frac{-i}{2\pi a}  \int \mathcal{D}\rho\mathcal{D}\overline{\rho}  e^{iS[\rho,\overline{\rho}]} \notag \\ &\times e^{\frac{i n_{2,1}}{\sqrt{2}} [\phi_{1}(0, \tau) - \phi_{1}(0,0) - \overline{\phi}_{1} (0, \tau) - \overline{\phi}_{1} (0,0) ]} \notag \\ &\times e^{\frac{i n_{2,2}}{\sqrt{2}} [\phi_{2}(0, \tau) - \phi_{2}(0,0) - \overline{\phi}_{2} (0, \tau) - \overline{\phi}_{2} (0,0) ]}\,,
   \end{align}
 with the action $S[\rho,\overline{\rho}]$ given in Eq.~\eqref{eq:action_43}.  Local GFs of this type are relevant for describing QPC geometries where quasiparticles tunnel between edges, see Sec.~\ref{sec:transport_co}, or  inter-mode scattering due to an isolated edge impurity. 

The calculation of the GFs closely parallels that for the FCS presented in Sec.~\ref{sec:FCS_co}. Proceeding in the same way, we come to a pair of coupled equations of motions for the densities analogous to Eq.~\eqref{eq:eom25_FCS} for the FCS. However, the source term is now different. Specifically, we obtain the following equations for the GF problem:
\begin{align}
\label{eq:eom43_GF}
&K \partial_t  \begin{pmatrix}
        \overline{\rho}_1 \\
        \overline{\rho}_{2} 
    \end{pmatrix} + \partial_x \begin{pmatrix}
        v_1 & u\\
        u & 3v_{2}
    \end{pmatrix}\begin{pmatrix}
        \overline{\rho}_1 \\
        \overline{\rho}_{2} 
    \end{pmatrix} = - j(x,t) \vec{n}_2 \,, 
\end{align}
with the source term $j(x,t)$ defined in Eq.~\eqref{eq:source_j}. 
We diagonalize the matrix stricture in Eq.~\eqref{eq:eom43_GF} by employing the transformation matrix \eqref{eq:Lambda_25} and pass to the frequency representation, which yields
\begin{align}
\label{eq:eom43_diag}
    &-i\omega  \begin{pmatrix}
        \overline{\rho}_+ \\
        \overline{\rho}_- 
    \end{pmatrix} + \partial_x \begin{pmatrix}
         v_+ && 0\\
        0 && v_-
    \end{pmatrix} \begin{pmatrix}
        \overline{\rho}_+ \\
        \overline{\rho}_- 
    \end{pmatrix}  = - j(x,\omega) \begin{pmatrix}
        n_{2,+}\\ n_{2,-}
    \end{pmatrix} \,,
\end{align}
where the rotated source components 
$n_{2,+}$ and $n_{2,-}$ are given by
\begin{align}
    \begin{pmatrix}
        n_{2,+} \\ n_{2,-}
    \end{pmatrix} = \Lambda^T \vec{n}_2 = \begin{pmatrix}
       \cos \gamma\, n_{2,1}+\frac{\sin \gamma}{\sqrt{3}} n_{2,2} \\[0.2cm] -\sin\gamma \,n_{2,1}+ \frac{\cos \gamma}{\sqrt{3}} n_{2,2}
    \end{pmatrix}.
\end{align}
A direct comparison of the diagonal equations of motion~\eqref{eq:eom43_diag} with the equations of motion for the FCS, Eq.~\eqref{eq:eom25_FCS}, shows that the two sets of equations are identical up to a replacement of the source terms,
\begin{align} \label{eq:FCStoGFs_43}
    \frac{\lambda }{2\pi} q_{\pm} \mapsto n_{2,\pm}\,. 
\end{align}
This correspondence allows us to apply the results obtained in Sec.~\ref{sec:FCS_co} to the present case, by implementing the substitution \eqref{eq:FCStoGFs_43}.   
With this substitution, we obtain the scattering phases for the GF problem [cf. the counting phases for FCS, Eq.~\eqref{eq:phases_43_FCS}]
\begin{subequations}
\label{eq:phases_25}
\begin{align}
    &\delta_{1, \tau}(t)=\delta_{1,+} w_{\tau}(t,-t_{1,+})+\delta_{1,-} w_{\tau}(t,-t_{1,-}),\\
    &\delta_{2, \tau}(t)=\delta_{2,+} w_{\tau}(t,-t_{2,+})+\delta_{2,-} w_{\tau}(t,-t_{2,-}),
\end{align}
\end{subequations}
where the amplitudes of the individual pulses are 
\begin{subequations}
\label{eq:phase_amplitudes_43_1}
\begin{align}
    \delta_{1,+} = 2\pi t_L n_{1,1} n_{2,+}\,,&\quad  
   \delta_{1,-} = 2\pi r_L n_{1,1} n_{2,-}, \\
   \delta_{2,+} = - \frac{2\pi r_{L} n_{1,2} n_{2,+}}{\sqrt{3}} \,, &\quad 
   \delta_{2,-} = \frac{2\pi t_{L} n_{1,2} n_{2,-}}{\sqrt{3}}\,, 
\end{align}
\end{subequations}
with the window function $w_\tau(t_1,t_2)$ given by Eq.~\eqref{eq:window_function}, the time offsets by Eq.~\eqref{eq:timeshifts43}, and the scattering amplitudes (for sharp interfaces) by Eq.~\eqref{eq:interfacetransmissionamplitudes}. 

Importantly, the phases $\delta_{1,\pm}$ and $\delta_{2,\pm}$ entering the expressions for the GFs (explicit results are given below) admit a direct physical interpretation as mutual braiding phases of the quasiparticle \eqref{eq:vertex_43} that tunnels between the edges with quasiparticles determining charge fluctuations (as revealed by the FCS) within the edge. We make it more explicit, considering for definiteness the case of sharp interfaces. The quasiparticles that are injected in the modes in the region I, see Eq.~\eqref{eq:excitations43}, are labeled  by the two-component vectors
\begin{align}
    \vec n_1^{(1)}\equiv (n_{1,1},0)\,,\qquad 
\vec n_1^{(2)}\equiv (0,n_{1,2})\,. 
\end{align}
Their injection drives out of equilibrium the $\nu_1= 1$ and $\nu_2=1/3$ modes, respectively. 
 Each of these excitations fractionalizes, upon crossing the interface to the region II, into two excitations corresponding to $\pm$ eigenmodes, see a discussion in Sec.~\ref{sec:43_FCS_sharp}
and Fig.~\ref{fig:pulsedynamics}(b,c). 
 The resulting fractionalized excitations
 are described by the following (interaction-dependent) vectors,
\begin{align}
    \vec{n}^{(a)}_{1,\pm} = (\Lambda^{-1})^T P_{\pm} \Lambda^T \vec{n}^{(a)}_1\,,\qquad a \in \{1,2\}\,, 
\end{align}
where $P_{\pm} = (1\pm \sigma_z)/2$ projects onto the $\pm$ eigenmode.
The mutual braiding phases of the 
(quantized) type-$\vec{n}_2$ with (non-quantized)  type-$\vec{n}^{(a)}_{1,\pm}$ quasiparticles read 
\begin{align}
\label{eq:43-frac-braiding-phases}
     2 \theta_{12,\pm}^{(a)} = 2\pi (\vec{n}_{1,\pm}^{(a)})^T K^{-1}\vec{n}_2 = \delta_{a,\pm}\,, \quad a \in \{1,2\}  \,.
\end{align}
These are precisely the phases $\delta_{1,\pm}$ and $\delta_{2,\pm}$,
Eq.~\eqref{eq:phase_amplitudes_43_1},
that we found when evaluating the GFs [see the determinant formula \eqref{eq:G1_final} for the GFs below].
It is worth noting that the sum of the phases for excitations that resulted from the fractionalization of an injected excitation $\vec{n}_1^{(a)}$ yields the mutual braiding phase between $\vec{n}_1^{(a)}$ and $\vec{n}_2$ excitations:
\begin{align}
\label{eq:braiding-phases-sum-rule}
    \delta_{a,+} + \delta_{a,-} = 2\theta_{12}^{(a)} = 2\pi (\vec{n}_1^{(a)})^T K^{-1} \vec{n}_2\,, 
\end{align}
which is a manifestation of the fact that the mutual braiding phase depends linearly on the excitation vector. In other words, the interaction-induced fractionalization of excitations leads to fractionalization of braiding phases. As we will see in Sec.~\ref{sec:transport_co}, these fractionalized braiding phases can be experimentally accessed by measuring Fano factors in the QPC geometry of Fig.~\ref{fig:QPC}.

As discussed above in Sec.~\ref{sec:FCS_co}, two distinct regimes can be identified depending on the relation between the system size $L$ and the relevant time scale $\tau$: the long-time (or, equivalently, short-length) limit $L/2 \ll v_{\pm}|\tau|$  and the short-time (long-length) limit $L/2 \gg v_{\pm}|\tau|$. (Here we assumed that the observation point $x_1=x_2$ is located in the central part of the region II, at a distance $\sim L/2$ from the left interface.)
The edge transport properties of interest, as described in Sec.~\ref{sec:tunneling}, are dominantly governed by the behavior of the GFs on relevant characteristic time scales. In equilibrium at temperature $T$, the dominant contribution to transport arises from times $\tau \sim 1/(k_B T)$. For dilute injections of anyons, which is the case of our main interest, the relevant time scale is instead $\tau \sim 1/(e_1^* V \mathcal{T})$. Accordingly, we approximate the GFs by asymptotic forms that are valid in that time range that dominantly contributes to the transport. Our main focus will be on GFs in the long-length limit, $|\tau| \ll L/(2v_{\pm})$, i.e., under the assumption that $L$ constitutes the largest length scale in the problem, since this is the regime where the intermode interaction crucially affects the results. We will also briefly comment on the short-length limit, in which the interaction becomes essentially irrelevant. 

In the long-$L$ limit, each determinant in the GFs splits into two parts~\cite{Gutman2010, Protopopov2013correlations} 
\begin{subequations}\label{eq:countingphases_43_GFs}
\begin{align}
\overline{\Delta}[\delta_{1, \tau}(t)]&\simeq \overline{\Delta}[\delta_{1,+}w_\tau(t,0)] \overline{\Delta}[\delta_{1,-}w_\tau(t,0)], \\
\overline{\Delta}[\delta_{2,\tau}(t)] &\simeq \overline{\Delta}[\delta_{2,+}w_\tau(t,0)]\overline{\Delta}[\delta_{2,-}w_\tau(t,0)],
\end{align}
\end{subequations}
with the phase amplitudes~\eqref{eq:phase_amplitudes_43_1}. 

Following the same procedure for the evaluation of the GFs as in Sec.~\ref{sec:GF_Laughlin_formalism} for the Laughlin edge and
using Eq.~\eqref{eq:countingphases_43_GFs} 
for the determinants in the long-length limit, we obtain the GFs  in the form
\begin{align}
\label{eq:G1_final}
    \mathcal{G}^{\gtrless}(\tau)  &\simeq \frac{\mp i}{2\pi a} \left(\frac{a}{ a \pm i v_{+} \tau}\right)^{\alpha} \left(\frac{a}{ a \pm i v_{-} \tau}\right)^{\beta} \nonumber 
    \\ &\times  \left\{\overline{\Delta}[\delta_{1,+}w_\tau(t,0)] \overline{\Delta}[\delta_{1,-}w_\tau(t,0)] \right\}^{1/n_{1,1}^2} \nonumber \\ & \times \left\{\overline{\Delta}[\delta_{2,+}w_\tau(t,0)] \overline{\Delta}[\delta_{2,-}w_\tau(t,0)] \right\}^{3/n_{1,2}^2}\,.
\end{align}
This is an extension of Eq.~\eqref{eq:GFgreater_gen_n} for a non-equilibrium Laughlin edge. 
Here, the (non-universal) power-law exponents $\alpha$ and $\beta$ are given by 
\begin{subequations} \label{eq:powerlawexp_43}
\begin{align} 
    \alpha &= \frac{1}{n_{1,1}^2}\sum_{s=\pm} \left(\frac{\delta_{1,s}}{2\pi}\right)^2 =  (t_Ln_{2,+})^2+(r_Ln_{2,-})^2\,, \\
    \beta &= \frac{3}{n_{1,2}^2}\sum_{s=\pm} \left(\frac{\delta_{2,s}}{2\pi}\right)^2 = (r_Ln_{2,+})^2+(t_Ln_{2,-})^2 \,.
\end{align}
\end{subequations}
These exponents satisfy
\begin{align} \label{eq:scalingdimenson_43}
\alpha + \beta =  n_{2,+}^2+n_{2,-}^2= n_{2,1}^{2} + \frac{n_{2,2}^{2}}{3}=\zeta,
\end{align}
with $\zeta$ being the equilibrium scaling dimension of the edge excitation~\eqref{eq:vertex_43}. This is in agreement with the known result that, for co-propagating modes, the equilibrium scaling dimension $\zeta$ is independent of the interaction strength $u$ \cite{Wen_Topological_1995}.

We also discuss the GFs in the short-length limit $L/2 \ll v_{\pm}|\tau|$.  [More accurately, to take into account also the case of a weak interaction, the condition should be written as $ |v_-^{-1} - v_+^{-1}| L/2 \ll |\tau|$.]
In this limit, the pulses in Eq.~\eqref{eq:phases_25} remain almost completely overlapping. In analogy with the derivation of the FCS in this limit, Sec.~\ref{sec:43_FCS_sharp}, we find
\begin{align}
\label{eq:G_43_short_length}
    \mathcal{G}^{\gtrless}(\tau)  &=\frac{\mp i}{2\pi a} \left(\frac{a}{ a \pm i v_{1} \tau}\right)^{n_{2,1}^2} \left(\frac{a}{ a \pm i v_{2} \tau}\right)^{n_{2,2}^2/3} \nonumber 
    \\ &\times  \left\{\overline{\Delta}[2\pi n_{1,1}n_{2,1}w_\tau(t,0)] \right\}^{1/n_{1,1}^2} \nonumber \\ & \times \left\{\overline{\Delta}[2\pi n_{1,2}n_{2,2}w_\tau(t,0)/3]  \right\}^{3/n_{1,2}^2}\,.
\end{align}
This is nothing but the result that we would obtain in the absence of interactions: The GFs~\eqref{eq:G_43_short_length} are equal (up to a constant prefactor) to a product of GFs~\eqref{eq:GFgreater_gen_n} of decoupled $\nu = 1$  and $\nu = 1/3$  edges. 
Physically, the fact that the interaction-induced fractionalization is not operative in this regime is explained as follows: When a quasiparticle experiences fractionalization at the I - II interface, two resulting fractionalized quasiparticles in region II arrive at the point where the GFs are studied with a time difference $ \sim |v_-^{-1} - v_+^{-1}| L/2$. When the GFs are studied on a much larger time scale $\tau$, these two quasiparticles cannot be considered as uncorrelated but rather combine coherently, thus acting as a single quasiparticle coming from region I. 

In connection to our analysis of the long-$L$ and short-$L$ limits, it is instructive to compare with works on dc transport in a conventional Luttinger-liquid wire attached to non-interacting leads \cite{Safi_Transport_1995,Maslov_Landauer_conductance_1995,Ponomarenko_Renormalization_1995}. There, the counterpart of our characteristic time $\tau$ is set by the frequency of the applied voltage, $\tau \sim 1/ \omega$. Thus, the dc limit corresponds to $\tau \to \infty$, i.e., to the short-$L$ limit, for any finite length $L$ of the interacting wire. Consequently, the dc transport is insensitive to the interactions. The interaction-induced fractionalization becomes manifest \cite{Safi_Transport_1995} only in the high-frequency regime, which thus corresponds to our long-$L$ limit.

We return now to the long-$L$ limit and analyze the implications of Eq.~\eqref{eq:G1_final} out of equilibrium. Before turning to the case 
of the double-step non-equilibrium distributions~\eqref{eq:doublestep_43} ``injected'' in region I, we briefly consider the case of partial equilibrium when the two modes in region I are separately in equilibrium, Eq.~\eqref{eq:n_T}, but with two different temperatures $T_1$ and $T_2$. In this case, we find~\cite{Supplemental_Material} that Eq.~\eqref{eq:G1_final} reduces to
\begin{align}
\label{eq:43GF_partial_eq}
    \mathcal{G}^{\gtrless}(\tau)  =\frac{\mp i}{2\pi a} &\left(\frac{a}{ a \pm i v_{+} \tau}\right)^{\alpha} \left(\frac{\pi T_1 \tau}{\sinh(\pi T_1 \tau)}\right)^{\alpha} \nonumber 
    \\ \times  &\left(\frac{a}{ a \pm i v_{-} \tau}\right)^{\beta} \left(\frac{\pi T_{2} \tau}{\sinh(\pi T_{2} \tau)}\right)^{\beta}\,,
\end{align}
with exponents $\alpha$ and $\beta$ from Eq.~\eqref{eq:powerlawexp_43}. In the global equilibrium case, $T_1=T_{2}=T$, Eq.~\eqref{eq:43GF_partial_eq} takes the familiar equilibrium form if one uses the identity  \eqref{eq:scalingdimenson_43} for the scaling exponents, $\alpha+\beta=\zeta$.

We consider now the non-equilibrium situation of our main interest, with the states of the incoming 1 and 1/3 modes characterized by double-step distributions \eqref{eq:doublestep_43} with $\mathcal{T}_1,\mathcal{T}_2 \ll 1$.
Since the mutual braiding phases are continuous functions of the interaction, we can safely assume that they are not equal to an integer multiple of $2\pi$. In view of this, the Szeg\H{o} approximation is sufficient to capture the leading contribution to the transport properties; see Sec.~\ref{sec:tun-noneq-Laughlin-edges} for a detailed discussion.

We further consider the two different interface types previously analyzed for the FCS: the sharp interfaces discussed in Sec.~\ref{sec:43_FCS_sharp} and the adiabatic interfaces of Sec.~\ref{sec:adiabatic_interfaces_43}. For sharp interfaces, the Szeg\H{o} approximation yields GFs 
\begin{align} \label{eq:GFs_doublesteup_43}
    &\mathcal{G}^{\gtrless}(\tau)  \simeq \frac{\mp i}{2\pi a} \left(\frac{a}{ a \pm i v_{+} \tau}\right)^{\alpha} \left(\frac{a}{ a \pm i v_{-} \tau}\right)^{\beta} \nonumber 
    \\ & \quad \times \exp\Big[-
 \frac{\mathcal{T}_1|e V_1 \tau|}{2\pi n_{1,1}}\sum_{s=\pm}(1-e^{-i \delta_{1,s} \text{sgn}(eV_1\tau)})\Big] \nonumber \\ & \quad  \times \exp\Big[-
 \frac{\mathcal{T}_2|e V_2 \tau|}{2\pi n_{1,2}}\sum_{s=\pm}(1-e^{-i \delta_{2,s} \text{sgn}(eV_2\tau)})\Big]\,. 
\end{align}
We see here explicitly how the fractionalized scattering phases \eqref{eq:phase_amplitudes_43_1}, which have a meaning of fractionalized braiding phases, Eq.~\eqref{eq:43-frac-braiding-phases}, determine the exponential decay rate (dephasing) and oscillations of the GFs with the time $\tau$.
For the adiabatic interfaces, we obtain instead 
\begin{align} \label{eq:GFs_doublestep_43_adiabatic}
   &\mathcal{G}^{\gtrless}(\tau)  \simeq \frac{\mp i}{2\pi a} \left(\frac{a}{ a \pm i v_{+} \tau}\right)^{\alpha} \left(\frac{a}{ a \pm i v_{-} \tau}\right)^{\beta} \nonumber 
    \\ & \quad \times \exp\Big[-
 \frac{\mathcal{T}_1|e V_1 \tau|}{2\pi n_{1,1}}(1-e^{-i 2\pi n_{1,1} n_{2,+} \text{sgn}(eV_1\tau)})\Big] \nonumber \\ & \quad  \times \exp\Big[-
 \frac{\mathcal{T}_2|e V_2 \tau|}{2\pi n_{1,2}}(1-e^{-i 2\pi n_{1,2} n_{2,-} \text{sgn}(eV_2\tau)/\sqrt{3}})\Big]\nonumber \\ & 
 \quad \times \exp[i e \tau \Big \{V_1 \mathcal{T}_1 (n_{2,+}-n_{2,1}) + V_2 \mathcal{T}_2 \big(\frac{n_{2,-}}{\sqrt{3}}- \frac{n_{2,2}}{3}\big ) \Big \} ]\,.  
\end{align}
The origin of the last factor is the same as of its counterpart in the result for the FCS, see Eq.~\eqref{eq:kappa_43_adiabatic_2_low}.

Having evaluated the quasiparticle GFs on a non-equilibrium interacting $\nu=4/3$ edge, we will now use these results in Sec.~\ref{sec:transport_co} to analyze the tunneling current and noise for a QPC connecting two such edges.

\subsection{Co-propagating edge modes: Transport properties}
\label{sec:transport_co}

We consider again the QPC setup shown in Fig.~\ref{fig:QPC}, now for tunneling between $\nu =4/3$ edges. In principle, one can consider each of the edges to be in a non-equilibrium state characterized by the GFs \eqref{eq:GFs_doublesteup_43} or \eqref{eq:GFs_doublestep_43_adiabatic}, in general with different parameters for the two edges. For the sake of physical transparency, we will focus here on the following ``minimal'' non-equilibrium setup: We will assume only the upper edge $u$ to be driven out of equilibrium,  while the lower edge $d$ is in equilibrium. Further, we will assume that only the mode $\nu_2= 1/3$ of the upper edge is driven out of equilibrium, and that this is achieved by dilute injection of elementary excitations with charge $e_{1,2}^* = e/3$ (i.e., $n_{1,2}=1$)
with the double-step distribution~\eqref{eq:doublestep_43} characterized by the voltage $V_u \equiv V$
and the transmission coefficient $\mathcal{T}_{2,u} \equiv \mathcal{T}$. 
Finally, we will assume that the transport between the edges occurs by tunneling of elementary (charge $e/3$) excitations of the $1/3$ modes, i.e., $(n_{2,1}, n_{2,2}) = (0,1)$. In the absence of interactions, the modes with $\nu_1=1$ do not play any role, and this setup reduces to that for tunneling between two $\nu=1/3$ Laughlin edges (one of which is out of equilibrium) as studied in Sec.~\ref{sec:tun-noneq-Laughlin-edges}. Thus, a comparison between the results for the interacting $\nu=4/3$ edges with those for the Laughlin edges will allow us to assess the impact of inter-channel interactions on tunneling transport properties.

In the absence of interactions (Laughlin-edge limit), we can directly use 
Eqs.~\eqref{eq:current_dilute_app},~\eqref{eq:noise_dilute_app}, and~\eqref{eq:gen_fano}.
from Sec.~\ref{sec:tun-noneq-Laughlin-edges}, with the parameters $V_0= V_d = 0$, $V_u = V$, $\mathcal{T}_u  = \mathcal{T}$, and $n_1= n_2 = 1$.  This yields for the tunneling current, noise, and Fano factor: 
\begin{subequations}
\begin{align} 
    \langle I_T \rangle &=\frac{e_2^* |\xi|^2}{2 \pi a v} \Big( \frac{\pi v}{a eV\mathcal{T} \sin(\pi \nu)} \Big)^{1-2\nu} 
     \notag \\& \times \frac{\cos[2\pi \nu (1-\nu)] \text{sgn}(eV)}{\Gamma(2 \nu) \cos(\pi \nu)} \,, \\
     \label{eq:ST_Laughlin}
    S_T &= \frac{(e_2^*)^2 |\xi|^2}{\pi a v} \Big( \frac{\pi v}{a eV\mathcal{T} \sin(\pi \nu)} \Big)^{1-2\nu} \notag \\& \times \frac{\sin[2\pi \nu (1-\nu)]}{\Gamma(2 \nu) \sin(\pi \nu)}\,, \\ F &= \nu \cot(\pi \nu) \tan[2\pi \nu (1-\nu)]\text{sgn}(V)\,,
    \label{eq:Fano_Laughlin_13}
\end{align}
\end{subequations}
with $\nu = 1/3$. We note that the scaling of the tunneling noise~\eqref{eq:ST_Laughlin} with the current on the edge, $S_T\sim \langle I_0 \rangle^{2\nu-1}$ [see Eq.~\eqref{eq:average_I} for the identification of the product $V\mathcal{T}$  in terms of the average edge current $\langle I_0 \rangle$], was found in Ref.~\cite{Levkivskyi2016Apr}.

For the model with inter-mode interactions $\gamma \ne 0$, we consider first the transport properties for sharp interfaces. In the long-$L$ limit, the tunneling current is evaluated by inserting
the non-equilibrium GFs \eqref{eq:GFs_doublesteup_43}, 
into the formula for the tunneling current~\eqref{eq:tunneling_current}. As we have just discussed, the parameters for the considered setup are
$V_{1,d}=V_{2,d}=V_{1,u}=0$, $V_{2,u}=V$, $\mathcal{T}_{2,u}=\mathcal{T}$, $n_{1,2}= 1$, and 
$(n_{2,1}, n_{2,2})= (0,1)$. With these parameters,  we obtain from Eq.~\eqref{eq:tunneling_current} the following result for the tunneling current: 
\begin{align} 
\label{eq:IT_integral_43}
    \langle I_T \rangle &= \frac{2ie_2^*|\xi|^2}{(2\pi a)^2} 
    \int_{-\infty}^{\infty} d\tau  \left(\frac{a}{a+ iv_+\tau}\right)^{2\alpha} \left(\frac{a}{a+ iv_-\tau}\right)^{2\beta}  \notag \\ &\quad \times e^{-(\gamma_{u,+}+ \gamma_{u,-})|\tau|}\sin(\tau (\omega_{u,+}+\omega_{u,-})) \,,
\end{align}
where the dephasing rates $\gamma_{u,\pm}$ and the oscillation frequencies $\omega_{u,\pm}$ are given by
\begin{align}
    \gamma_{u,\pm} &= \frac{|eV| \mathcal{T} (1-\cos(\delta_{2,\pm}))}{2\pi}\,, \\ 
    \omega_{u,\pm} &= \frac{eV \mathcal{T} \sin(\delta_{2,\pm})}{2\pi}\,,
\end{align}
with the phases
\begin{subequations}
\label{eq:phases_43_quasi}
\begin{align} 
    \delta_{2,+} &= -\frac{2\pi}{\sqrt{3}} r_L n_{2,+} = -\frac{2\pi}{3} r_L \sin\gamma \,, \\
    \delta_{2,-} &= \frac{2\pi}{\sqrt{3}} t_L n_{2,-} = \frac{2\pi}{3} t_L \cos\gamma\,. 
\end{align}
\end{subequations} 
We recall that $t_L = \cos \gamma$ and $r_L = - \sin \gamma$ for sharp interfaces, see Eq.~\eqref{eq:interfacetransmissionamplitudes}. Integrating Eq.~\eqref{eq:IT_integral_43} in the limit $a\to 0$, we find
\begin{align} \label{eq:IT_final_43}
    \langle I_T\rangle &= \frac{e_2^* a^{2\zeta} |\xi|^2}{2(\pi a)^2 v_+^{2\alpha} v_{-}^{2\beta}} \frac{\pi}{\Gamma(2 \zeta) \cos(\pi \zeta)} \notag \\& \times \frac{\sin \big[(1-2\zeta)\tan^{-1} (p)    \big]}{[(\omega_{u,+}+\omega_{u,-})^2 + (\gamma_{u,+}+\gamma_{u,-})^2]^{1/2-\zeta}}\,,
\end{align}
where we identified the scaling dimension $\zeta = \nu_2$ from Eq.~\eqref{eq:scalingdimenson_43}. The dimensionless parameter $p$ [cf. Eq.~\eqref{eq:bias_parameter}] is for the considered setup fully governed by the  $u$ edge (since the $d$ edge is at zero-temperature equilibrium),
\begin{align} \label{eq:dimensionlessparameter_43}
    p &=\frac{\omega_{u,+}+\omega_{u,-}}{\gamma_{u,+}+\gamma_{u,-}}=  \frac{\text{sgn}(eV)\sum_{s=\pm}\sin(\delta_{2,s})}{\sum_{s=\pm}\left[1-\cos(\delta_{2,s})\right]}\,.
\end{align}
In the same way, we evaluate the tunneling noise~\eqref{eq:tunnelin_noise},
\begin{align} \label{eq:noise_integral_43}
    S_T &= \frac{(e_2^*)^2 |\xi|^2}{(\pi a)^2v_+^{2\alpha} v_{-}^{2\beta}} \frac{\pi  a^{2\zeta}}{\Gamma(2 \zeta) \sin(\pi \zeta)} \notag \\ &\times \frac{\cos \big[(1-2\zeta) \tan^{-1} (p) \big]}{[(\omega_{u,+}+\omega_{u,-})^2 + (\gamma_{u,+}+\gamma_{u,-})^2]^{1/2-\zeta}}.
\end{align}
Using Eqs.~\eqref{eq:IT_final_43} and \eqref{eq:noise_integral_43}, we obtain the Fano factor
\begin{align}
\label{eq:Fano_43}
    F= \frac{e_2^*}{|e|} \cot(\pi \nu_2) \cot[(1-2\nu_2)\tan^{-1}(p)]\,. 
\end{align}

For adiabatic interfaces, the calculations are based on the GFs from Eq.~\eqref{eq:GFs_doublestep_43_adiabatic} and are performed in a similar way. 
The resulting Fano factor
has the same form \eqref{eq:Fano_43} but with a modified parameter $p$, 
\begin{align}
\label{eq:p-43-adiabatic}
    p &=\text{sgn}(eV) \frac{\sin [2\pi \nu_2 \cos \gamma]+ 2\pi \nu_2 (1-\cos \gamma)}{1-\cos [2\pi \nu_2 \cos\gamma]}\,.
\end{align} 

\begin{figure}[t!]
    \centering
\subfloat[][]{
\includegraphics[width=0.8\columnwidth]{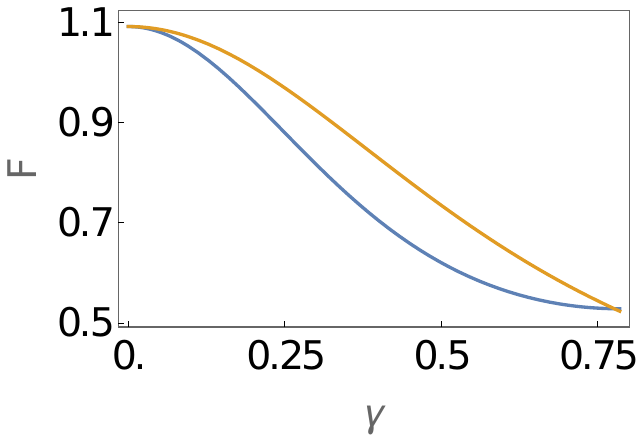}  
\put(-200,120){\large\textbf{(a)}}
}
\par\vspace{-3mm}
\subfloat[][]{
  \includegraphics[width=0.8\columnwidth]{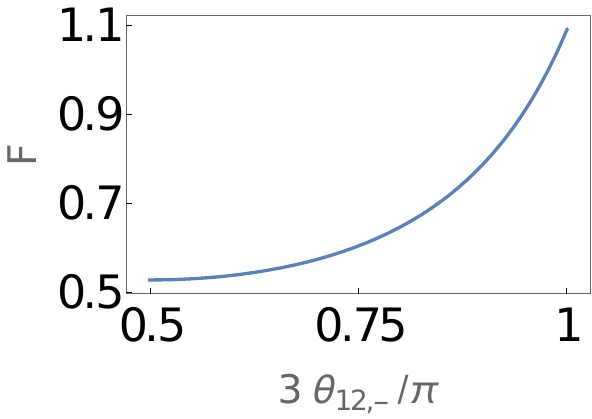}  
  \put(-200,120){\large\textbf{(b)}}
}
\caption{Fano factor $F$, Eq.~\eqref{eq:Fano_43}, for tunneling 
transport between two $\nu=4/3$ edges
in a setup considered in Sec.~\ref{sec:transport_co}.
In this setup, the upper edge $u$ is driven out of equilibrium by quasiparticle injection, under a voltage $V>0$, into the $\nu_2 = 1/3$ mode, while the lower edge $d$ is in equilibrium. 
(a) Fano factor as a function of the intermode interaction parameter $\gamma$ [see Eq.~\eqref{eq:Lambda_25}].
The results for sharp (blue) and adiabatic (orange) interfaces between the non-interacting and interacting regions are shown. For the considered setup, $F(\gamma) = F(-\gamma)$, so that only the region $\gamma \ge 0$  is presented. (b) Fano factor for the case of sharp interfaces as a function of the interaction-dependent mutual braiding phase $2\theta_{12, -}$, Eq.~\eqref{eq:delta_2_minus} (normalized to its non-interacting value $2\pi/3$), between a tunneling quasiparticle with charge $e/3$ and a fractionalized excitation of the $-$ eigenmode.   The braiding phase associated with a fractionalized excitation of the $+$ mode, $2\theta_{12,+}$ in Eq.~\eqref{eq:delta_2_plus}, can be determined from the sum rule~\eqref{eq:sumrulebraiding}. 
}
\label{fig:Fano_plot}
\end{figure}

In the non-interacting limit, $\gamma = 0$, the results for $p$ in both cases of sharp and adiabatic interfaces, 
Eqs.~\eqref{eq:dimensionlessparameter_43} and \eqref{eq:p-43-adiabatic}, straightforwardly reduce to $p = \text{sgn}(e_2^* V)\cot(\pi\nu_2)$, so that the Fano factor \eqref{eq:Fano_43} becomes identical to Eq.~\eqref{eq:Fano_Laughlin_13} for $\nu =\nu_2 = 1/3$ as expected. In the presence of interactions, the Fano factor $F$ exhibits a pronounced dependence on the interaction parameter $\gamma$ as shown in Fig.~\ref{fig:Fano_plot}\textcolor{blue}{(a)}. Furthermore, the values of the Fano factor are different for the cases of sharp and adiabatic interfaces. At the same time, the behavior is qualitatively the same in both cases: increasing the interaction strength leads to a reduction of the Fano factor. 

Let us emphasize the significance of the obtained result, focusing for definiteness on the case of sharp interfaces. As discussed in Sec.~\ref{sec:GFs_co}, the phases 
$\delta_{1,s}$ and $\delta_{2,s}$ entering the expressions for GFs 
are identified with mutual braiding phases between the fractionalized components of injected excitations and the tunneling quasiparticles. According to Eq.~\eqref{eq:43-frac-braiding-phases},
for injected quasiparticles 
described by $\vec{n}_1=(0,1)$, the fractionalization into excitations of the $+$ and $-$ eigenmodes gives rise to braiding phases [with respect to the tunneling quasiparticles $\vec{n}_2=(0,1)$] given by
\begin{subequations}
\begin{align}
\label{eq:delta_2_plus}
\delta_{2,+} &= 2\theta_{12,+} = \frac{2\pi}{3}\sin^2\gamma\,,\\
\delta_{2,-} &= 2\theta_{12,-} = \frac{2\pi}{3}\cos^2\gamma\,,
\label{eq:delta_2_minus}
\end{align}
\end{subequations}
in agreement with Eq.~\eqref{eq:phases_43_quasi}. 
These braiding phases satisfy the sum rule
\begin{align} \label{eq:sumrulebraiding}
2\theta_{12,+}+2\theta_{12,-}=2\theta_{12}=2\pi\nu_2\,,
\end{align}
which is a particular case of Eq.~\eqref{eq:braiding-phases-sum-rule}.
These braiding phases enter the experimentally observable Fano factor. As shown in Fig.~\ref{fig:Fano_plot}\textcolor{blue}{(b)}, measuring the Fano factor as a function of $\gamma$ therefore provides a direct access to the braiding phases $\theta_{12,+}$ and $\theta_{12,-}$. Thus, the Fano factor serves as a probe of the braiding statistics between interaction-induced fractionalized excitations and tunneling anyons.

In this connection, it is instructive to discuss also dependence of the Fano factor on the length $L$. We recall that the system is in the long-length or short-length regime depending on the relation between the characteristic time $\tau$ and 
the propagation times $(L/2+x_0)v^{-1}_{\pm}$ (or, more precisely, their difference), see a discussion in the beginning of Sec.~\ref{sec:43_FCS_sharp}.
Here $L/2 + x_0$ is the 
distance  from the central QPC to the left interface; it is equal to $L/2$ if the QPC is in the center of the interacting region. For the dilute injection of anyons that we consider here, the characteristic time $\tau$ that enters this comparison is  $\tau \sim 1/(e_1^* V \mathcal{T})$ as discussed in Sec.~\ref{sec:GFs_co} above Eq.~\eqref{eq:countingphases_43_GFs}; This time can be expressed in terms of the average current $\langle I_0\rangle$ on the edge by using Eq.~\eqref{eq:average_I}.
The above results for the Fano factor hold in the long-length regime. In the opposite, short-length, limit, the non-interacting results are recovered 
(see Sec.~\ref{sec:GFs_co}), so that 
the Fano factor takes the form \eqref{eq:Fano_Laughlin_13} with $\nu=1/3$, revealing the braiding phase $2\theta_{12}=2\pi/3$ between the injected and the tunneling quasiparticles. Thus, changing the length $L/2 + x_0$ (or, alternatively, the injected current $\langle I_0\rangle$), one can probe the crossover between the two braiding limits.  

This completes our analysis, in the framework of non-equilibrium bosonization, of  FQH edges with co-propagating modes. We turn now to the case of FQH edges with counter-propagating modes.

\subsection{Counter-propagating edge modes: Non-equilibrium action}
\label{sec:Counter_prop}

The topological order described by Eq.~\eqref{eq:KMat2x2} admits edges with counter-propagating modes, i.e., with $\nu_1 \nu_2<0$. We study below the prototypical example of this class, namely the $\nu=2/3$ edge~\cite{Johnson_Composite_edge_1991,Kane_Randomness_1994,Protopopov_transport_2_3_2017,Spanslatt2021} composed of  $\nu_1=1$ and $\nu_2=-1/3$ modes. The corresponding $K$-matrix and the charge vector $\vec{q}$ are given by
\begin{align}
\label{eq:K23}
K =
\begin{pmatrix}
1 & 0\\
0 & -3
\end{pmatrix}\,, \quad \vec{q} = \begin{pmatrix}
    1 \\ 1
    \end{pmatrix}\,.
\end{align}
While the interacting $\nu=2/3$ edge is governed by the same Hamiltonian as the $\nu = 4/3$ edge, Eq.~\eqref{eq:H25}, its physical properties differ significantly due to the counter-propagating character of the modes, which is encoded in the following 
density commutation relations:
\begin{subequations}
\label{eq:23_CCR}
\begin{align}
&[\rho_1(x),\rho_1(y)]=-\frac{i}{2\pi}\partial_x\delta(x-y),\\
    &[\rho_{2}(x),\rho_{2}(y)]=+\frac{1}{3}\frac{i}{2\pi}\partial_x\delta(x-y).
\end{align}
\end{subequations}
The signs in these equations correspond to opposite chiralities of the two modes, $\eta_1=-\eta_2=+1$, and the densities are given by $\rho_1=+\partial_x \phi_1/(2\pi)$ and $\rho_2=-\partial_x \phi_2/(2\pi)$.

To diagonalize the Hamiltonian~\eqref{eq:H25}, we introduce the transformation matrix $\Lambda$,
\begin{align}
\label{eq:Lambda_23}
   \Lambda=\begin{pmatrix}
        \cosh\gamma & -\sinh\gamma\\
        -\frac{\sinh\gamma}{\sqrt{3}} & \frac{\cosh\gamma}{\sqrt{3}} 
    \end{pmatrix}, \quad \tanh (2\gamma) = \frac{2}{\sqrt{3}}\frac{u}{v_1+v_2}.
\end{align}
This transformation diagonalizes and rescales the $K$-matrix and preserves its signature (i.e., edge-mode chiralities):
$\Lambda^{T} K \Lambda = \mathrm{diag}(1,-1)\equiv\sigma_z$. In the resulting eigenmode basis, the Hamiltonian and the eigenmode fields take the same form as those obtained for the $\nu=4/3$ edge, see Eqs.~\eqref{eq:H_diag} and~\eqref{eq:eigenmode43}, respectively. However, the corresponding eigenmode velocities are now
\begin{align}
\label{eq:vpm_23}
v_\pm
= v_{1,2}\cosh^{2}\gamma
+ v_{2,1}\sinh^{2}\gamma
- \frac{u}{\sqrt{3}}\sinh(2\gamma).
\end{align}
Stability of the $\nu=2/3$ edge is ensured if the condition $u^2 < 3 v_1 v_2$ is satisfied, such that $v_\pm > 0$. 

We will now study the $\nu=2/3$ edge driven out of equilibrium by injections of the quasiparticle excitations~\eqref{eq:excitations43}. The injections in the $\nu_1=1$ and $\nu_2=1/3$ modes occurs in region I and region III, respectively, see Fig.~\ref{fig:complex_setup}\textcolor{blue}{(b)}. The Keldysh non-equilibrium action for $\nu=2/3$ remains identical in form to that of the $\nu=4/3$ edge, Eq.~\eqref{eq:action_43}. In Secs.~\ref{sec:FCS_counter} and \ref{sec:GFs_counter}, we will use this action to compute the FCS and the quasiparticle GFs for the $\nu=2/3$ edge. Subsequently, in Sec.~\ref{sec:transport_counter}, we  
will extend the analysis to tunneling transport between two $\nu=2/3$ edges.

\subsection{Counter-propagating edge modes: Full counting statistics}
\label{sec:FCS_counter}

Derivation of the FCS for the $\nu=2/3$ edge largely parallels that for the $\nu=4/3$ edge discussed in Sec.~\ref{sec:Co_prop}. Thus, we focus here on features specific to the counter-propagating nature of the $\nu=2/3$ edge and on key results. 
Details of the calculations are presented in the Supplemental Material~\cite{Supplemental_Material}. In this section and in Sec.~\ref{sec:GFs_counter} and \ref{sec:transport_counter} below, we focus on the long-length limit, $L \gg v_\pm \tau$, where the interaction-induced fractionalization is operative. As for the case of co-propagating modes, we will consider two types of interfaces: sharp and adiabatic. 

\subsubsection{Sharp interfaces}

We begin with the case of sharp interfaces as illustrated in Fig.~\ref{fig:complex_setup}\textcolor{blue}{(b)}. As for the $\nu=4/3$ edge, the total charge to the left of the point $x_0$ at time $t$ is given by Eq.~\eqref{eq:Q_43}. 
In the long-length limit $L \gg v_{\pm} \tau$, evaluation of the counting phases yields the following results for the cumulants of charge fluctuations~\eqref{eq:FCS_cumulants},
\begin{align} \label{eq:23chargefluctuations_sharp_main}
     \langle\langle(\delta Q)^k \rangle\rangle = \sum_{s=\pm}\sum_{m=0}^\infty 
     \Big[&(n_{1,1}q^{\text{(p)}}_{1,s,m})^{k-1}  \langle Q_{1,s, m}\rangle  \notag \\+ &(n_{1,2}q^{\text{(p)}}_{2,s,m})^{k-1}  \langle Q_{2,s, m}\rangle \Big],
\end{align}
with 
\begin{subequations}
\label{eq:23_individual_pulses_main}
\begin{align}
    &q^{\text{(p)}}_{1,+,m} =e q_+\frac{\tanh(\gamma)^{2m}}{\cosh(\gamma)},\\
    &q^{\text{(p)}}_{1,-,m} =eq_- \frac{\tanh(\gamma)^{2m+1}}{\cosh(\gamma)},\\
    &q^{\text{(p)}}_{2,-,m} =  -\frac{eq_-}{\sqrt{3}} \frac{\tanh(\gamma)^{2m}}{\cosh(\gamma)},\\
    &q^{\text{(p)}}_{2,+,m} = -\frac{eq_+}{\sqrt{3}}\frac{\tanh(\gamma)^{2m+1}}{\cosh(\gamma)},
\end{align}
\end{subequations}
and the average charges traversing the point $x_0$ during the time interval $\tau$,
\begin{subequations}
\label{eq:phases_FCS_23_average_charges}
\begin{align}
    &\langle Q_{1,+,m} \rangle = +q_+ \frac{\tanh(\gamma)^{2m}}{\cosh(\gamma)} \frac{e^2 \mathcal{T}_1 V_1 \tau}{2\pi},\\
    &\langle Q_{1,-,m} \rangle =+ q_-  \frac{\tanh(\gamma)^{2m+1}}{\cosh(\gamma)} \frac{e^2 \mathcal{T}_1 V_1 \tau}{2\pi},\\
    &\langle Q_{2,-,m} \rangle = -q_- \frac{\tanh(\gamma)^{2m}}{\cosh(\gamma)} \frac{e^2 \mathcal{T}_2 V_2 \tau}{2\pi \sqrt{3}} ,\\
    &\langle Q_{2,+,m} \rangle = - q_+ \frac{\tanh(\gamma)^{2m+1}}{\cosh(\gamma)} \frac{e^2 \mathcal{T}_2 V_2 \tau}{2\pi \sqrt{3}} .
\end{align}
\end{subequations}
In Eqs.~\eqref{eq:23_individual_pulses_main}-\eqref{eq:phases_FCS_23_average_charges}, the interaction-dependent eigenmode charges are given by $(q_+,q_-)^T=\Lambda^T \vec{q}$ with $\vec{q}$ and $\Lambda$ from Eq.~\eqref{eq:K23} and Eq.~\eqref{eq:Lambda_23}, respectively.

Equation~\eqref{eq:23chargefluctuations_sharp_main} is the FCS of a superposition of independent Poissonian processes characterized by fractional charges $n_{1,1}q^{\text{(p)}}_{1,\pm,m}$ and $n_{1,2}q^{\text{(p)}}_{2,\pm,m}$. The physical picture behind this result is as follows. When the mode $\nu_1=1$ in the region I is driven out of equilibrium, it hosts a dilute train of quasiparticles with charge $n_{1,1}e$. These quasiparticles experience multiple scattering events at the two interfaces, each time with fractionalization into transmitted and reflected components. This process leads to an infinite series of excitations detected in the FCS in the interacting region II. The charges  $n_{1,1}q^{\text{(p)}}_{1,+,m}$ correspond to the excitations that have undergone an even number of reflections and thus move from left to right. Likewise, the charges $n_{1,1}q^{\text{(p)}}_{1,-,m}$ in the FCS
correspond to the excitations that experienced an odd number of reflections. Thus, they move from right to left and their actual charge is the opposite, i.e.,
$- n_{1,1}q^{\text{(p)}}_{1,-,m}$. In the same way, the charges $n_{1,2}q^{\text{(p)}}_{2,\pm,m}$ in FCS correspond to multiple fractionalization of quasiparticles $n_{1,2}e/3$ from a non-equilibrium mode $\nu_2=-1/3$ in region III. Specifically, these fractionalization processes generate, in the interacting region, quasiparticles with charges 
$- n_{1,2}q^{\text{(p)}}_{2,-,m}$ moving from right to left and quasiparticles with charges  $n_{1,2}q^{\text{(p)}}_{2,+,m}$ moving from left to right.

It is easy to check that the charges $n_{1,1} q^{\text{(p)}}_{1,s,m}$ and $n_{1,2} q^{\text{(p)}}_{2,s,m}$ entering the result~\eqref{eq:23chargefluctuations_sharp_main} for the cumulants add up to $n_{1,1} e$ and $-n_{1,2} e/3$, respectively,
\begin{align}
\label{eq:totalinjectedcharge_23}
    \sum_{s=\pm}\sum_{m=0}^\infty q^{\text{(p)}}_{1,s,m}=  e\,, \quad 
    \sum_{s=\pm}\sum_{m=0}^\infty q^{\text{(p)}}_{2,s,m}= - \frac{e}{3}\,.
\end{align}
This is a manifestation of charge conservation: the charge $n_{1,1} e$ injected in the $\nu_1=1$ mode in region I should eventually go out via the $\nu_1=1$ mode in region III. The situation with the charge $n_{1,2} e/3$
in the  $\nu_2=-1/3$ mode is similar, except that it is injected in region III and eventually leaves to region I, which explains the minus sign in Eq.~\eqref{eq:totalinjectedcharge_23}. 
 
\subsubsection{Adiabatic interfaces}
For adiabatic interfaces, the excitations in the 
$\nu_1 = 1$ mode (regions I and III) are smoothly connected to the $+$ eigenmode in the interacting region (region II). Likewise, the excitations in the mode 
$\nu_2=-1/3$ are smoothly connected to the 
$-$ eigenmode. In both cases the smooth pulses are accompanied by an extended background contribution. Further details can be found in Supplemental Material~\cite{Supplemental_Material}. The high-frequency contribution of the counting pulses with temporal width $\tau$ have the charges 
\begin{subequations}
 \label{eq:adiabatic_charges_23}
 \begin{align}
   q_{1,h}^{\text{(p)}} &=q_+ e \,, 
    \label{eq:adiab_charge_23_1}
    \\
   q_{2,h}^{\text{(p)}} &=-\frac{q_- e}{\sqrt{3}}\,.
   \label{eq:adiab_charge_23_2}
 \end{align}
 \end{subequations}

In the long-length limit, the total average charge (first cumulant, $k=1$) passing across the point $x_0$ during the time interval $\tau$ evaluates in the adiabatic limit to
\begin{align}
\label{eq:average_charge_23_adiabatic}
\langle Q \rangle &=  \langle  Q_1\rangle + \langle Q_2\rangle  \,, \nonumber \\ \text{with }\langle Q_{1}\rangle  &= \frac{\mathcal{T}_1 e^2 V_1 \tau}{2\pi}\,, \quad 
    \langle Q_{2}\rangle  = -\frac{1}{3}\frac{\mathcal{T}_2 e^2 V_2 \tau}{2\pi}\,.
\end{align}
As expected, $\langle Q \rangle$ is independent of interactions and obeys charge conservation, i.e., it equals the average injected charge during the same time interval. The higher cumulants ($k>1$) evaluate to
\begin{align} \label{eq:23chargefluctuations}
     \langle\langle(\delta Q)^k \rangle\rangle &= 
    (n_{1,1} q_{1,h}^{\text{(p)}})^{k-1}  \langle Q_{+}\rangle +  (n_{1,2} q_{2,h}^{\text{(p)}})^{k-1}  \langle Q_{-}\rangle \nonumber \\ \text{with }\langle Q_{+}\rangle  &= \frac{q_+\mathcal{T}_1 e^2 V_1 \tau}{2\pi}\,, \quad 
    \langle Q_{-}\rangle  = -\frac{q_-\mathcal{T}_2 e^2 V_2 \tau}{2\pi \sqrt{3}}\,. 
\end{align}
Together with Eq.~\eqref{eq:average_charge_23_adiabatic}, 
these cumulants describe a superposition of two independent Poissonian processes carrying charges $n_{1,1} q_{1,h}^{\text{(p)}} $ and $n_{1,2} q_{2,h}^{\text{(p)}} $. The physical interpretation is analogous to that of the FCS for a co-propagating edge with adiabatic interfaces, see Sec.~\ref{sec:adiabatic_interfaces_43}.
The difference is that, in the present case, the $\nu_2=-1/3$ mode  propagates from right to left, which explains the minus signs in Eqs.~\eqref{eq:adiab_charge_23_2}
and \eqref{eq:average_charge_23_adiabatic}.

\subsection{Counter-propagating edge modes: Green's functions}
\label{sec:GFs_counter}

We turn now to the analysis of GFs for the $\nu=2/3$ edge, see Fig.~\ref{fig:complex_setup}\textcolor{blue}{(b)},
with non-equilibrium distribution functions injected in regions I and III having the double-step form~\eqref{eq:doublestep_43}, with 
$\mathcal{T}_1,\mathcal{T}_2\ll 1$.
Details of the calculations are presented in the Supplemental Material~\cite{Supplemental_Material}.

For sharp interfaces, the GFs 
of a $\vec{n}_2$-type excitation
[see Eq.~\eqref{eq:vertex_43}]
evaluate within the Szeg\H{o} approximation to
\begin{align} \label{eq:GFs_double_step_23}
    &\mathcal{G}^{\gtrless}(\tau)  \simeq \frac{\mp i}{2\pi a} \left(\frac{a}{ a \pm i v_{+} \tau}\right)^{\alpha} \left(\frac{a}{ a \pm i v_{-} \tau}\right)^{\beta} \nonumber 
    \\ &  \times \exp\Big[-
 \frac{\mathcal{T}_1|e V_1 \tau|}{2\pi n_{1,1}}\sum_{s=\pm}\sum_{m=0}^\infty(1-e^{-i \delta_{1,s,m} \text{sgn}(eV_1\tau)})\Big] \nonumber \\ &   \times \exp\Big[-
 \frac{\mathcal{T}_2|e V_2 \tau|}{2\pi n_{1,2}}\sum_{s=\pm}\sum_{m=0}^\infty(1-e^{-i \delta_{2,s,m} \text{sgn}(eV_2\tau)})\Big]\,,
\end{align}
 with the phases
\begin{subequations}
\label{eq:gen_phases_23}
\begin{align}
    &\delta_{1,+,m} =+2\pi n_{1,1} n_{2,+} \frac{\tanh(\gamma)^{2m}}{\cosh(\gamma)},  \\ 
    &\delta_{1,-,m} = +2\pi n_{1,1}  n_{2,-} \frac{\tanh(\gamma)^{2m+1}}{\cosh(\gamma)},\\
\label{eq:gen_phases_23_2_even}
    &\delta_{2,-,m} =-\frac{2\pi n_{1,2}n_{2,-}}{\sqrt{3}}\frac{\tanh(\gamma)^{2m}}{\cosh(\gamma)},\\
\label{eq:gen_phases_23_2_odd}
    &\delta_{2,+,m} = -\frac{2\pi n_{1,2} n_{2,+}}{\sqrt{3}}\frac{\tanh(\gamma)^{2m+1}}{\cosh(\gamma)}.
\end{align}
\end{subequations}
In analogy with the discussion in Sec.~\ref{sec:GFs_co}, these phases have a meaning of the mutual braiding
phases between the fractionalized quasiparticles revealed by the FCS of Sec.~\ref{sec:FCS_counter} and the $\vec{n}_2$-type excitation. 
For each of the modes, the phases 
 add up to the mutual braiding phases between the injected $\vec{n}_1$-type excitations and the $\vec{n}_2$-type excitations,  
\begin{subequations}
\begin{align}
    \sum_{m=0}^{\infty} \sum_{s=\pm} \delta_{1,s,m} &= 2\pi\, n_{1,1}n_{2,1}\,, \\ \sum_{m=0}^{\infty} \sum_{s=\pm}  \delta_{2,s,m} &= \frac{2\pi n_{1,2}n_{2,2}}{3}\,. 
\end{align}
\end{subequations}
The coefficients $n_{2,+}$ and $n_{2,-}$ entering the expressions~\eqref{eq:gen_phases_23}  for the phases  are given by
\begin{align}
\label{eq:n2pm_23}
    \begin{pmatrix}
        n_{2,+} \\ n_{2,-}
    \end{pmatrix} = \Lambda^T\sigma_z \vec{n}_2 = \begin{pmatrix}
       \cosh \gamma\, n_{2,1}+\frac{\sinh \gamma}{\sqrt{3}} n_{2,2} \\[0.2cm] -\sinh\gamma \,n_{2,1}- \frac{\cosh \gamma}{\sqrt{3}} n_{2,2}
    \end{pmatrix}.
\end{align} 

Furthermore, the exponents $\alpha$ and $\beta$ entering Eq.~\eqref{eq:GFs_double_step_23} for the GFs read
\begin{subequations}
\label{eq:alpha_beta_23}
\begin{align}
    \alpha&=\frac{1}{n_{1,1}^2}\sum_{s=\pm}\sum_{m=0}^\infty \left(\frac{\delta_{1,s,m}}{2\pi}\right)^2=\frac{t_L^2\left[n_{2,+}^2+r_R^2 n_{2,-}^2\right]}{1-r_L^2r_R^2},\\
    \beta&=\frac{3}{n_{1,2}^2}\sum_{s=\pm}\sum_{m=0}^\infty \left(\frac{\delta_{2,s,m}}{2\pi}\right)^2=\frac{t_R^2\left[n_{2,-}^2+r_L^2 n_{2,+}^2\right]}{1-r_L^2r_R^2}.
  \end{align}
\end{subequations}  
  These exponents satisfy the sum rule  
\begin{align}
    \label{eq:23sumrule}
&\alpha+\beta = n_{2,+}^2+n_{2,-}^2 \notag \\ &= (n_{2,1}^2+\frac{n_{2,2}^2}{3})\cosh(2\gamma)+\frac{2n_{2,1} n_{2,2}\sinh(2\gamma)}{\sqrt{3}}=\zeta \,,
\end{align}
where $\zeta$ is the non-universal (interaction-dependent) ~\cite{Wen_Topological_1995} scaling dimension of the excitation~\eqref{eq:vertex_43} on the $\nu=2/3$ edge. 

For the case of adiabatic interfaces, the counterpart of Eq.~\eqref{eq:GFs_double_step_23} reads
\begin{align} \label{eq:GFs_doublestep_23_adiabatic}
   &\mathcal{G}^{\gtrless}(\tau)  \simeq \frac{\mp i}{2\pi a} \left(\frac{a}{ a \pm i v_{+} \tau}\right)^{\alpha} \left(\frac{a}{ a \pm i v_{-} \tau}\right)^{\beta} \nonumber 
    \\ & \quad \times \exp\Big[-
 \frac{\mathcal{T}_1|e V_1 \tau|}{2\pi n_{1,1}}(1-e^{-i 2\pi n_{1,1} n_{2,+} \text{sgn}(eV_1\tau)})\Big] \nonumber \\ & \quad  \times \exp\Big[-
 \frac{\mathcal{T}_2|e V_2 \tau|}{2\pi n_{1,2}}(1-e^{-i 2\pi n_{1,2} n_{2,-} \text{sgn}(eV_2\tau)/\sqrt{3}})\Big]\nonumber \\ & 
 \quad \times \exp[i e \tau \Big \{V_1 \mathcal{T}_1 (n_{2,+}-n_{2,1}) - V_2 \mathcal{T}_2 \big(\frac{n_{2,-}}{\sqrt{3}}+ \frac{n_{2,2}}{3}\big ) \Big \} ]\,,  
\end{align}
with exponents $\alpha$ and $\beta$ from Eq.~\eqref{eq:alpha_beta_23} and $n_{2,\pm}$ from Eq.~\eqref{eq:n2pm_23}. 

\subsection{Counter-propagating edge modes: Transport properties}
\label{sec:transport_counter}

With the non-equilibrium GFs at hand, we compute now transport observables for tunneling between two $\nu=2/3$ edges coupled by a QPC, see Fig.~\ref{fig:QPC}. For definiteness, and for the sake of comparison, we will assume a setup analogous to that considered in Sec.~\ref{sec:transport_co} for the tunneling between $\nu=4/3$ edges. Specifically, we assume that only the $u$ edge is driven out of equilibrium, and this is done by injection of 
minimal quasiparticles (with $n_{1,2} =1$, i.e., with the charge $e_{1,2}^* = e/3$) in the mode $\nu_2 = -1/3$.
The corresponding distribution function is characterized by $V_{2,u}=V$ and $\mathcal{T}_{2,u}=\mathcal{T} \ll 1$.
The $d$ edge is at zero-temperature equilibrium, i.e., $V_{1,u} = V_{1,d} = V_{2,d}=0$.  We further assume that
the same type of minimal excitations tunnel at the QPC, corresponding to $\vec{n}_2=(n_{2,1}, n_{2,2}) = (0,1)$.

\begin{figure}[t!]
    \centering
\subfloat[][]{
\includegraphics[width=0.8\columnwidth]{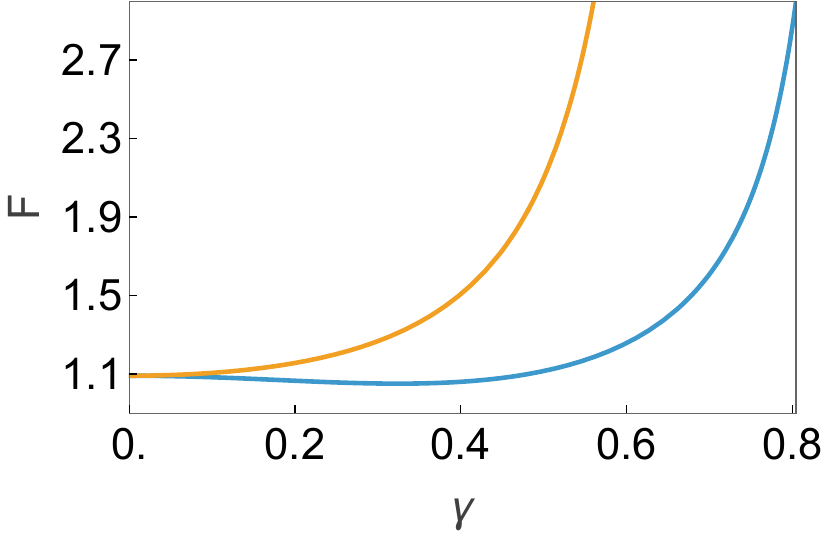}  
\put(-200,120){\large\textbf{(a)}}
}
\par\vspace{-3mm}
\subfloat[][]{
\includegraphics[width=0.8\columnwidth]{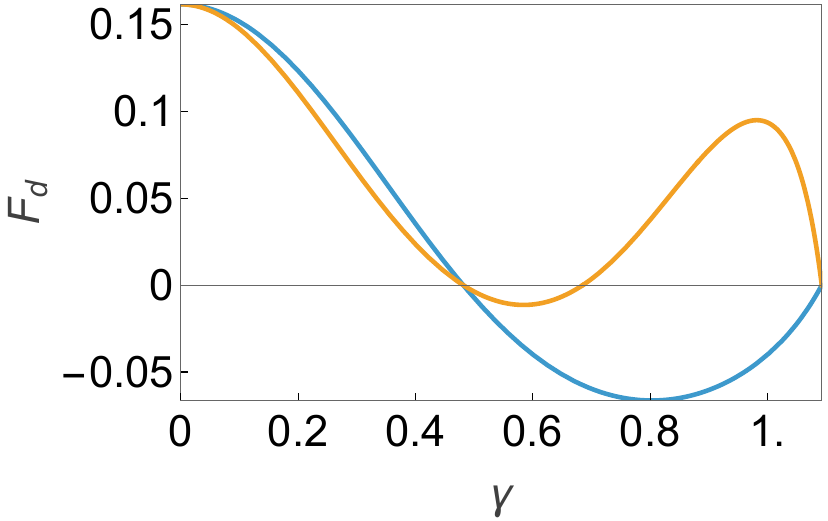}  
  \put(-200,120){\large\textbf{(b)}}
}
\caption{Fano factor $F$ [Eq.~\eqref{eq:gen_fano}] and differential Fano factor $F_d$ [Eq.~\eqref{eq:Fano_diff_eval}] for tunneling transport between two $\nu=2/3$ edges as described  in Sec.~\ref{sec:transport_counter}: The upper edge $u$ is driven out of equilibrium by quasiparticle injection, under a voltage $V>0$, into the $\nu_2 = -1/3$ mode, while the lower edge $d$ is in equilibrium. (a) Fano factor as a function of the inter-mode interaction parameter $\gamma$ [see Eq.~\eqref{eq:Lambda_23}]. The results for sharp (blue) and adiabatic (orange) interfaces are shown. (b) The differential Fano factor (at zero constant bias, $V_0=0$) as a function of $\gamma$ for the validity range $0\leq \gamma \lesssim 1.09$. For the considered setup, $F$ and $F_d$ are even function of $\gamma$, so that only the range $\gamma > 0$ (repulsive interaction) is presented. The point $\gamma=0$ corresponds to tunneling transport between Laughlin $\nu=1/3$ edges, see Sec.~\ref{sec:transport-zeta-less-12} and \ref{sec:diff_FF}.
}
\label{Fig_fano23} 
\end{figure}

We find then that the Fano factor 
can be written in the same form as Eq.~\eqref{eq:gen_fano} with the scaling dimension \eqref{eq:23sumrule}, and
a modified expression for the parameter $p$. Specifically, for the case of sharp interfaces, when the non-equilibrium GFs of the $u$ edge are given by Eq.~\eqref{eq:GFs_double_step_23}, we obtain
\begin{align} \label{eq:dimensionlessparameter_23}
    p = \frac{\text{sgn}(eV) \sum_{s=\pm}\sum_{m=0}^\infty\sin(\delta_{2,s,m})}{\sum_{s=\pm}\sum_{m=0}^\infty\left[1-\cos(\delta_{2,s,m})\right]}\,,
\end{align}
with the interaction-dependent phase amplitudes $\delta_{2,\pm,m}$ given by Eq.~\eqref{eq:gen_phases_23_2_even}-\eqref{eq:gen_phases_23_2_odd}. 
For adiabatic interfaces [with the GFs of the $u$ edge obtained from Eq.~\eqref{eq:GFs_doublestep_23_adiabatic}], we find
\begin{align}\label{eq:dimensionlessparameter_23_adiabatic}
   p &=\text{sgn}(eV) \frac{\sin [2\pi |\nu_2| \cosh \gamma]+2\pi |\nu_2| (1-\cosh \gamma)}{1-\cos [2\pi \nu_2 \cosh\gamma]}\,.
\end{align}

Inserting the expressions for $p$ [Eqs.~\eqref{eq:dimensionlessparameter_23} and \eqref{eq:dimensionlessparameter_23_adiabatic}] and the scaling dimension [Eq.~\eqref{eq:23sumrule}] into Eq.~\eqref{eq:gen_fano}, we compute the Fano factor $F$ for both sharp and adiabatic interfaces. We recall that the expression for the Fano factor in Eq.~\eqref{eq:gen_fano} is valid only when the scaling dimension $\zeta$ satisfies $\zeta<1$, 
see Sec.~\ref{sec:tun-noneq-Laughlin-edges},
which for the $\nu=2/3$ edge amounts to a constraint on the interaction strength. By plugging $n_{2,1}=0$ and $n_{2,2}=1$ into Eq.~\eqref{eq:23sumrule}, we find that $\zeta<1$ constraints the interaction parameter to $ |\gamma| < \gamma_c = 0.5 \cosh^{-1}3 \approx 0.88$. 

The resulting $F$ is plotted as a function of interaction parameter $\gamma$ in Fig.~\ref{Fig_fano23}\textcolor{blue}{(a)}. For sharp interfaces, the Fano factor $F$ [Eq.~\eqref{eq:gen_fano}], depends only weakly on $\gamma$ at small $\gamma$, but increases rapidly and eventually diverge as $\gamma$ approaches the boundary $\gamma_c\approx0.88$ of the validity regime. By contrast, for adiabatic interfaces, $F$ increases already at small $\gamma$ and continues to grow, eventually diverging as $\gamma$ approaches $\gamma_c$. The divergent behavior of the Fano factor originates from the singularity of the tunneling noise $S_T$ [Eq.~\eqref{eq:noise_dilute_app}] as $\zeta \rightarrow 1$. These features stand in stark contrast to those observed for the $\nu=4/3$ edge; see Fig.~\ref{fig:Fano_plot}\textcolor{blue}{(a)}.

As discussed in Sec.~\ref{sec:diff_FF}, the range of applicability of our analysis can be extended up to $\zeta<3/2$ by considering the differential Fano factor $F_d$. In the present case, the condition $\zeta<3/2$
corresponds to interactions in the range $|\gamma| = \gamma_c <0.5 \cosh^{-1}4.5 \approx 1.09$.
We compute $F_d$ by using
Eq.~\eqref{eq:Fano_diff_eval},
with $\zeta$ given by Eq.~\eqref{eq:alpha_beta_23} and with the tunneling charge $e_2^* = |\nu_2|e= e/3$. At the point of vanishing bias offset $V_0=0$
(see Sec.~\ref{sec:diff_FF} for details), the parameter $p$ that enters Eq.~\eqref{eq:Fano_diff_eval} is that of Eq.~\eqref{eq:dimensionlessparameter_23} and~\eqref{eq:dimensionlessparameter_23_adiabatic} for sharp and adiabatic interfaces, respectively. The resulting  $F_d$ is shown as a function of $\gamma$ in Fig.~\ref{Fig_fano23}\textcolor{blue}{(b)}. We see that, for both types of interfaces, $F_d$ exhibits a strong dependence on $\gamma $. Specifically, $F_d$ decreases when $\gamma$ increases from the non-interacting value $\gamma = 0$ and changes sign  at $\gamma = 0.5 \cosh^{-1}1.5 \approx 0.48$ corresponding to $\zeta = 1/2$. For adiabatic interfaces, as $\gamma$ increases further, $F_d$ exhibits an upturn and crosses zero again at $\gamma \approx 0.69$, which corresponds to the point where $p$ in Eq.~\eqref{eq:dimensionlessparameter_23_adiabatic} vanishes. For both types of interfaces, the differential tunneling conductance $\partial \langle I_T\rangle/ \partial V_0$ [Eq.~\eqref{eq:current_dilute_app_deriv}] diverges as $\gamma$ approaches the boundary $\gamma_c\approx1.09$ of the validity regime, leading to vanishing $F_d$.

Let us remind the reader that the above results are obtained in the dilute limit, $\mathcal{T}_{u,d} \ll 1$, with corrections to $F$ due to finite values of $\mathcal{T}_{u,d}$ having a relative smallness
$ \sim \mathcal{T}_{u,d}^{2-2\zeta}\ll 1$, see the discussion in the end of Sec.~\ref{sec:transport-zeta-less-12} and the Supplemental Material \cite{Supplemental_Material}. The role of these corrections thus
increases when $\zeta$ approaches unity. By the same token, corrections to $F_d$ have a relative smallness $ \sim \mathcal{T}_{u,d}^{3-2\zeta}\ll 1$, and their significance grows when $\zeta$ approaches 3/2. 

\section{Summary and Outlook}
\label{sec:Summary_Conclusions}
We have developed a non-equilibrium bosonization theory for Abelian FQH edges, including single-mode (Laughlin) edges and complex edges with co-propagating and counter-propagating modes. Our main physical interest is in the properties of a FQH edge driven out of equilibrium by a dilute injection of quasiparticles (which may have fractional or integer charge) and in the tunneling transport between such edges.
Our formalism is based on the Keldysh action for the non-equilibrium chiral Luttinger liquid, see Sec.~\ref{sec:formalism}. It extends the non-equilibrium bosonization formalism for conventional Luttinger liquids \cite{Gutman2010} by incorporating crucial properties of anyons on FQH edges: fractional charge and fractional statistics. 
The corresponding Keldysh action is not Gaussian but the non-Gaussian part depends only on the quantum component of the density field in view of the quadratic character of the bosonized Hamiltonian. This part contains information about all orders of charge fluctuations on the edge and is expressed in terms of a Fredholm functional determinant. To benchmark the theory, we have verified  that it correctly reproduces the average current and the FCS of anyons on the non-equilibrium Laughlin edge and that it also reduces to the known Gaussian action in equilibrium.

With this framework for bosonized non-equilibrium FQH edges, we have then derived in Sec.~\ref{sec:GF_Laughlin}
the GFs of generic excitations on a Laughlin edge, $\nu=1/m$, driven out of equilibrium by a dilute quasiparticle injection. The long-time asymptotics of the GFs is provided by the generalized Fisher-Hartwig formula for the Fredholm determinant (which is of Toeplitz form), which yields the results in terms of a sum over branches of the complex logarithm.  In this context, a parameter of crucial importance  is the phase $\delta_0 = 2\theta_{12} = 2\pi \nu n_1n_2$, which characterizes the mutual braiding between the excitations with charge $n_1\nu e$ injected to drive the edge out of equilibrium and the excitations
with charge $n_2\nu e$ for which the GFs are studied. When $2\theta_{12}$ is not equal to a multiple integer of $2\pi$, and in the weak injection, or dilute, limit $\mathcal{T} \ll 1$,
the generalized Fisher-Hartwig formula is well-approximated by the contribution of the ``main'' branch, leading to the Szeg\H{o} approximation. 
When $2\theta_{12}$ is  a multiple integer of $2\pi$ (which is the case, e.g., when $n_1=1$ and $n_2=m$, or vice versa), the Szeg\H{o} approximation misses completely the non-equilibrium physics, and one has to use the full generalized Fisher-Hartwig formula, which is then exact and contains only a finite number of branches.  

In Sec.~\ref{sec:tunneling}, we have used the GF results to calculate the tunneling current and noise in a QPC connecting two Laughlin edges, with one or both of them being out of equilibrium. This has allowed us to evaluate  dimensionless quantities that characterize this transport setup and can be measured experimentally: the Fano factor $F$ and the differential Fano factor $F_d$. These results are illustrated in Fig.~\ref{fig:Fanovsbraidingangles}. Whenever a comparison is possible, our results for the Laughlin edge are consistent with earlier theoretical findings (obtained by different methods) in  Refs.~\cite{Levkivskyi2016Apr,Rosenow2016Apr, Lee2020, Morel2022fractionalization, Lee2022non-abelian, Lee2023May,Iyer2023,Schiller2023anyon,Zhang2025Mar}
and with the experimental results in
Refs.~\cite{Bartolomei2020Apr,Lee2023May,Glidic2024Aug}.

Importantly, our approach permits a direct extension to complex (multi-mode) FQH edges. This is done
in Sec.~\ref{sec:multiple-mode-edges}, where we considered two-mode edges with co-propagating and counter-propagating 
modes and tunneling transport between such edges. (An extension to edges with a larger number of modes is straightforward.) For definiteness, we focused on the $\nu=4/3$ and $\nu=2/3$ edges as paradigmatic examples for each of the two classes, explicitly incorporating effects of inter-mode interactions.
For each of these edges driven out of equilibrium, we first calculated the FCS which revealed the interaction-induced fractionalization of anyonic excitations. In contrast to the topological, quantized  fractionalization of quasiparticles that can tunnel via the FQH bulk, this fractionalization is not quantized and depends continuously on the interaction strength. The interaction-induced fractionalization of anyons manifests itself in the results for the GFs that we evaluate subsequently. Specifically, it leads to a fractionalization of the mutual braiding phase $2\theta_{12}$. We used the results for the GFs to calculate the tunneling transport and noise via a QPC, again putting particular focus on the Fano factors $F$ and $F_d$. The dependence of the Fano factors on the interaction parameter $\gamma$ is shown in
Figs.~\ref{fig:Fano_plot} and
\ref{Fig_fano23} for tunneling between $\nu=4/3$ edges and between $\nu=2/3$ edges, respectively. The point $\gamma=0$ on these plots corresponds to the non-interacting limit, when the results become identical to those for Laughlin edges. Our findings illustrated in Figs.~\ref{fig:Fano_plot} and
\ref{Fig_fano23} show that the interaction-induced fractionalization of the braiding phases strongly affects the Fano factor characterizing the non-equilibrium physics of complex FQH edges.

Our non-equilibrium FQH-edge bosonization theory paves the way for future developments and applications. We end by briefly discussing some relevant directions. 

 \begin{itemize}

   \item[{i)}] An important case of a complex edge with two co-propagating modes 
   is the $\nu=2/5$ edge, for which the elementary (charge-$e/5$) quasiparticles produce $\zeta = 3/5$. 
 Tunneling transport for non-equilibrium $\nu=2/5$ edges was experimentally studied in
Refs.~\cite{Ruelle2023Mar,Glidic2023Mar,Lee2023May}, emphasizing a need for a controllable analytic theory. Our approach, as presented in Sec.~\ref{sec:Co_prop}--\ref{sec:transport_co} 
for edges with co-propagating modes, is directly applicable to the $\nu=2/5$ edge with inter-mode interactions and to tunneling transport of charge-$e/5$ quasiparticles between $\nu=2/5$ edges.

 \item[{ii)}] The calculations of the tunneling current, noise, and Fano factor $F$ in this work are applicable for scaling dimensions $\zeta < 1$, which amounts to the involved time integrals being governed by long times. (An exception is the case of the mutual braiding phase being an integer multiple of $2\pi$). Calculating these transport observables for larger $\zeta$ requires an analysis of the GFs at short times. It remains to be seen whether this can be done analytically. An alternative possibility is to evaluate the corresponding Toeplitz determinants numerically. 
 
  \item[{iii)}] When studying transport noise in this paper, we have focused 
  on the noise of the tunneling current. Another important--and experimentally relevant--question concerns cross-correlations noise of the currents emanating from the QPC~\cite{Rosenow2016Apr,Bartolomei2020Apr,Idrisov2022}. Calculating this noise by means of the non-equilibrium FQH bosonization formalism is an interesting prospect for future work.

  \item[{iv)}]  Finally, a very  important direction is to generalize our non-equilibrium formalism to non-Abelian FQH edges that host, in addition to chiral bosonic modes, also Majorana modes.
 \end{itemize}

\begin{acknowledgments}
We thank Igor Gornyi for instructive discussions and acknowledge funding from the Swedish Vetenskapsr\r{a}det via Project No.~2023-04043 (C.S.) and from the Deutsche Forschungsgemeinschaft (DFG, German Research Foundation)--- Projects No.~559260877 (J.P.) and No.~320540272 (A.D.M.).
\end{acknowledgments}

\bibliography{Refs.bib}

\clearpage
\newpage

\onecolumngrid
\setcounter{equation}{0}
\setcounter{section}{0}
\setcounter{figure}{0}
\setcounter{table}{0}
\setcounter{page}{1}
\renewcommand{\theequation}{S\arabic{equation}}
\renewcommand{\thefigure}{S\arabic{figure}}
\renewcommand{\bibnumfmt}[1]{[S#1]}
\renewcommand{\thesection}{S\Alph{section}}
\renewcommand{\thesubsection}{\Roman{subsection}}

\bigskip

\begin{center}
\large{\bf Supplemental Material to ``Non-equilibrium bosonization of fractional quantum Hall edges''\\}
\end{center}
\begin{center}
Christian Sp\r{a}nsl\"{a}tt,$^{1}$ Jinhong Park,$^{2,3}$, and Alexander D. Mirlin$^{3,4}$
 \\
{\it $^1$Department of Engineering and Physics, Karlstad University, Karlstad, Sweden \\ 
$^2$Department of Physics, Konkuk University, Seoul 05029, Republic of Korea\\
$^3$Institut f\"{u}r Nanotechnologie, Karlsruhe Institute of Technology, 76021 Karlsruhe, Germany \\$^4$Institut f\"{u}r Theorie der Kondensierte Materie, Karlsruhe Institute of Technology, 76128 Karlsruhe, Germany }\\

(Dated: \today)

\end{center}

In this Supplemental Material, we provide details of derivations that are presented in a shorter form in the main text. Specifically, Sec.~\ref{app:determinant_calcs} contains details of calculations of Toeplitz determinants and Green's functions, Sec.~\ref{app:current_noise} provides details of the analysis of the tunneling current and noise,
while in Sec.~\ref{sec:Counter_SM} we present technical details of the calculations 
in Secs.~\ref{sec:Counter_prop}-\ref{sec:transport_counter} of the main text, which are devoted to the non-equilibrium bosonization of FQH edges with counter-propagating modes.

\section{Toeplitz determinants and Green's functions}
\label{app:determinant_calcs}
In this Section, we provide details of the analysis of determinants in the main text, largely following Refs.~\cite{Gutman2010,Gutman2011non-equilibrium}. The considered determinants are of the form
\begin{align} \label{eq:Fredholmdet_App}
    &\Delta[\delta_{\tau}(t)] = \text{Det} \big[ 1 + (e^{-i\delta_{\tau}(t)} - 1) f (\epsilon) \big ],\\
    & \delta_{\tau}(t) = \delta_0 \left[\Theta(-t)-\Theta(-t-\tau)\right],
    \end{align}
and belong to the class of Fredholm functional determinants. Here, the time $t$ and the energy $\epsilon$ are to be understood as canonically conjugate variables, i.e., $[\epsilon, t] = i$, which makes the calculation of the determinant, in general, highly non-trivial. However, the problem is simplified by the fact that the
scattering phase $\delta_\tau(t)$
is non-zero only in a finite time interval $\tau$ and, furthermore, is constant (equal to $\delta_0$) in this interval. The determinant then acquires so-called Toeplitz form. 
Such a determinant arises in the process of evaluating Green's functions of quasiparticles with charge $e^*_2=n_2\nu e$ for a Laughlin edge driven out of equilibrium by injection of quasiparticles with charge $e^*_1=n_1\nu e$, see  Sec.~\ref{sec:GF_Laughlin} of the main text. The phase amplitude $\delta_0$ is then equal to the mutual braiding phase for these two types of quasiparticles, $\delta_0 = 2\pi \nu n_1 n_2$.

 In equilibrium, the determinant becomes a Gaussian functional of $\delta_\tau(t)$ and can be evaluated exactly as discussed below. After this discussion, we focus on the non-equilibrium situation of our main interest: a double-step distribution function $f(\epsilon)$. In this case, the large-$\tau$ asymptotics of the determinant can be found within the Szeg\H{o} approximation and, more accurately, within the generalized Fisher-Hartwig formula. When the phase $\delta_0$ is $2\pi$ (or a multiple integer of $2\pi$), the generalized Fisher-Hartwig formula becomes exact, as we also discuss below.

\subsection{Zero temperature equilibrium}
\label{app:Zero_T_determinant}
We first consider the case of a zero temperature distribution function
\begin{align}
\label{eq:zeroT_dist_app}
    f(\epsilon) = \Theta(-\epsilon) \equiv f_0(\epsilon) \,. 
\end{align}
The functional determinant in Eq.~\eqref{eq:Fredholmdet_App} then evaluates to~\cite{Gutman2010}
\begin{align}
\label{eq:det_zero_T_app}
    \Delta[\delta_\tau(t)]_{T=0, V=0} = \frac{e^{-i \delta_0 \tau \Lambda/(2\pi)}}{(1+\tau^2\Lambda^2)^{\frac{1}{2}\left(\frac{\delta_0}{2\pi}\right)^2}}\,,
\end{align} 
where $\Lambda=v/a$ is the frequency UV cutoff. 

We shift now the Fermi-level with the potential $e_1^*V$, i.e., take 
\begin{align}
\label{eq:zeroT_dist_V_app}
    f(\epsilon) = 
    f_0(\epsilon - e_1^*V) \equiv
    \Theta(-\epsilon+e_1^*V )\,. 
\end{align}
The functional determinant \eqref{eq:Fredholmdet_App} acquires then, in comparison with Eq.~\eqref{eq:det_zero_T_app}, an additional phase factor:
\begin{align}
    \label{eq:det_zero_T_finiteV_app}
    \Delta[\delta_\tau(t)]_{T=0, V}  = e^{-i \delta_0 \tau e_1^* V/(2\pi)} \Delta[\delta_\tau(t)]_{T=0, V=0}\,. 
\end{align}
Normalizing the determinant to its value at zero temperature and zero voltage and raising to a power $1/n_1^2 \nu$ according to our formula for the Green's function (Sec.~\ref{sec:GF_Laughlin} of the main text) and using $\delta_0 = 2\pi \nu n_1 n_2$, we get
\begin{align}
\label{appeq:det_constant_bias}
  \left\{\overline{\Delta}[\delta_{\tau}(t)]_{T=0,V}\right\}^\frac{1}{n_1^2\nu} \equiv \Big\{\frac{ \Delta[\delta_\tau(t)]_{T=0, V}}{ \Delta[\delta_\tau(t)]_{T=0, V=0}} \Big \}^{\frac{1}{n_1^2\nu}}= 
    \left[e^{-in_1\nu e (\delta_0/2\pi) V\tau}\right]^\frac{1}{n_1^2\nu}
    =
    e^{-ie_2^* V \tau}\,,
\end{align}
which is the expected oscillatory dependence of the Green's  function on $e_2^* V$.

\subsection{Finite temperature equilibrium}
\label{app:Finite_T_determinant}

For the case of an equilibrium distribution function at finite temperature $T$ with the Fermi level at $e_1^* V$,
\begin{align} \label{eq:equilibriumdist}
  f(\epsilon) =  f_T(\epsilon - e_1^*V ) \equiv \frac{1}{e^{(\epsilon- e_1^* V)/T}+1}. 
\end{align}
the determinant becomes~\cite{Gutman2010}
\begin{align}
\label{eq:det_finite_T_app}
    \Delta[\delta_\tau(t)]_{T,V} = \left(\frac{\pi T \tau}{\sinh(\pi T \tau)}\right)^{\left(\frac{\delta_0}{2\pi}\right)^2}\times\frac{e^{-i \delta_0  \tau \Lambda/(2\pi)}}{(1+\tau^2 \Lambda^2)^{\frac{1}{2}\left(\frac{\delta_0}{2\pi}\right)^2}}  e^{-i \delta_0 \tau e_1^* V/(2\pi)} .
\end{align}
Similarly to the zero temperature case above, a shift of the Fermi level produces an overall phase factor $e^{-ie_1^* (\delta_0/2\pi) V \tau}$ in Eq.~\eqref{eq:det_finite_T_app}. The determinant normalized to its value at zero temperature and zero voltage 
and raised to the power $1/n_1^2 \nu$ then becomes
\begin{align}
\label{appeq:det_constant_bias_finitetemp}
  \left\{\overline{\Delta}[\delta_{\tau}(t)]_{T,V}\right\}^\frac{1}{n_1^2\nu} \equiv \Big\{\frac{ \Delta[\delta_\tau(t)]_{T, V}}{ \Delta[\delta_\tau(t)]_{T=0, V=0}} \Big \}^{\frac{1}{n_1^2\nu}}= \left(\frac{\pi T \tau}{\sinh(\pi T \tau)}\right)^{n_2^2 \nu}
   e^{-ie_2^* V \tau}\,. 
\end{align}

\subsection{Double step distribution}

We consider now the case of our main interest: the non-equilibrium double-step distribution function
\begin{align} \label{appeq:doublestepdistfun}
    &f(\epsilon) = \mathcal{T}f_{0}(\epsilon-e_1^* V) + (1-\mathcal{T})f_{0}(\epsilon).
\end{align}
Here, $0 <\mathcal{T} < 1$ is the probability for charge $e_1^* = n_1 e \nu$ (with $n_1$ an integer) quasiparticles to tunnel from another edge at potential $V$, and $f_0(\epsilon)$ is the zero-temperature zero-voltage Fermi function,  Eq.~\eqref{eq:zeroT_dist_app}. Below, we use two approaches ---the Szeg\H{o} approximation and the  generalized Fisher-Hartwig conjecture--- to compute the Fredholm (Toeplitz) determinant \eqref{eq:Fredholmdet_App}  when the distribution function takes the form~\eqref{appeq:doublestepdistfun}. This calculation follows to a large extent Ref.~\cite{Gutman2011non-equilibrium}. We will be particularly  interested in the dilute limit, $\mathcal{T} \ll 1$.

\subsubsection{Szeg\H{o} approximation} \label{appsec:szegoappro}
The Szeg\H{o} approximation applies to determinants generated by a smooth and periodic function on the form $g(z) = \exp\{U(z)\}$. This function has as its domain the unit circle parameterized by $z = e^{i\varphi}$ with polar angle $\varphi \in [-\pi, \pi]$. The function $U(z)$ can then be described by its Fourier modes,
\begin{align}
\label{eq:V_FM}
    U(z) = \sum_{k=-\infty}^{\infty} U_k e^{i \varphi k}\,, \quad \text{where }U_k = \int_{-\pi}^{\pi} \frac{d \varphi}{2\pi} U(z) e^{-i \varphi k}\,.
\end{align}
The continuous function $g(z)$ generates a Toeplitz matrix $g_N$ with elements $\{(g_N)_{ij} \equiv  (g_N)_{i-j}\}$, obtained by discretizing the polar angle as $\varphi_{i-j} = -\pi + 2\pi (i-j)/ N$ with integer $1\leq i,j \leq N$, i.e., $(g_N)_{i-j} = \exp \big [{U(e^{i\varphi_{i-j}})}\big ]$. Here, $N$ is the size of the discretized matrix.

According to the Szeg\H{o} approximation, the large-$N$ asymptotic behavior of the determinant $\det (g)$ can be obtained by computing $\det (g_N)$ as 
\begin{align} \label{eqapp:szegoapprox}
\det (g) \sim \det(g_N) \approx \exp (N U_0 +  \sum_{k=1}^N k U_k U_{-k}) \,, 
\end{align}
where $U_k$ are the Fourier modes in Eq.~\eqref{eq:V_FM}.

To apply the Szeg\H{o} approximation to determinants~\eqref{eq:Fredholmdet_App},  we first define the periodic, smooth function $g(\epsilon)$ as 
\begin{align}
\label{eq:g_symbol}
    g (z = e^{i \pi \epsilon/\Lambda})  = \big[1+ (e^{- i \delta_0}-1) f(\epsilon) \big] e^{-i \epsilon \delta_0/(2\Lambda)}\,. 
\end{align}
The energy variable $\epsilon$ is here restricted to $\epsilon \in [-\Lambda, \Lambda]$, where $\Lambda$ is a ultraviolet energy cutoff. Since the generic distribution function satisfies $f(\epsilon \to 0)\to 0$ and $f(\epsilon \to -\infty)\to 1$,  the final phase factor in Eq.~\eqref{eq:g_symbol} ensures that the function $g(\epsilon)$ satisfies the periodic boundary condition $g(-\Lambda) = g(\Lambda)$. The energy $\epsilon$ is further parameterized by the angle $\varphi = \pi \epsilon / \Lambda$ which is discretized into $N$ segments. The same discretization scheme is applied to its conjugate variable, the time $t \in [0, \tau]$, which is divided into $N = \tau / \Delta t$ intervals with the unit time step $\Delta t = \pi / \Lambda$. Here, we assume $\tau>0$; the result for $\tau<0$ follows from the property $\Delta[\delta_{\tau}(t)] = (\Delta[\delta_{-\tau}(t)])^*$. 
Applying then the Szeg\H{o} approximation~\eqref{eqapp:szegoapprox} in the case of the double-step distribution~\eqref{appeq:doublestepdistfun}, we obtain the long-$\tau$ asymptotic behavior of the determinant of $g$, 
\begin{align}
    \Delta[\delta_{\tau}(t)] \sim \det (g_N) \approx  e^{-i \frac{\delta_0 \tau \Lambda}{2\pi}}e^{i e^*V \beta_1 \tau} (\Lambda \tau )^{-(\beta_0^2 + \beta_1^2)} \Big(\frac{\Lambda}{|e_1^*V|} \Big)^{-2\beta_0 \beta_1}\,,
\end{align}
where $\beta_0$ and $\beta_1$ are given by 
\begin{align} \label{appeq:beta1beta0}
   \beta_ 0 = - \beta_1 - \frac{\delta_0}{ 2\pi}\,, \quad \beta_1  = - \frac{i}{2\pi} \text{sgn}(e_1^* V) \ln [ 1+ \mathcal{T} (e^{-i \delta_0 \text{sgn}(e_1^* V)} -1 ) ]\,. 
\end{align}
The property $\Delta[\delta_{\tau}(t)] = \Delta^*[\delta_{-\tau}(t)]$ allows us to extend the result to the case of $\tau<0$. Combining the obtained results for $\tau>0$ and $\tau<0$, we find that 
\begin{align}
    \Delta[\delta_{\tau}(t)] \approx  e^{-i \frac{\delta_0 \tau \Lambda}{2\pi}}e^{i e_1^*V \beta_1 \tau} (\Lambda |\tau| )^{-(\beta_0^2 + \beta_1^2)} \Big(\frac{\Lambda}{|e_1^*V|} \Big)^{-2\beta_0 \beta_1}\,.
\end{align}
Here, the extended $\beta_0$ and $\beta_1$ that cover the whole region of $\tau$ are given by
\begin{align} \label{appeq:beta1beta0_ext}
   \beta_ 0 = - \beta_1 - \frac{\delta_0}{ 2\pi}\,, \quad \beta_1  = - \frac{i}{2\pi} \text{sgn}(e_1^* V\tau) \ln [ 1+ \mathcal{T} (e^{-i \delta_0 \text{sgn}(e_1^* V \tau)} -1 ) ]\,. 
\end{align}
Normalizing the determinant to its value for the zero-temperature equilibrium distribution function~\eqref{eq:det_zero_T_app}, we then have
\begin{align}    \label{eq:normalizedDetFullSzego} 
\overline{\Delta}[\delta_{\tau}(t)] \equiv \frac{\Delta[\delta_{\tau}(t)]}{\Delta[\delta_{\tau}(t)]_{T=0, V=0}} \approx e^{i e^*V \beta_1 \tau}  \Big(\frac{1}{|e_1^*V \tau|} \Big)^{-2\beta_0 \beta_1}\,. 
\end{align}
We recall that the Szeg\H{o} approximation applies to the large-$N$ asymptotic regime, which translates into the condition of long time $\tau$. 
Since the characteristic energy scale is $e_1^* V$, the condition is $|e_1^* V \tau| \gg 1$.

We now approximate $\overline{\Delta}[\delta_{\tau} (t)]$ in the weak tunneling regime (i.e., the dilute limit) $\mathcal{T} \ll 1$. In this regime, $\beta_1$ is small and can be approximated as 
\begin{align}
\label{eq:beta_1_approx}
    \beta_1 \approx - \frac{i\,\text{sgn}(e_1^* V\tau)\mathcal{T}}{2\pi} (e^{-i\delta_0 \text{sgn}(e_1^* V \tau)} - 1)\,, 
\end{align}
and hence the power-law exponent, $-2\beta_0\beta_1$, in Eq.~\eqref{eq:normalizedDetFullSzego} can be neglected. Using Eq.~\eqref{eq:beta_1_approx} in Eq.~\eqref{eq:normalizedDetFullSzego}, we obtain 
\begin{align}
    \overline{\Delta}[\delta_{\tau}(t)] \approx  \exp \Big[-\frac{
|e_1^* V\tau| \mathcal{T}}{2\pi}   (1- e^{- i\delta_0 \text{sgn}(e V\tau)})  \Big]
= \exp \Big[- \frac{|e_1^* \langle I_0 \rangle \tau|}{e^2\nu}  (1- e^{- i\delta_0 \text{sgn}(\langle I_0 \rangle  \tau)})  \Big]\,,
\end{align}
where, in the final equality, we used that the product $\mathcal{T}V$ is proportional to the average current $\langle I_0 \rangle = \nu e^2 V \mathcal{T}/(2\pi)$ on the edge. Using this result for the determinant in the expression for the Laughlin Green's functions, given in Eq.~\eqref{eq:GFgreater_gen_n} in the main text, we obtain 
\begin{align}
\label{eq:SM-Szego-GF}
    \mathcal{G}^{\gtrless}(\tau) =  
     \mp \frac{i}{2\pi a} \Big(\frac{a}{a \pm i v\tau} \Big)^{n_2^2\nu} \exp [- \frac{|\langle I_0 \rangle  \tau|}{|e_1^*|}   (1- e^{- i\delta_0\text{sgn}(\langle I_0 \rangle  \tau)}) ] \,,
\end{align}
where $e_1^* = n_1 e \nu$ and $\delta_0 = 2 \pi n_1 n_2 \nu $, which
is  Eq.~\eqref{eq:GF_Poisson_gen} of the main text.

For the case of $n_1= n_2=1$ and thus $e_1^* = \nu e$ and $\delta_0 = 2\pi \nu $, we find 
\begin{align} \label{appeq:LaughlinGF}
   \mathcal{G}^{\gtrless}(\tau) =  
     \mp \frac{i}{2\pi a} \Big(\frac{a}{a \pm i v\tau} \Big)^\nu \exp [- \frac{|\langle I_0 \rangle \tau|}{|e_1^*|} (1- e^{- 2i \pi \nu \text{sgn}(\langle I_0 \rangle  \tau)}) ]\,,
\end{align}
which is  Eq.~\eqref{eq:GF_Poisson} of the main text. This equation is one of key results of Ref.~\cite{Rosenow2016Apr}, which our theory thus reproduces within the Szeg\H{o} approximation. 

\subsubsection{Generalized Fisher-Hartwig conjecture}

We next use the extended form of the Fisher-Hartwig conjecture to compute the Fredholm determinant  with the double-step distribution function~\eqref{appeq:doublestepdistfun}. As in Sec.~\ref{appsec:szegoappro}, we first focus on the case of $\tau > 0$, and then generalize it to time $\tau <0$ by using the relation $\overline{\Delta}[\delta_{\tau} (t)] = (\overline{\Delta}[\delta_{-\tau} (t)])^*$. The application of the extended Fisher-Hartwig formula will not only confirm the results of the Szeg\H{o} approximation, but also goes beyond the Szeg\H{o} approximation by taking into account the different logarithm branches of $\beta_1$ in Eq.~\eqref{appeq:beta1beta0_ext}. 

The Fisher-Hartwig conjecture deals with a class of generating functions $g(z)$ taking the form,  
\begin{align}
\label{eq:generating_function}
    g(z) = e^{U(z)} z^{\sum_{j=0}^{m} \beta_j} \prod_{j=0}^{m} |z - z_{j} |^{2 \alpha_j} h_{z_j, \beta_j} (z) z_{j}^{-\beta_j}\,,
\end{align}
where $U(z)$ is a smooth function and
\begin{align}
   h_{z_j, \beta_j} (z) = \left\{ \begin{matrix} e^{i \pi \beta_j}\,, \quad \quad -\pi < \text{arg}\, z < \varphi_j \\ e^{- i\pi \beta_j }\,, \quad \quad \varphi_j < \text{arg} \, z < \pi\,, \end{matrix} \right. 
\end{align}
where $z$ belongs to the unit circle parameterized as $z = e^{i \varphi}$. The function $g(z)$ has $m+1$ singularities at points $z = z_j = e^{i \varphi_j}$; $\{\varphi_j\}$ takes values in increasing order according to $-\pi < \varphi_0 < \varphi_1<\cdots < \varphi_m < \pi$. 
The asymptotic (large-$N$) behavior of a Toeplitz determinant with $m+1$ Fisher-Hartwig singularities is given by \cite{Gutman2011non-equilibrium}
\begin{align} \label{appeq:FisherHartwig}
    \Delta_N \simeq e^{N U_0} \sum_{k_0 + \cdots + k_m =0} \prod_{j = 0}^{m} z_j^{k_j N} \left [
    N^{- \sum_{j=0}^m \beta_j^2} \prod_{0 \leq j < l \leq m} |z_j - z_l |^{2 \beta_j \beta_l} \prod_{j =0}^m G(1+ \beta_j) G (1-\beta_j)
    \right ]_{\beta_j \rightarrow \beta_j +k_j}\,,
\end{align}
where $G(z)$ is the Barnes $G$-function.

The double-step distribution~\eqref{appeq:doublestepdistfun} belongs to the class of generating functions~\eqref{eq:generating_function}, and the corresponding generating function reads 
\begin{align} \label{appeq:generating}
   g(z) = \big[1+ (e^{-i \delta_0} -1) f(\epsilon(z))\big] e^{-i \delta_0 \epsilon (z) / (2 \Lambda) } =  \big[1+ (e^{-i \delta_0} -1) f(\epsilon(z))\big] z^{-\frac{\delta_0}{2\pi}} 
\end{align}
with $z= e^{i \epsilon \pi/\Lambda}$. We begin with the case of $e_1^*V >0$. 
The singularities of the double-step distribution at $\epsilon = 0$ and $\epsilon= e_1^*V$ lead to the following two singularities of $g(z)$:
\begin{align} \label{appeq:singularities}
   \epsilon_0 = 0\,, \quad \epsilon_1 = e_1^*V\quad\Rightarrow \quad  z_0 = 1 \,, \quad z_1=  e^{i \pi e_1^{*} V / \Lambda}\,. 
\end{align}
Comparing Eqs.~\eqref{appeq:generating} and~\eqref{appeq:singularities} to the general form of the generating function~\eqref{eq:generating_function}, we identify 
\begin{align}
    \alpha_j = 0\,,\quad   \beta_1 = - \frac{i}{2\pi} \ln [ 1+ \mathcal{T} (e^{-i \delta_0} -1 ) ]\,, 
    \quad \beta_ 0 = - \beta_1 - \frac{\delta_0}{ 2\pi}\,, \quad U = -\frac{i \delta_0}{2} + \frac{i \pi e_1^* V \beta_1}{\Lambda}\,. 
\end{align}
The direct application of the Fisher-Hartwig conjecture \eqref{appeq:FisherHartwig} then yields the following result for the determinant 
\begin{align} \label{eq:HartwigFisherneq}
       \Delta [\delta_{\tau}(t)] \simeq e^{-i \tau \Lambda \frac{\delta_0}{2 \pi}} e^{i e_1^* V \beta_1 \tau} \sum_{k = -\infty}^{\infty } & \Big [\left(\frac{\tau \Lambda}{\pi} \right)^{- (\beta_0+k)^2 - (\beta_1 -k)^2} \left(\frac{\pi e_1^* V}{\Lambda} \right)^{2 (\beta_0+k) (\beta_1 - k)}  e^{-i k e_1^* V \tau}\nonumber \\ & \times G(1 + \beta_0 + k) G(1-\beta_0-k) G(1 + \beta_1 - k) G(1-\beta_1 + k) \Big]\,, 
\end{align}
in the long-time limit $e_1^* V \tau \gg 1$. Similarly, for the case $e_1^* V <0$, the singularities of $g(z)$ correspond to 
\begin{align} \label{appeq:singularities_negative}
   \epsilon_0 = e_1^*V\,, \quad \epsilon_1 =0 \quad \Rightarrow \quad z_0 =  e^{i \pi e_1^{*} V / \Lambda} \,, \quad z_1=1\,. 
\end{align}
We then identify 
\begin{align}
    \alpha_j = 0\,,\quad   \beta_0 = \frac{i}{2\pi} \ln [ 1+ \mathcal{T} (e^{i \delta_0} -1 ) ]\,, 
    \quad \beta_1 = - \beta_0 - \frac{\delta_0}{ 2\pi}\,, \quad U = -\frac{i \delta_0}{2} + \frac{i \pi e_1^* V \beta_0}{\Lambda}\,,
\end{align}
resulting in the long-time asymptotics of the determinant 
\begin{align} \label{eq:HartwigFisherneq_neg}
       \Delta [\delta_{\tau}(t)] \simeq e^{-i \tau \Lambda \frac{\delta_0}{2 \pi}} e^{i e_1^* V \beta_0 \tau} \sum_{k = -\infty}^{\infty } & \Big [\left(\frac{\tau \Lambda}{\pi} \right)^{- (\beta_0+k)^2 - (\beta_1 -k)^2} \left(\frac{\pi |e_1^* V|}{\Lambda} \right)^{2 (\beta_0+k) (\beta_1 - k)}  e^{i k e_1^* V \tau}\nonumber \\ & \times G(1 + \beta_0 + k) G(1-\beta_0-k) G(1 + \beta_1 - k) G(1-\beta_1 + k) \Big]\,.
\end{align}
Combining the results~\eqref{eq:HartwigFisherneq} and~\eqref{eq:HartwigFisherneq_neg} for the cases $e_1^* V >0$ and $e_1^* V<0$, we find 
\begin{align}
\label{eq:HartwigFisherneq_tot}
        \Delta [\delta_{\tau}(t)] \simeq e^{-i \tau \Lambda \frac{\delta_0}{2 \pi}} e^{i e_1^* V \beta_1 \tau} \sum_{k = -\infty}^{\infty } & \Big [\left(\frac{\tau \Lambda}{\pi} \right)^{- (\beta_0+k)^2 - (\beta_1 -k)^2} \left(\frac{\pi |e_1^* V|}{\Lambda} \right)^{2 (\beta_0+k) (\beta_1 - k)}  e^{-i k e_1^* V \tau}\nonumber \\ & \times G(1 + \beta_0 + k) G(1-\beta_0-k) G(1 + \beta_1 - k) G(1-\beta_1 + k) \Big]\,, 
\end{align}
with 
 \begin{align}
     \beta_1 = - \frac{i\, \text{sgn}(e_1^* V)}{2\pi} \ln [ 1+ \mathcal{T} (e^{-i \delta_0 \text{sgn}(e_1^* V)} -1 ) ]\,, 
    \quad \beta_ 0 = - \beta_1 - \frac{\delta_0}{ 2\pi}\,. 
\end{align}
Importantly, the extended version of the Fisher-Hartwig formula 
contains a sum over the logarithmic branches $k$, which yields leading asymptotics for terms with all harmonics of oscillations $e^{-i k e_1^* V \tau}$
\cite{Gutman2011non-equilibrium,Protopopov2012Luttinger}.
Accounting for contributions from all possible logarithm branches in the determinant represents the key extension of the Fisher-Hartwig formula beyond the Szeg\H{o} approximation. 
When Fourier-transformed to the energy representation, this translates to singularities at the set of energies $ke_1^* V$. Emergence of these singularities is physically due to processes which transfer quasiparticles between the two Fermi-edges of the double-step distribution. 

For $\mathcal{T}=0$, only the term with $k=0$ in the sum survives and we get the result for the zero-temperature equilibrium distribution function,
\begin{align} \label{eq:HartwigFishereq}
   \Delta [\delta_{\tau}(t)]_{T=0, V=0} = e^{ - i \tau \Lambda \frac{\delta_0}{2 \pi}} \left(\frac{\tau \Lambda}{\pi} \right)^{- \frac{\delta_0^2}{(2 \pi)^2}} G( 1+\frac{\delta_0}{2 \pi })
    G( 1-\frac{\delta_0}{2 \pi})\,. 
\end{align}
Normalizing the determinant for the double-step distribution to its equilibrium, zero-temperature value, we get for the  long-time asymptotics of the normalized determinant
$\overline{\Delta} [\delta_{\tau}(t)]$
by using Eqs.~\eqref{eq:HartwigFishereq} and~\eqref{eq:HartwigFisherneq}, 
\begin{align} 
\label{eq:HartwigFisher}
     \overline{\Delta} [\delta_{\tau}(t)]  \equiv \frac{\Delta [\delta_{\tau}(t)]}{\Delta [\delta_{\tau}(t)]_{T=0, V=0}} \simeq \frac{e^{i e_1^* V \beta_1 \tau}}{{G( 1+\frac{\delta_0}{2 \pi })
    G( 1-\frac{\delta_0}{2 \pi})}} & \sum_{k = -\infty}^{\infty } \Big [ \left(\frac{1}{|e_1^* V \tau|} \right)^{- 2 (\beta_0 +k) (\beta_1 -k)}  e^{-i k e_1^* V \tau}\nonumber \\ & \times G(1 + \beta_0 + k) G(1-\beta_0-k) G(1 + \beta_1 - k) G(1-\beta_1 + k) \Big]\,. 
\end{align}
This result for time $\tau>0$ can be extended to the case of $\tau<0$, resulting in the general expression for the long-time asymptotics ($|e_1^* V \tau| \gg 1$) of the determinant covering both $\tau>0$ and $\tau < 0$ cases, 
\begin{align} \label{eq:HartwigFisher2longfull}
     \overline{\Delta} [\delta_{\tau}(t)] = \frac{e^{i e_1^* V \beta_1(\tau) \tau}}{G( 1+\frac{\delta_0}{2 \pi })
    G( 1-\frac{\delta_0}{2 \pi})} & \sum_{k = -\infty}^{\infty }  \Big [ \left(\frac{1}{|e_1^* V \tau|} \right)^{- 2 (\beta_0(\tau) + k) (\beta_1(\tau) -k)}  e^{-i k e_1^* V \tau}\nonumber \\ & \times G(1 + \beta_0(\tau) + n) G(1-\beta_0(\tau)-n) G(1 + \beta_1(\tau) - k) G(1-\beta_1(\tau) + k) \Big]\,, 
\end{align}
with $\beta_0 (\tau) = \beta_0^* (-\tau)$, $\beta_1 (\tau) = \beta_1^* (-\tau)$, and
\begin{align} \label{appeq:beta1full}
    \beta_1(\tau) = - \frac{i}{2\pi} \text{sgn}(e_1^* V \tau) \ln [ 1+ \mathcal{T} (e^{-i \delta_0 \text{sgn}(e_1^*V \tau)} -1 ) ]\,, \quad \beta_0 (\tau) = -\frac{\delta_0}{2\pi} - \beta_1 (\tau)\,.
\end{align}
The symmetry relation $\overline{\Delta}[\delta_{\tau} (t)] = (\overline{\Delta}[\delta_{-\tau} (t)])^*$ 
is preserved by this asymptotic form, in view of the following property of the Barnes $G$-function: $[G(z)]^* = G(z^*)$. 

The contribution of each of the branches (labeled by $k$) is characterized by  power-law exponents
$- 2 (\beta_0(\tau) + k) (\beta_1(\tau) -k)$, which depend on $\delta_0$ and on $k$. The dominant contributions at large $|\tau|$ are characterized by the smallest values of the real part of this exponent. As discussed below in more detail, it is sufficient, in most cases of our interest, to keep the contributions of the $k=0$ and $k=1$ branches, as they will give the dominating contributions.   We get for these contributions:
\begin{align} \label{eq:zeronsector}
    \overline{\Delta}_{k=0} [\delta_{\tau}(t)]   =  e^{i e_1^* V \beta_1 \tau} \frac{G(1+\beta_0) G(1-\beta_0) G(1+\beta_1) G(1-\beta_1)}{G[1 - \delta_0/(2 \pi)] G[1 + \delta_0/(2 \pi)] } \Big (\frac{1}{|e_1^* V \tau|} \Big)^{-2 \beta_0 \beta_1}\,, 
\end{align}
and 
\begin{align} \label{eq:firstsector}
    \overline{\Delta}_{k=1} [\delta_{\tau}(t)]   =  e^{i e_1^* V \beta_1 \tau} e^{-i e_1^* V \tau} \frac{G(2+\beta_0) G(-\beta_0) G(\beta_1) G(2-\beta_1)}{G[1 - \delta_0/(2 \pi)] G[1 + \delta_0/(2 \pi)] } \Big (\frac{1}{|e_1^* V \tau|} \Big)^{-2 (\beta_0+1)(\beta_1-1)}\,, 
\end{align}
respectively. Adding these two contributions yields 
\begin{align} 
\label{eq:zerofirstsector}
    \overline{\Delta}_{k=0,1}\equiv \overline{\Delta}_{k=0}+\overline{\Delta}_{k=1}  &=  \overline{\Delta}_{k=0} \Big[ 1+  e^{-i e_1^* V \tau} \frac{G(2+\beta_0) G(-\beta_0) G(\beta_1) G(2-\beta_1)}{G(1+\beta_0) G(1-\beta_0) G(1+\beta_1) G(1-\beta_1)} \Big (\frac{1}{|e_1^* V \tau|} \Big)^{2(\beta_0 - \beta_1 +1)} \Big ]\,, \nonumber \\ 
    &= \overline{\Delta}_{k=0} \Big[ 1+  e^{-i e_1^* V \tau} \frac{\Gamma(1+\beta_0) \Gamma(1-\beta_1)}{\Gamma(-\beta_0) \Gamma(\beta_1)}  \Big (\frac{1}{|e_1^* V \tau|} \Big)^{2(\beta_0 - \beta_1 +1)} \Big ]\,, \nonumber \\ 
     &= \overline{\Delta}_{k=0} \Big[ 1+  e^{-i e_1^* V \tau} \Big(\frac{\Gamma(1-\beta_1)}{\Gamma(-\beta_0)}\Big)^2 \frac{\sin(\pi \beta_1)}{\sin(\pi\beta_1+\frac{\delta_0}{2})}  \Big (\frac{1}{|e_1^* V \tau|} \Big)^{2(\beta_0 - \beta_1 +1)} \Big ]\,,
\end{align}
where $\Gamma(z)$ is the Gamma function and we have used the identities $G(z+1) = G(z) \Gamma(z)$ and $\Gamma(z) \Gamma(1-z) = \pi / \sin(\pi z)$. 

One case when keeping the $k=0$ and $k=1$ contributions is sufficient---but also necessary!---is the case of a phase amplitude $\delta_0 = 2\pi$. This value of $\delta_0$ arises for the free-fermion Green's function and, in the FQH context, for a Laughlin $\nu = 1/m$ edge in the case $n_1=1$ and $n_2=m$ or, vice versa, $n_1=m$ and $n_2=1$. Let us consider for definiteness free fermions, $\nu=n_1=n_2=1$ and $e_1^* = e$. 
Setting $\delta_0 = 2\pi - \tilde{\delta}_0$ and taking the limit $\tilde{\delta}_0 \rightarrow 0$, one obtains from Eqs.~\eqref{eq:zeronsector}-\eqref{eq:zerofirstsector},
\begin{align} \label{appeq:freefermionGF}
    \overline{\Delta}[\delta_\tau (t)] = 1 - \mathcal{T} + \mathcal{T} e^{-i e V \tau} \quad \longrightarrow \quad    \mathcal{G}^{\gtrless}(\tau) = \mp  \frac{i}{2\pi a} \frac{a}{ (a \pm iv\tau)} 
  \big [1- \mathcal{T}+ \mathcal{T}e^{-i eV \tau} \big] \,,
\end{align}
which is the exact result for the GFs for free fermions with a double-step distribution function with arbitrary $\mathcal{T}$. 
Contributions of other branches (with $k \ne 0, 1$) vanish identically for $\delta_0 = 2\pi$.
We note that, importantly, the Szeg\H{o} approximation 
is not sufficient in this case, since it fully misses the non-equilibrium physics, as is clear from setting $\delta_0=2\pi$ in Eq.~\eqref{appeq:LaughlinGF}.
The result for the determinant with $\delta_0=2\pi$
is used in the end of Sec.~\ref{sec:GF-asymptotic} of the main text for the analysis of the GFs for a Laughlin $\nu = 1/m$ edge in the case $n_1=1$ and $n_2=m$ or, alternatively, $n_1=m$ and $n_2=1$, and in
Sec.~\ref{sec:Andreev_setups} of the main text for the analysis of transport between two such edges. 

We consider now the generic case of $\delta_0$ away from $2\pi$ (and from integer multiples of $2\pi$). We focus here on the weak-tunneling regime (which is characteristic for tunneling of fractionally-charged FQH quasiparticles via the FQH bulk) with $\mathcal{T} \ll 1$. In this regime, 
$\beta_1$ in Eq.~\eqref{appeq:beta1full} is approximated as 
\begin{align}
\label{eq:SM_beta1}
    \beta_1 \approx -\frac{i}{2\pi}  \text{sgn}(e_1^*V\tau) \mathcal{T} (e^{-i \delta_0\, \text{sgn}(e_1^* V \tau)}-1)\quad \Rightarrow \quad\,|\beta_1|\ll 1\,.
\end{align}
Given that $|\beta_1|\ll 1$ in the weak tunneling regime, the dominant logarithm branches (from the point of view of the infrared scaling) are determined by the value of $\beta_0 \approx - \delta_0/2\pi$. In particular, if $\delta_0$ lies in the range $\delta_0 \in (-4\pi \text{Re}\beta_1, 4\pi (1-\text{Re}\beta_1))\approx (0,4\pi)$, the $k=0$ and $k=1$ branches are the dominant ones. 
The range $0 < \delta_0 < 4\pi$ covers essentially all values of interest for the purpose of this work, so that we restrict ourselves to keeping only the $k=0$ and $k=1$ terms as in Eqs.~\eqref{eq:zeronsector} and~\eqref{eq:firstsector}. We further use 
the condition~\eqref{eq:SM_beta1} and expand the Barnes-$G$ functions in $\beta_1$ up to linear order: 
\begin{align} \label{eq:zerothorderexpansion}
    \frac{G(1+\beta_0) G(1-\beta_0) G(1+\beta_1) G(1-\beta_1)}{G[1 - \delta_0/(2 \pi)] G[1 + \delta_0/(2 \pi)] }  &= \frac{G[1 - \delta_0/(2 \pi) - \beta_1 ] G[1+ \delta_0/(2 \pi) + \beta_1] G(1+\beta_1) G(1-\beta_1)}{G[1 - \delta_0/(2 \pi)] G[1 + \delta_0/(2 \pi)] } 
    \nonumber \\ & \approx 1 - \beta_1 \frac{\delta_0}{2\pi}\Big(2-\psi (1- \frac{\delta_0}{2\pi}) - \psi (1+\frac{\delta_0}{2\pi}) \Big)\,,
\end{align}
where $\psi(z)=\Gamma'(z)/\Gamma(z)$ is the di-Gamma function and we used $G(1) = 1$. The $k=0$ contribution is then approximated as 
\begin{align}  \label{eq:zerothweaktun}
    \overline{\Delta}_{k=0} [\delta_{\tau}(t)]  &\approx 
    \exp \Big[\frac{\mathcal{T}|e_1^* V \tau|}{2\pi} (e^{-i\delta_0 \text{sgn}(e_1^*V\tau)} - 1)\Big] \Big(\frac{1}{|e_1^*V \tau|} \Big)^{-\frac{\delta_0 \mathcal{T}}{\pi^2} \sin \frac{\delta_0}{2} \exp [-i \frac{\delta_0}{2} \text{sgn}(e_1^* V \tau)]} \nonumber \\ 
    & \times \Big [1+ \mathcal{T}\frac{\delta_0}{2\pi^2} e^{-i \text{sgn}(e_1^* V\tau) \delta_0/2} \sin \big(\frac{\delta_0}{2} \big) \Big(2-\psi (1- \frac{\delta_0}{2\pi}) - \psi (1+\frac{\delta_0}{2\pi}) \Big) \Big]\,
\end{align}
In the same way the total contribution of the $k=0$ and $k=1$ terms, 
Eq.~\eqref{eq:zerofirstsector}, is  approximated as 
\begin{align} \label{eq:zerofirstweaktun}
    \overline{\Delta}_{k=0,1} [\delta_{\tau}(t)]  &\underset{|\beta_1| \ll 1}{\approx} \overline{\Delta}_{k=0} [\delta_{\tau}(t)]  \Big[ 1+ e^{-i e_1^* V \tau} \frac{\pi \beta_1}{\big(\Gamma(\frac{\delta_0}{2\pi})\big)^2} \frac{1}{\sin(\frac{\delta_0}{2})} \Big( \frac{1}{|e_1^*V\tau|} \Big)^{2(1- \frac{\delta_0}{2\pi})} \Big] \nonumber \\
    &= \overline{\Delta}_{k=0} [\delta_{\tau}(t)]  \Big [1- \mathcal{T}e^{-i e_1^* V \tau} e^{-i \frac{\delta_0}{2} \text{sgn}(e_1^* V\tau)}\frac{1}{\big(\Gamma(\frac{\delta_0}{2\pi})\big)^2}   \Big( \frac{1}{|e_1^*V\tau|} \Big)^{2(1- \frac{\delta_0}{2\pi})}\Big]\,. 
\end{align}
Combining Eqs.~\eqref{eq:zerothweaktun} and~\eqref{eq:zerofirstweaktun} and inserting them into the general GF expression, Eq.~\eqref{eq:GFgreater_gen_n} of the main text, we obtain the following result for the GFs for $\mathcal{T} \ll 1$:
\begin{align} \label{eq:Greenfunzerofirstweaktun}
  \mathcal{G}^\gtrless(\tau) &\approx  \mp \frac{i}{2\pi a} \frac{a^{n_2^2\nu}}{ (a \pm i v\tau)^{n_2^2\nu}} \left\{ \overline{\Delta}_{k=0,1} [\delta_{\tau}(t)]\right\}^{\frac{1}{n_1^2\nu}} \nonumber \\ 
  &= \mp  \frac{i}{2\pi a} \frac{a^{n_2^2\nu}}{ (a \pm  i v \tau)^{n_2^2\nu}}  \exp [- \frac{|\langle I_0  \rangle \tau| }{|e_1^*|}  (1- e^{- i\delta_0 \text{sgn}(\langle I_0\rangle \tau)}) ] \Big(\frac{1}{|e_1^*V \tau|} \Big)^{-\frac{2n_2}{\pi n_1} \mathcal{T} \sin \frac{\delta_0}{2} \exp [i \frac{\delta_0}{2} \text{sgn}(\langle I_0 \rangle \tau)]}
  \nonumber \\ & \quad \quad \times
  \Big [1+ \mathcal{T}\frac{\delta_0}{2\pi^2} e^{-i \delta_0 \text{sgn}(\langle I_0\rangle \tau)/2} \sin \big(\frac{\delta_0}{2} \big) \Big(2-\psi (1- \frac{\delta_0}{2\pi}) -\psi (1+\frac{\delta_0}{2\pi}) \Big) \Big]^{1/{n_1^2\nu}}  \nonumber \\ & \quad \quad \times \Big [1- \mathcal{T}e^{-i e_1^* V \tau} e^{-i \frac{\delta_0}{2} \text{sgn}(\langle I_0\rangle \tau)}\frac{1}{\big(\Gamma(\frac{\delta_0}{2\pi})\big)^2}   \Big( \frac{1}{|e_1^*V\tau|} \Big)^{2(1- \frac{\delta_0}{2\pi})}\Big]^{1/{n_1^2\nu}} \nonumber \\
  &\underset{\mathcal{T}\ll 1}{\approx} \mp \frac{i}{2\pi a} \frac{a^{n_2^2\nu}}{ (a \pm i v\tau)^{n_2^2\nu}}  \exp [- \frac{|\langle I_0 \rangle \tau| }{|e_1^*|}  (1- e^{- i\delta_0 \text{sgn}(\langle I_0\rangle \tau)}) ] \Big(\frac{1}{|e_1^*V \tau|} \Big)^{-\frac{2n_2}{\pi n_1} \mathcal{T} \sin \frac{\delta_0}{2}  \exp [i \frac{\delta_0}{2} \text{sgn}(\langle I_0\rangle \tau)]}  \nonumber \\ 
  & \quad \quad \times
  \Big [1+ \frac{\mathcal{T}}{n_1^2 \nu} e^{-i \frac{\delta_0}{2} \text{sgn}(\langle I_0\rangle \tau)} \Big \{\frac{\delta_0}{2\pi^2} \sin \big(\frac{\delta_0}{2} \big) \Big(2-\psi (1- \frac{\delta_0}{2\pi}) - \psi (1+\frac{\delta_0}{2\pi}) \Big) - e^{-i e_1^*V \tau}  \frac{1}{\big(\Gamma(\frac{\delta_0}{2\pi})\big)^2} \Big( \frac{1}{|e_1^*V\tau|} \Big)^{2(1- \frac{\delta_0}{2\pi})}\Big \} \Big]\,. 
\end{align}
This formula, which is  Eq.~\eqref{eq:GreenfunFullHartwig} of the main text, is our key result for the non-equilibrium quasiparticle GFs on the Laughlin edge. Let us recall that here $n_1$ and $n_2$ characterize the excitations whose injection drives the system out of equilibrium and the excitations for which the GFs are calculated, respectively, and $\delta_0 = 2\pi\nu n_1 n_2$. The first three factors in the final expression in Eq.~\eqref{eq:Greenfunzerofirstweaktun} 
yield the Szeg\H{o} approximation
\eqref{eq:SM-Szego-GF}.

\section{Tunneling current and noise}
\label{app:current_noise}

In this section, we 
provide details of the analysis of tunneling current and noise in Sec.~\ref{sec:transport-zeta-less-12}. We begin by
directly evaluating the integrals for the tunneling current and noise in the Szeg\H{o} approximation [Eqs.~\eqref{eq:current_dilute_app} and \eqref{eq:noise_dilute_app} of the main text]. To do so, it is useful to consider the following integral:  
\begin{align}
    \mathcal{I} [\omega, \gamma] \equiv \int_{-\infty}^{\infty} d\tau \Big(\frac{1}{a + iv\tau} \Big)^{2\zeta} e^{-i \omega \tau} e^{-\gamma |\tau|}\,. 
\end{align}
This integral can be decomposed into separate integrals along the positive and negative time axes:
\begin{align} \label{eq:Integrals}
   \mathcal{I} [\omega, \gamma]  = \int_{0}^{\infty}  d\tau \Big(\frac{1}{a + iv\tau} \Big)^{2\zeta} e^{-i \omega \tau} e^{-\gamma \tau} + \int_{-\infty}^{0}  d\tau \Big(\frac{1}{a + iv\tau} \Big)^{2\zeta} e^{-i \omega \tau} e^{\gamma \tau} \equiv \mathcal{I}_0 [i\omega + \gamma, v] + \mathcal{I}_0 [-i\omega + \gamma, -v]\,, 
\end{align}
where we introduced
\begin{align}
   \mathcal{I}_0 [i\omega + \gamma, v] \equiv \int_{0}^{\infty} d\tau \Big(\frac{1}{a + iv\tau} \Big)^{2\zeta} e^{-i \omega \tau} e^{-\gamma \tau}\,. 
\end{align}
When $0< \zeta < 1/2$, the integral $\mathcal{I}_0$ converges in the limit of vanishing ultraviolet regularization, $a \rightarrow 0$. Thus, for $0< \zeta < 1/2$, we can safely neglect $a$ in the denominator for the integral $\mathcal{I}_0$, and the integral then evaluates as   
\begin{align} \label{eq:integraldelta12}
    \mathcal{I}_0 [i \omega + \gamma, v] &\xrightarrow{a\rightarrow0} \int_{0}^{\infty} d\tau  \frac{1}{(iv\tau)^{2\zeta}} e^{-i \omega \tau} e^{- \gamma \tau}  = \frac{2}{v} \Gamma(1-2\zeta) \Big(\frac{i\omega+\gamma}{v} \Big)^{2\zeta-1} e^{-i\pi \zeta}\,, \quad \text{for }0< \zeta<1/2\,.
\end{align} 
To extend the analysis of this integral to a broader range $0 < \zeta<1$ and to explore the effect of finite $a$, we integrate by parts, which yields
\begin{align} \label{eq:integrationbypartsI0}
   \mathcal{I}_0  [i \omega + \gamma, v] &= \int_{0}^{\infty} d\tau  (\frac{1}{a + iv\tau})^{2\zeta} \, e^{-i \omega \tau} e^{- \gamma \tau} = -\frac{i }{1-2\zeta}  \frac{a}{v}\int_{0}^{\infty} d\tau \frac{d}{d\tau}\Big(\frac{1}{(a+i v\tau)^{2\zeta-1}}\Big) e^{- i \omega \tau } e^{-\gamma \tau} \nonumber \\ 
    & = 
    - \frac{i}{1-2\zeta} \frac{a (i\omega+\gamma)}{v}\int_{0}^{\infty} d\tau \frac{1}{(a+iv\tau)^{2\zeta-1}} e^{-i \omega \tau} e^{-\gamma \tau} 
  +  \frac{i}{1-2 \zeta} \frac{a^{1-2\zeta}}{v} 
    \nonumber \\
     & \simeq \frac{2}{v} \Gamma(1-2\zeta) \Big(\frac{i\omega+\gamma}{v} \Big)^{2\zeta-1} e^{-i\pi \zeta}+ \frac{i}{1-2 \zeta} \frac{a^{1-2\zeta}}{v}  \,. 
\end{align}
In the third equality of Eq.~\eqref{eq:integrationbypartsI0}, we use the fact that the exponent, $2\zeta - 1$, in the integrand falls into the regime $0< 2\zeta-1 <1$, so that we can use the integral formula \eqref{eq:integraldelta12} directly. Integrating once more by parts in the second line of Eq.~\eqref{eq:integrationbypartsI0}, we determine a correction of the order $a^{2-2\zeta}$ and also extend the validity range up to $0<\zeta<3/2$:
\begin{align}
\label{eq:I_0-second-iteration}
\mathcal{I}_0[i\omega + \gamma, v] &=\frac{2}{v} \Gamma(1-2\zeta) \Big(\frac{i\omega+\gamma}{v} \Big)^{2\zeta-1} e^{-i\pi \zeta}+\frac{i}{1-2 \zeta} \frac{a^{1-2\zeta}}{v} +\frac{1}{(1-2 \zeta)(2-2\zeta)} \frac{a^{2-2\zeta}}{v} \Big(\frac{i\omega+\gamma}{v} \Big)\,.
\end{align}
In principle, this iterative procedure can be continued further, generating a series in the small parameter $a (i\omega + \gamma)/v \ll 1$ but Eq.~\eqref{eq:I_0-second-iteration} is sufficient for our purposes. 
Substituting Eq.~\eqref{eq:I_0-second-iteration} into Eq.~\eqref{eq:Integrals}, we find
\begin{align}
    \mathcal{I}[\omega, \gamma] &= \frac{4}{v} \Gamma(1-2\zeta) \text{Im} \Big[ \Big(\frac{\omega-i\gamma}{v} \Big)^{2\zeta-1} \Big]+ \frac{1}{(1-2 \zeta)(1-\zeta)} \frac{a^{2-2\zeta}}{v} \Big(\frac{\gamma}{v} \Big)  \,,
\qquad \text{for } 0 < \zeta < \frac{3}{2}\,.
\end{align}
Using this integral formula, we evaluate the integrals for the current and noise for tunneling between non-equilibrium Laughlin edges in the Szeg\H{o} approximation
[Eqs.~\eqref{eq:current_dilute_app} and \eqref{eq:noise_dilute_app} in the main text], resulting in 
\begin{align} \label{tunnelingcurrent_app}
     \langle I_T\rangle & =  \frac{2ie_2^*|\xi|^2}{(2\pi a)^2} 
    \int_{-\infty}^{\infty} d\tau  \left(\frac{a}{a+ iv\tau}\right)^{2\zeta}  e^{-(\gamma_u+\gamma_d)|\tau|}\sin(\tau(\omega_u-\omega_d)) \notag \\ &= \frac{e_2^*|\xi|^2 a^{2\zeta}}{(2\pi a)^2} \big( \mathcal{I}[-(\omega_u -\omega_d), \gamma_u + \gamma_d]  - \mathcal{I}\big[\omega_u -\omega_d,  \gamma_u + \gamma_d \big]\big)
   \notag \\ &=\frac{e_2^* a^{2\zeta} |\xi|^2}{2(\pi a)^2 v^{2\zeta}} \frac{\pi}{\Gamma(2 \zeta) \cos(\pi \zeta)}  \frac{\sin \big[(1-2\zeta)\tan^{-1} (p)    \big]}{(|\omega_u-\omega_d|^2 + (\gamma_u+\gamma_d)^2)^{1/2-\zeta}}\,,
\end{align}
and 
\begin{align}
     S_T &= \frac{4(e_2^*)^2|\xi|^2}{(2\pi a)^2}
    \int_{-\infty}^{\infty} d\tau  \left(\frac{a}{a+ iv\tau}\right)^{2\zeta}   e^{-(\gamma_u+\gamma_d)|\tau|}\cos(\tau(\omega_u-\omega_d))\notag \\ 
    &= \frac{2(e_2^*)^2|\xi|^2a^{2\zeta}}{(2\pi a)^2}
    \big( \mathcal{I}[-(\omega_u -\omega_d), \gamma_u + \gamma_d]  +\mathcal{I}\big[\omega_u -\omega_d,  \gamma_u + \gamma_d \big]\big) \notag \\ 
    & = \frac{(e_2^*)^2 |\xi|^2}{(\pi a)^2}\frac{a}{v} \Big[\frac{\pi}{\Gamma(2 \zeta) \sin(\pi \zeta)} \Big(\frac{a}{v} \Big)^{2\zeta-1} \frac{\cos \big[(1-2\zeta) \tan^{-1} (p) \big]}{(|\omega_u-\omega_d|^2 + (\gamma_u+\gamma_d)^2)^{1/2-\zeta}} + \Big(\frac{a}{v} \Big) \frac{\gamma_u + \gamma_d}{(1-2\zeta)(1-\zeta)} \Big].
    \label{tunneling_noise_app}
\end{align}
The result 
\eqref{tunnelingcurrent_app}
for the tunneling current $\langle I_T\rangle$ and the first term in 
Eq.~\eqref{tunneling_noise_app} for
the noise $S_T$ are presented in Eqs.~\eqref{eq:current_dilute_app} and \eqref{eq:noise_dilute_app} in the main text. As we see, the relative correction for the noise 
is of the order $[a (\gamma_u + \gamma_d)/v]^{2-2\zeta}$ and is thus small for $\zeta < 1$, as stated in the main text. The correction to the tunneling current turns out to be zero to this order but will appear in the next order of the expansion.

Now, we discuss a contribution of the terms in 
the last form of Fisher-Hartwig formula \eqref{eq:Greenfunzerofirstweaktun} for the Green's function [Eq.~\eqref{eq:GreenfunFullHartwig} of the main text] 
that are beyond the Szeg\H{o} approximation. We recall that we work under a general assumption of dilute limit, $\mathcal{T}_{u,d} \ll 1$. The most important correction to tunneling current and noise comes from the last term in the curly bracket. 
Substituting into Eqs.~\eqref{eq:tunneling_current} and \eqref{eq:tunnelin_noise}, we observe that the time integral is now governed by relatively short times,
$\tau \sim [e_1^* V_{u,d}]^{-1}$. The Fisher-Hartwig (and Szeg\H{o}) formulas, which are controllable at long times, are at the borderline of applicability on this time scale. However, we can still use the Fisher-Hartwig formula to estimate the magnitude of the correction. Performing this estimate for the correction to tunneling current and noise induced by a correction to the Green's function of the edge $u$,  we find a contribution that scales  $\propto  \mathcal{T}_u V_u^{2\zeta-1}$. Comparing to the Szeg\H{o}-approximation result $\propto  (\mathcal{T}_u V_u)^{2\zeta-1}$, we see that the correction is parametrically small 
(due to $\mathcal{T}_u \ll 1$) for $\zeta < 1$. 
Thus, the conditions $\mathcal{T}_{u,d}\ll 1$ and $\zeta < 1$ allow us to neglect the corrections to the Szeg\H{o} approximations.

It is worth mentioning that the correction that we have just discussed---which originates from another branch in the Fisher-Hartwig formula---physically corresponds to anyons that tunnel from the edge $u$ to the egde $d$ carrying an energy $e_1^* V_u$. Processes of this type were termed ``direct'' in Refs.~\cite{Lee2023May, Schiller2023anyon,Zhang2025Mar}.

\section{Non-equilibrium bosonization of an edge with counter-propagating modes}
\label{sec:Counter_SM}

In this section, we present technical details of calculations to non-equilibrium bosonization of a FQH edge with counter-propagating modes, Secs.~\ref{sec:Counter_prop}-\ref{sec:transport_counter} of the main text.

The FQH topological order is described by the $K$-matrix and the charge vector $\vec{q}$.
We focus on edges with two modes,
\begin{align}
\label{appeq:KMat2x2}
    K = \begin{pmatrix}
    \nu_1^{-1} & 0\\
    0 & \nu_2^{-1}
    \end{pmatrix}, \quad \vec{q}=\begin{pmatrix}
        1\\1
    \end{pmatrix}.
\end{align}
The case of counter-propagating modes corresponds to  $\nu_1 \nu_2<0$. More specifically, we study the prototypical example of this class, namely the $\nu=2/3$ state~\cite{Johnson_Composite_edge_1991,Kane_Randomness_1994,Protopopov_transport_2_3_2017,Spanslatt2021}, characterized by $\nu_1=1$ and $\nu_2=-1/3$. Accordingly, the $K$-matrix and the charge vector $\vec{q}$ are given as
\begin{align}
\label{appeq:K23}
K =
\begin{pmatrix}
1 & 0\\
0 & -3
\end{pmatrix}\,, \quad \vec{q} = \begin{pmatrix}
    1 \\ 1
    \end{pmatrix}\,.
\end{align}
The corresponding Hamiltonian, including interactions, is
\begin{align}
\label{appeq:H25}
    H &=\pi \int dx \left( v_1\rho_{1}^2+3v_{2}\rho_{2}^2+2u(x)\rho_{1}\rho_{2}\right),
\end{align}
with associated density commutation relations
\begin{subequations}
\label{appeq:23_CCR}
\begin{align}
&[\rho_1(x),\rho_1(y)]=-\frac{i}{2\pi}\partial_x\delta(x-y),\\
    &[\rho_{2}(x),\rho_{2}(y)]=+\frac{1}{3}\frac{i}{2\pi}\partial_x\delta(x-y).
\end{align}
\end{subequations}
The signs in these equations correspond to opposite chiralities of the two modes, $\eta_1=-\eta_2=+1$ and the densities are given as $\rho_1=+\partial_x \phi_1/(2\pi)$ and $\rho_2=-\partial_x \phi_2/(2\pi)$.

To diagonalize the Hamiltonian~\eqref{appeq:H25} while preserving the edge modes' chiralities, we introduce the transformation matrix $\Lambda$,
\begin{align}
\label{appeq:Lambda_23}
   \Lambda=\begin{pmatrix}
        \cosh\gamma & -\sinh\gamma\\
        -\frac{\sinh\gamma}{\sqrt{3}} & \frac{\cosh\gamma}{\sqrt{3}} 
    \end{pmatrix}, \quad \tanh (2\gamma) = \frac{2}{\sqrt{3}}\frac{u}{v_1+v_2}.
\end{align}
This transformation rescales the $K$-matrix, preserving  its signature:
$\Lambda^{T} K \Lambda = \mathrm{diag}(1,-1)\equiv\sigma_z$. In the resulting eigenmode basis, the Hamiltonian takes the form
\begin{align}
\label{appeq:H_diag}
    H=\pi \int dx \left(v_+ \rho^2_++v_- \rho^2_-\right),
\end{align}
with eigenmodes
\begin{align} \label{appeq:eigenmode43}
 \begin{pmatrix}
        \rho_+\\
        \rho_-
    \end{pmatrix} =\Lambda^{-1}\begin{pmatrix}
        \rho_1\\
        \rho_{2}
    \end{pmatrix}.
\end{align}
The corresponding eigenmode velocities are 
\begin{align}
\label{appeq:vpm_23}
v_\pm
= v_{1,2}\cosh^{2}\gamma
+ v_{2,1}\sinh^{2}\gamma
- \frac{u}{\sqrt{3}}\sinh(2\gamma).
\end{align}
The stability of the $\nu=2/3$ edge is ensured if the condition $u^2 \leq 3 v_1 v_2$ is satisfied, such that $v_\pm \geq 0$. 

We study the $\nu=2/3$ edge driven out of equilibrium by injections of the quasiparticle excitations created by operators 
\begin{align} \label{appeq:excitations43}
 \psi_{n_{1,1}}^{\dagger} \sim e^{-i n_{1,1} \phi_1} \,, \qquad \psi_{n_{1,2}}^{\dagger} \sim e^{-i n_{1,2} \phi_2}\,.
\end{align}
The injections in the $\nu_1=1$ and $\nu_2=-1/3$ modes occur in region I and region III, respectively, see Fig.~\ref{fig:complex_setup}\textcolor{blue}{(b)} in the main text. 

The Keldysh non-equilibrium action for $\nu=2/3$ is identical in form to that of the $\nu=4/3$ edge (see Sec.~\ref{sec:Co_prop} of the main text),
\begin{align}
\label{appeq:action_43}
    e^{iS[\rho,\overline{\rho}]} &= e^{- i \rho_1\, \Pi_{{1,a}}^{-1} \overline{\rho}_1} \Delta [\Pi_{1,a}^{-1}\overline{\rho}_1]^{1/(n_{1,1})^2}\times e^{- i 3 \rho_{2}\, \Pi_{2,a}^{-1} \overline{\rho}_{2}} \Delta [3\Pi_{2,a}^{-1}\overline{\rho}_{2}]^{3/(n_{1,2})^2} \times e^{-i\int dt dx 2\pi u(x)\left( \overline{\rho}_1\rho_{2} + \overline{\rho}_{2}\rho_1  \right)}.
\end{align}
However, the counterpropagating character of the modes of the $\nu=2/3$ edge leads to very essential differences. 

\subsection{Full counting statistics}
\label{sec:FCS_counter_SM}

The total charge $Q$ to the left of a point $x_0$, and the FCS generating function, $\kappa(\lambda,x_0,\tau)$, are given by
\begin{align}
\label{appeq:Q_43}
    Q(x_0,t) =e \int_{-\infty}^{x_0}dx\left[\rho_1(x,t) + \rho_{2}(x,t) \right]\,, 
\end{align}
and 
\begin{align}
\label{appeq:FCS_25}
    \kappa(\lambda,x_0,\tau) =\left\{\overline{\Delta}[\delta_{1, \tau}(t)]\right\}^{1/(n_{1,1})^2}\left\{\overline{\Delta}[\delta_{2,\tau}(t)]\right\}^{3/(n_{1,2})^2}\,.
\end{align}
Our goal in the following is to compute the counting phases entering the two determinants in the generating function~Eq.~\eqref{appeq:FCS_25}. 
To this end, we proceed in the same way as described in the main text for a Laughlin edge and for and edge with co-propagating modes. We integrate out the classical density fields in the action and obtain the equations for the quantum components. These equations take the form 
\begin{align}
\label{appeq:eom25_FCS}
&K \partial_t  \begin{pmatrix}
        \overline{\rho}_1 \\
        \overline{\rho}_{2} 
    \end{pmatrix} + \partial_x \begin{pmatrix}
        v_1 & u\\
        u & 3v_{2}
    \end{pmatrix}\begin{pmatrix}
        \overline{\rho}_1 \\
        \overline{\rho}_{2} 
    \end{pmatrix} = -\frac{\lambda }{2\pi}j(x-x_0,t) \vec{q}\,,
\end{align}
with the $K$-matrix from Eq.~\eqref{appeq:K23}. The source term is given by
 \begin{align}
\label{appeq:source_j}
    j(x,t) \equiv  \frac{1}{\sqrt{2}}\delta(x)\left(\delta(t-\tau)-\delta(t)\right).
\end{align}
 For the $\nu=2/3$ edge, the diagonalization of these equations proceeds with the rotation matrix $\Lambda$ given by Eq.~\eqref{appeq:Lambda_23}, resulting in the following equation (in the frequency representation):
\begin{align}
\label{appeq:eoms_23_diag_SM}
    &-i\omega\sigma_z  \begin{pmatrix}
        \overline{\rho}_+ \\
        \overline{\rho}_- 
    \end{pmatrix} + \partial_x \begin{pmatrix}
         v_+ && 0\\
        0 && v_-
    \end{pmatrix} \begin{pmatrix}
        \overline{\rho}_+ \\
        \overline{\rho}_- 
    \end{pmatrix}   =-\frac{\lambda }{2\pi}j(x-x_0,\omega)\begin{pmatrix}
      q_+ \\ q_-
    \end{pmatrix}.
\end{align}
Here, the frequency representation of the source term $j(x,\omega)$ is given as
\begin{align}
\label{appeq:j_x_omega}
     j(x,\omega) = \frac{1}{\sqrt{2}}\delta(x)\left(e^{i\omega \tau}-1\right),
\end{align}
and the eigenmode charges $q_\pm$ for the $\nu=2/3$ edge are
\begin{align}
\label{appeq:eigencharges_23_SM}
    \begin{pmatrix}
        q_+\\
        q_-
    \end{pmatrix} = \Lambda^T \vec{q} = \begin{pmatrix}
       \cosh \gamma-\frac{\sinh \gamma}{\sqrt{3}} \\ \frac{\cosh \gamma}{\sqrt{3}}-\sinh \gamma
    \end{pmatrix},
\end{align}
where we used $\Lambda$ from Eq.~\eqref{appeq:Lambda_23} and $\vec{q}$ from Eq.~\eqref{appeq:KMat2x2}. 
Similarly to the co-propagating edge, we need to find the \textit{advanced} solution of Eq.~\eqref{appeq:eoms_23_diag_SM}. However, compared to the co-propagating edge, the edge with counter-propagating modes hosts incoming density fields both in region I and in region III, see Fig.~\ref{fig:complex_setup}\textcolor{blue}{(b)} in the main text. The outgoing fields in these regions are zero due to the advanced nature of our sought solution. 

Focusing on the charge measurement point $x_0$ located in region II, i.e., $|x_0|<L/2$, the calculation of the generating function proceeds similarly to the calculation for the co-propagating edge in Sec.~\ref{sec:FCS_co} of the main text. In the non-interacting regions I and III, where $u(x)=0$ and there is no source, we obtain the following chiral, plane-wave solutions
\begin{subequations}
\label{appeq:planewave23_sol}
\begin{align}
\bar\rho_{1}(x,\omega)
&=
\begin{cases}
A^{\text{I}}_{1}\,e^{ i\omega x / v_{1}}, & \qquad  x<-\tfrac{L}{2},\\
0, & \qquad x>\tfrac{L}{2},
\end{cases}
\label{eq:planewave23a}
\\[4pt]
\bar\rho_{2}(x,\omega)
&=
\begin{cases}
0, & \qquad x<-\tfrac{L}{2},\\
A^{\text{III}}_{2}\,e^{ -i\omega x / v_{2}}, & \qquad  x>\tfrac{L}{2},
\end{cases}
\label{eq:planewave23b}
\end{align}
\end{subequations}
where $A^{\text{I,III}}_{1/2}$ are constants. In the interacting region II containing the source, we find the solutions
\begin{align}
\label{eq:planewave23_sol_source}
\bar\rho_{\pm}(x,\omega)
&=
e^{\pm i\omega x / v_{\pm}}
\big(
A^{\text{II}}_{\pm}
- \mathcal{J}_{\pm} \big),
\quad |x|<\tfrac{L}{2}\,,
\end{align}
where $A_{\pm}^{\text{II}}$ are constants and $\mathcal{J}_{\pm}$ take the form 
\begin{align}
\label{appeq:extended_source}
    \mathcal{J}_{\pm} = \Theta(x-x_0)\,
\frac{\lambda q_{\pm}}{2\pi\sqrt{2}\,v_{\pm}}\,
e^{- i\omega \eta_{\pm} x_0 / v_{\pm}}\,
\big(e^{i\omega \tau}-1\big)\,,
\end{align}
with $\eta_\pm =\pm 1$.

The solutions~\eqref{appeq:planewave23_sol}-\eqref{eq:planewave23_sol_source} are matched by the scattering matrices
\begin{align}
\label{appeq:scattering_matrices}
    S_L \equiv \begin{pmatrix}
        t_L &-r_L \\
        r_L & t_L
    \end{pmatrix}, \quad S_R \equiv \begin{pmatrix}
        t_R &r_R \\
        -r_R & t_R
    \end{pmatrix},
\end{align}
connecting incoming and outgoing modes at each interface according to 
\begin{subequations} \label{eq:scatteringmatrix_counter}
\begin{align} 
\begin{pmatrix}
v_+ \overline{\rho}_+(\omega)\\
\sqrt{3}\,v_2\overline{\rho}_2(\omega)
\end{pmatrix}
&=
S_L
\begin{pmatrix}
v_1 \overline{\rho}_1(\omega)\\
v_- \overline{\rho}_-(\omega)
\end{pmatrix},
\label{eq:SL_bc_counter}
\\
\begin{pmatrix}
v_1\overline{\rho}_1(\omega)\\
v_- \overline{\rho}_-(\omega)
\end{pmatrix}
&=
S_R
\begin{pmatrix}
v_+ \overline{\rho}_+(\omega)\\
v_2 \sqrt{3}\,\overline{\rho}_2(\omega)
\end{pmatrix}.
\label{eq:SR_bc_counter}
\end{align}
\end{subequations}
Solving the scattering problem for $A^{\text{I}}_1$ and $A^{\text{III}}_2$, we find  the incoming solutions 
\begin{subequations}
\label{eq:rho_sol_23_omega}
\begin{align}
    \overline{\rho}_1^{}(x,\omega) &= \frac{\lambda t_L}{2\pi\sqrt{2}v_1} e^{i\frac{\omega}{v_1}(\frac{L}{2}+x)}  (e^{i\omega \tau}-1) \times \frac{(q_+ e^{-i\frac{\omega}{v_+}(\frac{L}{2}+x_0)} +q_-r_R e^{i\frac{\omega}{v_-}(\frac{L}{2}+x_0)} e^{-2iL \omega/\bar{v}} )}{1-r_R r_Le^{-2i\omega L/\bar{v}}}, \quad \text{for } x<-\frac{L}{2},\\
    \overline{\rho}^{}_2(x,\omega) &= -\frac{\lambda t_R}{2\pi\sqrt{6}v_2} e^{-i\frac{\omega}{v_2}(x-\frac{L}{2})} (e^{i\omega \tau}-1)\times \frac{(q_- e^{-i\frac{\omega}{v_-}(\frac{L}{2}-x_0)} +q_+ r_L e^{i\frac{\omega}{v_+}(\frac{L}{2}-x_0)} e^{-2iL \omega/\bar{v}} )}{1-r_R r_Le^{-2i\omega L/\bar{v}}}, \quad \text{for }x>\frac{L}{2},  
\end{align}
\end{subequations}
where we defined the average velocity 
\begin{align}
\label{eq:average_v}
    \overline{v}\equiv 2(v_+^{-1}+v_-^{-1})^{-1}
\end{align}
of the eigenmodes during a complete round trip in the interacting region II.

In the time representation, Eqs.~\eqref{eq:rho_sol_23_omega} take the form of infinite sums of sharp density spikes 
\begin{subequations}
\label{eq:densities_23_sol_time_FCS}
\begin{align}
   &\overline{\rho}_1(x,t)=-\frac{\lambda t_L}{2\pi\sqrt{2}v_1} \sum_{m=0}^\infty (r_R r_L)^{m} \Big [q_+ \Upsilon_\tau(t,\frac{x-x_0}{v_1}+t_{1,+,m}) + q_- r_R \Upsilon_\tau(t,\frac{x-x_0}{v_1} +t_{1,-,m}) \Big ],\\
  &\overline{\rho}_2(x,t)=\frac{\lambda t_R}{2\pi\sqrt{6}v_2} \sum_{m=0}^\infty (r_R r_L)^m \Big[ q_- \Upsilon_\tau(t, - \frac{x-x_0}{v_2}+t_{2,-,m}) + q_+ r_L \Upsilon_\tau(t,- \frac{x-x_0}{v_2} +t_{2,+,m}) \Big ],
\end{align}
\end{subequations}
where we defined the function $\Upsilon_\tau(t_1,t_2)$,
\begin{align}
    \Upsilon_\tau(t_1,t_2) = \delta(t_1-t_2) -\delta(t_1 - t_2 -\tau)\,,
\end{align}
and introduced the time offsets
\begin{subequations}
\label{eq:23_time_offsets}
\begin{align}
    t_{1,+,m}&\equiv \Big(x_0 + \frac{L}{2} \Big) \Big(\frac{1}{v_1} - \frac{1}{v_+} \Big) -\frac{2m L}{\bar{v}}\,, \\ 
    t_{1,-,m}&\equiv \Big(x_0 + \frac{L}{2} \Big) \Big(\frac{1}{v_1} + \frac{1}{v_-}\Big)-\frac{2(m+1)L}{\bar{v}}\,, \\ t_{2,-,m} &\equiv \Big( \frac{L}{2} - x_0 \Big) \Big(\frac{1}{v_2} - \frac{1}{v_-}\Big) -\frac{2mL}{\bar{v}},\\ 
    t_{2,+,m} &\equiv\Big( \frac{L}{2} - x_0 \Big) \Big(\frac{1}{v_2} + \frac{1}{v_+}\Big) -\frac{2(m+1)L}{\bar{v}}\,. 
\end{align}
\end{subequations}
Finally, to obtain the counting phases entering the generating function, Eq.~\eqref{appeq:FCS_25}, 
we integrate the density solutions~\eqref{eq:densities_23_sol_time_FCS} according to
\begin{subequations}
\label{eq:counting_phases_gen_23}
\begin{align}
    \delta_{1,\tau}(t) &= 2\pi\sqrt{2}\, n_{1,1}
    \lim_{t'\to -\infty}\int^{v_1(t'+t)}_{x_0} dx'\,\overline{\rho}_{1}(x',t')\,, \\
    \delta_{2,\tau}(t) &= -2\pi\sqrt{2}\, n_{1,2}
    \lim_{t'\to -\infty}\int^{- v_2(t'+t)}_{x_0} dx'\,\overline{\rho}_{2}(x',t')\,.
\end{align}
\end{subequations}
Performing the integrations, 
we find that the phases take the form of trains of rectangular pulses,
\begin{subequations}
\label{eq:phases_23_FCS_kid_SM}
\begin{align}
    &\delta_{1,\tau}(t) = \sum_{s=\pm}\sum_{m=0} ^\infty  \delta_{1,s,m}w_{\tau}(t,-t_{1,s,m})  , \\
    &\delta_{2,\tau}(t) =\sum_{s=\pm} \sum_{m=0}^\infty \delta_{2,s,m}w_{\tau}(t,-t_{2,s,m}).
\end{align}
\end{subequations}
Here, $w_\tau(t_1,t_2)$ is given by 
\begin{align}   \label{appeq:window_function}
        w_\tau(t_1,t_2) \equiv \Theta(t_2-t_1)-\Theta(t_2-t_1-\tau)\,,
    \end{align}
the time offsets $t_{1,\pm, m}$ and $t_{2,\pm, m}$ are given by Eq.~\eqref{eq:23_time_offsets}, and the phase amplitudes read
\begin{subequations}
\label{eq:phases_FCS_23_kiddle}
\begin{align}
    &\delta_{1,+,m} =\lambda q_+ n_{1,1}t_Lr_L^mr^m_R   ,\\
    &\delta_{1,-,m} = \lambda q_- n_{1,1} t_Lr_L^mr^{m+1}_R ,\\
    &\delta_{2,-,m} = - \frac{\lambda q_-}{\sqrt{3}} n_{1,2} t_R r_L^mr^m_R   ,\\
    &\delta_{2,+,m} = - \frac{\lambda q_+}{\sqrt{3}} n_{1,2} t_R r_L^{m+1}r^{m}_R\,. 
\end{align}
\end{subequations}

To better understand the fractionalized physics underlying these results, we define the counting pulses
\begin{subequations} \label{eq:totalcharge_xandt_23}
\begin{align}
    q_{1} (x,t) &= e\frac{2\pi \sqrt{2}}{\lambda} \int_{-\infty}^{x} dx' \overline{\rho}_1(x',t)\,, \\
    q_{2} (x,t) &= e\frac{2\pi \sqrt{2}}{\lambda} \int_{x}^{\infty} dx' \overline{\rho}_2(x',t)\,. 
\end{align}
\end{subequations}
These counting pulses, viewed as functions of $t$, describe the dynamics of the fractionalized charge pulses induced by inter-mode interactions. The counting pulses of each mode take the following form in the corresponding non-interacting region (I and III, respectively): 
\begin{align}
    q_1(x,t) &= \sum_{s=\pm}\sum_{m=0}^{\infty} q_{1,s,m}^{\text{(p)}} w_{\tau} (t, -t_{1,s,m})\,, \quad \text{for }x<-\frac{L}{2}\,,\\
    q_2(x,t) &= \sum_{s=\pm}\sum_{m=0}^{\infty} q_{2,s,m}^{\text{(p)}} w_{\tau} (t, -t_{2,s,m})\,, \quad \text{for }x>\frac{L}{2}\,,
\end{align}
with amplitudes $q_{1,\pm, m}^{\text{(p)}}$ and $q_{2,\pm, m}^{\text{(p)}}$  related to the counting phases~\eqref{eq:phases_FCS_23_kiddle}, 
\begin{align} \label{eq:phases_FCS_23_relation}
   q_{1,s,m}^{\text{(p)}}  = \frac{e\delta_{1,s,m}}{\lambda n_{1,1}}\,, \quad  q_{2,s,m}^{\text{(p)}} = \frac{e\delta_{2,s,m}}{\lambda n_{1,2}}. 
\end{align}

For the counter-propagating edge, the time-reversed pulse dynamics is qualitatively different from that of the co-propagating edge (illustrated in Fig.~\ref{fig:pulsedynamics}\textcolor{blue}{(a)} in the main text).
For the counter-propagating edge, pulses injected in each mode at $x=x_0$, corresponding to charges $e$ and $e/3$, fractionalize into the two eigenmodes, $+$ and $-$, and propagate in opposite directions. Upon reaching the I-II and II-III interfaces, these fractionalized pulses are partially transmitted or reflected, leading to further fractionalization. This process is repeated multiple times as the pulses undergo successive reflections and transmissions, 
until they eventually arrive in the mode $\nu_1=1$ in region I or the mode $\nu_2=1/3$ in region III. As a result, infinite sets of pulses with different amplitudes arrive in regions I and III. In mode $\nu_1=1$, the amplitudes of the pulses arriving region I, resulting from the initial fractionalization into the $+$ and $-$ eigenmodes at $x=x_0$, are $q_{1,+,m}^{\text{(p)}}$ and $q_{1,-,m}^{\text{(p)}}$. Similarly, in the mode $\nu_2=-1/3$, the pulses with amplitudes $q_{2,+,m}^{\text{(p)}}$ and $q_{2,-,m}^{\text{(p)}}$ arrive in region III.

The scattering amplitudes entering the counting phases depend on the sharpness of the spatial variation of the inter-mode interaction. For counter-propagating edges, a sharp variation leads to the scattering amplitudes
\begin{align}
\label{eq:t_r_counter_sharp_SM}
t_R=t_L = \frac{1}{\cosh\gamma},\qquad
r_R=r_L = \tanh\gamma,
\end{align}
which follow from the generic scattering matrices~\eqref{appeq:scattering_matrices} 
combined with the counter-propagating rotation matrix~\eqref{appeq:Lambda_23}.
In the opposite, adiabatic limit of a smooth interaction profile, the interfaces are perfectly transmitting, with scattering amplitudes given by
\begin{align} \label{appeq:interfacetransmissionamplitudes_adaibatic}
t_L = t_R = 1\,, \quad r_L = r_R = 0\,.
\end{align}
In the following, we analyze the FCS separately for sharp and adiabatic interfaces.

\subsubsection{Sharp interfaces}

In the long-time (short-length) limit, $|\tau| \gg L/\bar{v}$ [with $\bar{v}$ in Eq.~\eqref{eq:average_v}], the pulses~\eqref{eq:phases_23_FCS_kid_SM} in each mode overlap  strongly. Under the sharp-interface conditions in Eqs.~\eqref{eq:t_r_counter_sharp_SM}, the counting phases~\eqref{eq:phases_FCS_23_kiddle} simplify to 
\begin{subequations}
\label{eq:counting_phases_sum_rule}
\begin{align}
    \delta_{1,\tau}(t)&\simeq w_\tau(t,0)\sum_{s=\pm}\sum_{m=0}^{\infty}\delta_{1,s,m} = +\lambda n_{1,1} w_\tau(t,0) ,\\
    \delta_{2,\tau}(t)&\simeq w_\tau(t,0)\sum_{s=\pm}\sum_{m=0}^{\infty}\delta_{2,s,m} = -\frac{\lambda}{3} n_{1,2}w_\tau(t,0)\,,
\end{align}
\end{subequations}
where the counting phases and hence the FCS become insensitive to the inter-channel interaction. Physically, the short-length limit does not resolve the interaction-induced charge  fractionalization. 

In the opposite, short-time (long-length) limit, $|\tau| \ll L/\overline{v}$, the coherence between the pulses can be neglected and each determinant in the generating function, Eq.~\eqref{appeq:FCS_25} 
splits into an infinite product of determinants~\cite{Gutman2010, Protopopov2013correlations}:
\begin{subequations}
\label{eq:det_split_23_FCS}
\begin{align}
    \overline{\Delta}[\delta_{1,\tau}(t)] & \simeq \prod_{s=\pm}\prod_{m=0}^\infty \overline{\Delta}[\delta_{1,s,m}w_\tau(t,0)], \\
    \overline{\Delta}[\delta_{2,\tau}(t)] & \simeq \prod_{s=\pm}\prod_{m=0}^\infty \overline{\Delta}[\delta_{2,s,m}w_\tau(t,0)],
\end{align}
\end{subequations}
with phases given by  Eq.~\eqref{eq:phases_FCS_23_kiddle}.  The corresponding generating function $\kappa(\lambda,x_0,\tau)$ is therefore that of an infinite product of independent pulse contributions,
\begin{align}
\label{eq:kappa_23_sharp_1}
    \kappa(\lambda,x_0,\tau) \simeq \prod_{s=\pm}\prod_{m=0}^\infty \kappa_{1,s,m}(\lambda,x_0,\tau)\kappa_{2,s,m}(\lambda,x_0,\tau).
\end{align}
For the dilute quasiparticle injections described by the distribution functions
\begin{eqnarray}
f_1(\epsilon) &=& \mathcal{T}_1 \Theta(-\epsilon+e_{1,1}^* V_1) + (1-\mathcal{T}_1)\Theta(-\epsilon), \nonumber \\
f_2(\epsilon) &=& \mathcal{T}_2 \Theta(-\epsilon+e_{1,2}^* V_2) + (1-\mathcal{T}_2)\Theta(-\epsilon), 
\label{appeq:doublestep_43}
\end{eqnarray}
 
the Szeg\H{o} approximation yields for each of the factors
\begin{subequations}
\label{eq:kappa_23_sharp_2}
\begin{align}
 \kappa_{1,s,m}(\lambda,x_0,\tau)
    &\simeq
    \exp\!\left[
    - \frac{e V_1 \mathcal{T}_1\tau}{2\pi n_{1,1}}
    \left(1- e^{- i \delta_{1,s,m}}\right)
    \right]\,, \\ 
    \kappa_{2,s,m}(\lambda,x_0,\tau)
    &\simeq
    \exp\!\left[
    - \frac{e V_2 \mathcal{T}_2\tau}{2\pi n_{1,2}}
    \left(1- e^{- i \delta_{2,s,m}}\right)
    \right]\,, 
\end{align}
\end{subequations}
with $\mathcal{T}_1,\mathcal{T}_2 \ll 1$. The generating function~\eqref{eq:kappa_23_sharp_1} produces the following results for the cumulants of charge fluctuations,
\begin{align} \label{eq:23chargefluctuations_sharp}
     \langle\langle(\delta Q)^k \rangle\rangle = \sum_{s=\pm}\sum_{m=0}^\infty 
     \Big[&(n_{1,1}q^{(p)}_{1,s,m})^{k-1}  \langle Q_{1,s, m}\rangle  + (n_{1,2}q^{(p)}_{2,s,m})^{k-1}  \langle Q_{2,s, m}\rangle \Big],
\end{align}
with the individual pulse charges
\begin{subequations}
\label{eq:23_individual_pulses}
\begin{align}
    &q^{(p)}_{1,+,m} =e q_+\frac{\tanh(\gamma)^{2m}}{\cosh(\gamma)},\\
    &q^{(p)}_{1,-,m} =eq_- \frac{\tanh(\gamma)^{2m+1}}{\cosh(\gamma)},\\
    &q^{(p)}_{2,-,m} =  -\frac{eq_-}{\sqrt{3}} \frac{\tanh(\gamma)^{2m}}{\cosh(\gamma)},\\
    &q^{(p)}_{2,+,m} = -\frac{eq_+}{\sqrt{3}}\frac{\tanh(\gamma)^{2m+1}}{\cosh(\gamma)},
\end{align}
\end{subequations}
and the average charges arriving at $x_0$ during $\tau$,
\begin{subequations}
\label{eq:phases_FCS_23_average_charges_SM}
\begin{align}
    &\langle Q_{1,+,m} \rangle = +q_+ \frac{\tanh(\gamma)^{2m}}{\cosh(\gamma)} \frac{e^2 \mathcal{T}_1 V_1 \tau}{2\pi},\\
    &\langle Q_{1,-,m} \rangle =+ q_-  \frac{\tanh(\gamma)^{2m+1}}{\cosh(\gamma)} \frac{e^2 \mathcal{T}_1 V_1 \tau}{2\pi},\\
    &\langle Q_{2,-,m} \rangle = -q_- \frac{\tanh(\gamma)^{2m}}{\cosh(\gamma)} \frac{e^2 \mathcal{T}_2 V_2 \tau}{2\pi \sqrt{3}} ,\\
    &\langle Q_{2,+,m} \rangle = - q_+ \frac{\tanh(\gamma)^{2m+1}}{\cosh(\gamma)} \frac{e^2 \mathcal{T}_2 V_2 \tau}{2\pi \sqrt{3}} .
\end{align}
\end{subequations}
Equation~\eqref{eq:23chargefluctuations_sharp} is the FCS of a superposition of independent Poissonian processes characterized by fractional charges $n_{1,1}q^{(p)}_{1,\pm,m}$ and $n_{1,2}q^{(p)}_{2,\pm,m}$. The physical picture behind this result is as follows. When the mode $\nu_1=1$ in region I is driven out of equilibrium, it hosts a dilute train of quasiparticles with charge $n_{1,1}e$. These quasiparticles experience multiple scattering events at the two interfaces, each time with an associated fractionalization into transmitted and reflected components. This process leads to an infinite series of excitations in the interacting region II detected by the FCS. The charges  $n_{1,1}q^{(p)}_{1,+,m}$ correspond to the excitations that have experienced an even number of reflections and thus move from left to right. Likewise, the charges $n_{1,1}q^{(p)}_{1,-,m}$ in the FCS
correspond to excitations that have experienced an odd number of reflections. Thus, they move from right to left and their actual charge is the opposite, i.e.,
$- n_{1,1}q^{(p)}_{1,-,m}$.
In the same way, the charges $n_{1,2}q^{(p)}_{2,\pm,m}$ in FCS correspond to multiple fractionalization of quasiparticles $n_{1,2}e/3$ from a non-equilibrium mode $\nu_2=-1/3$ in region III. Specifically, these fractionalization processes generate, in the interacting region, quasiparticles with charges 
$- n_{1,2}q^{(p)}_{2,-,m}$ moving from right to left and quasiparticles with charges  $n_{1,2}q^{(p)}_{2,+,m}$ moving from left to right.

In each of the edge modes, the charges carried by the individual pulses add up to the total injected charge,
\begin{subequations} \label{eq:totalinjectedcharge_23_SM}
\begin{align}
    \sum_{s=\pm}\sum_{m=0}^\infty q^{(p)}_{1,s,m}=  e\,, \quad 
    \sum_{s=\pm}\sum_{m=0}^\infty q^{(p)}_{2,s,m}= - \frac{e}{3}\,,
\end{align}
\end{subequations}
which is a manifestation of charge conservation.

\subsubsection{Adiabatic interfaces}

The analysis of adiabatic interfaces for $\nu=2/3$ follows closely the same steps as for $\nu=4/3$, see Sec.~\ref{sec:adiabatic_interfaces_43} in the main text. While the high-frequency components of the pulses incoming from regions I and III are perfectly transmitted into region II, i.e., with the adiabatic amplitudes, ~\eqref{appeq:interfacetransmissionamplitudes_adaibatic}, 
there are inevitably low-frequency components for which the interfaces appear sharp. For these components, the amplitudes are given in Eq.~\eqref{eq:t_r_counter_sharp_SM}. Taking into account both high- and low-frequency sectors, each determinant in the generating function~\eqref{appeq:FCS_25} 
splits into high- and low-frequency determinants according to 
\begin{subequations}
\label{appeq:FCS_43_long_time_int_region_adiabatic}
\begin{align}
\overline{\Delta}[\delta_{1,\tau} (t)] \simeq \overline{\Delta}_h[\delta_{1,\tau} (t)] \overline{\Delta}_l[\delta_{1,\tau} (t)]\,, \\ 
\overline{\Delta}[\delta_{2,\tau} (t)] \simeq \overline{\Delta}_h[\delta_{2,\tau} (t)] \overline{\Delta}_l[\delta_{2,\tau} (t)]\,.
\end{align}
\end{subequations}

For the $\nu=2/3$ edge, the two high-frequency determinants take the form 
\begin{subequations}
\label{appeq:HF_43}
\begin{align}
   \overline{\Delta}_h[\delta_{1,\tau} (t)] \simeq  \overline{\Delta} [n_{1,1}q_{1,h}^{\text{(p)}}\lambda w_\tau(t,0)]\,, \\
    \overline{\Delta}_h[\delta_{2,\tau} (t)] \simeq   \overline{\Delta} [n_{1,2}q_{2,h}^{\text{(p)}}\lambda w_\tau(t,0)]\,.
\end{align}
\end{subequations}
with the charges
\begin{subequations}
\label{eq:adiabatic_charges_23_SM}
\begin{align}
  q_{1,h}^{\text{(p)}} &=q_+ e \,, \\
  q_{2,h}^{\text{(p)}} &=-\frac{q_- e}{\sqrt{3}}\,.
\end{align}
\end{subequations}
At high frequencies, the counter-propagating eigenmodes $+$ and $-$ are thus smoothly connected to the $\nu_1=1$ mode in region I and the $\nu_2=-1/3$ mode in region III, respectively.

The low-frequency determinants take the form
\begin{subequations}
\label{eq:det_split_23_FCS_low_adiabatic}
\begin{align}
    \overline{\Delta}_l[\delta_{1,\tau}(t)] & \simeq \prod_{s=\pm}\prod_{m=0}^\infty \overline{\Delta}[q^{(p)}_{1,s,m}w_{\Delta x/\bar{v}}(t,0)], \\ \overline{\Delta}_l[\delta_{2,\tau}(t)] & \simeq \prod_{s=\pm}\prod_{m=0}^\infty \overline{\Delta}[q^{(p)}_{2,s,m}w_{\Delta x/\bar{v}}(t,0)], 
\end{align}
\end{subequations}
with $q^{(p)}_{1,s,m}$ and $q^{(p)}_{1,s,m}$ from Eq.~\eqref{eq:23_individual_pulses}. 

For the non-equilibrium situation described by the distribution functions~\eqref{appeq:doublestep_43},
the generating function factorizes into  high- ($h$) and low-frequency ($l$) components according to
\begin{align}
\label{eq:kappa_23_adiabatic_1}
    \kappa(\lambda,x_0,\tau)
    &\simeq \kappa_{1, h}(\lambda,x_0,\tau) \kappa_{2, h}(\lambda,x_0,\tau) \prod_{s=\pm}\prod_{m=0}^\infty
    \kappa_{1,s,m,l}(\lambda,x_0,\tau)  \kappa_{2,s,m,l}(\lambda,x_0,\tau)\,.
\end{align}
Here, the $h$ factors are produced from the counting pulses with the amplitudes~\eqref{eq:adiabatic_charges_23_SM}, 
\begin{subequations}\label{eq:kappa_23_adiabatic_2_high}
\begin{align}
 \kappa_{1,h}(\lambda,x_0,\tau)
    &=
    \exp\!\left[
    - \frac{e V_1 \mathcal{T}_1\tau}{2\pi n_{1,1}}
    \left(1- e^{- i \lambda n_{1,1} q_{1,h}^{\text{(p)}}}\right)
    \right]\,, \\ 
    \kappa_{2, h}(\lambda,x_0,\tau)
    &=
    \exp\!\left[
    - \frac{e V_2 \mathcal{T}_2\tau}{2\pi n_{1,2}}
    \left(1- e^{- i \lambda n_{1,2} q_{2,h}^{\text{(p)}}}\right)
    \right]\,, 
\end{align}
\end{subequations}
while the individual $l$ factors read  
\begin{subequations}
\label{eq:kappa_23_adiabatic_2_low}
\begin{align}
 \kappa_{1,s,m,l}(\lambda,x_0,\tau)
    &=
    \exp\!\left[-
    \frac{ie^2 V_1 \mathcal{T}_1\lambda\tau}{2\pi}
    (q^{(p)}_{1,s,m}-q_+) \right]\,, \\ 
    \kappa_{2,s,m,l}(\lambda,x_0,\tau)
    &=
   \exp\!\left[-
    \frac{ie^2 V_2\mathcal{T}_2\lambda\tau}{2\pi} \Big(  q^{(p)}_{1,s,m}+\frac{q_-}{\sqrt{3}}\Big)
    \right]\,,
\end{align}
\end{subequations}
with charge pulse amplitudes from Eq.~\eqref{eq:23_individual_pulses}. The low-frequency contribution for the $\nu=2/3$ edge is thus a superposition of an infinite sum of fractionalized pulses, with small amplitudes and broadly spread in time. 

Using the conservation rules~\eqref{eq:totalinjectedcharge_23_SM}, we find the total low-frequency contributions
\begin{subequations}
\label{eq:kappa_23_adiabatic_2_low_tot}
\begin{align}
 \prod_{s=\pm}\prod_{m=0}^\infty\kappa_{1,s,m,l}(\lambda,x_0,\tau)
    &=
    \exp\!\left[-
    \frac{ie^2 V_1 \mathcal{T}_1\lambda\tau}{2\pi}
    (1-q_+) \right]\,, \\ 
    \prod_{s=\pm}\prod_{m=0}^\infty\kappa_{2,s,m,l}(\lambda,x_0,\tau)
    &=
   \exp\!\left[
    \frac{ie^2 V_2\mathcal{T}_2\lambda\tau}{2\pi} \Big(  \frac{1}{3}-\frac{q_-}{\sqrt{3}}\Big)
    \right]\,.
\end{align}
\end{subequations}

From Eqs.~\eqref{eq:kappa_23_adiabatic_1}-\eqref{eq:kappa_23_adiabatic_2_low_tot}, we obtain all cumulants of charge fluctuations. 
The total average charge (first cumulant, $k=1$) passing across the point $x_0$ during the time interval $\tau$ is found as
\begin{align}
\label{eq:average_charge_23_adiabatic_SM}
\langle Q \rangle &=  \langle  Q_1\rangle + \langle Q_2\rangle  \,, \nonumber \\ \text{with }\langle Q_{1}\rangle  &= \frac{\mathcal{T}_1 e^2 V_1 \tau}{2\pi}\,, \quad 
    \langle Q_{2}\rangle  = -\frac{1}{3}\frac{\mathcal{T}_2 e^2 V_2 \tau}{2\pi}\,.
\end{align}
It is independent of interactions and obeys charge conservation (being equal to the average injected charge during the same time interval) as expected. Similarly to the co-propagating edge, the low-frequency components contribute only to the average charge and are also essential to ensure charge conservation. The higher cumulants ($k>1$) evaluate to
\begin{align} \label{eq:23chargefluctuations_SM}
     \langle\langle(\delta Q)^k \rangle\rangle &= 
    (n_{1,1} q_+e)^{k-1}  \langle Q_{+}\rangle +  (n_{1,2} \frac{q_-}{\sqrt{3}} e)^{k-1}  \langle Q_{-}\rangle \nonumber \\ \text{with }\langle Q_{+}\rangle  &= \frac{q_+\mathcal{T}_1 e^2 V_1 \tau}{2\pi}\,, \quad 
    \langle Q_{-}\rangle  = -\frac{q_-\mathcal{T}_2 e^2 V_2 \tau}{2\pi \sqrt{3}}\,.
\end{align}
Equations~\eqref{eq:average_charge_23_adiabatic_SM}-\eqref{eq:23chargefluctuations_SM} admit an interpretation analogous to the FCS of the co-propagating edge with adiabatic interfaces.
However, in the present case, the $\nu_2=-1/3$ mode is injected from region III (i.e., propagating from right to left) and therefore contributes to the FCS with a sign opposite to that for the $\nu_1=1$ mode. 

For vanishing interactions, $\gamma=0$, we have $eq_+n_{1,1}=en_{1,1}$ and $eq_-n_{1,2}\sqrt{3}=en_{1,2}/3$, as expected.

The $\nu=2/3$ edge further admits a charge-neutral decoupling at the Kane-Fisher-Polchinski (KFP)  point~\cite{Kane_Randomness_1994}, for which the interactions take the special value $\gamma=\gamma_{\rm KFP}$, with 
\begin{align}
\label{eq:gamma_KFP}
    \gamma_{\rm KFP}\equiv \frac{1}{2}\text{tanh}^{-1}\left(\frac{\sqrt{3}}{2}\right)\approx 0.66.
\end{align}
At the KFP point, the eigenmode charges [see Eq.~\eqref{appeq:eigencharges_23_SM}] are given by $q_+ = \sqrt{2/3}$ and $q_-=0$, with only the charge mode contributing to the FCS.

\subsection{Green's functions}
\label{sec:GFs_counter_SM}

Here, we compute, for the non-equilibrium $\nu=2/3$ edge, the GFs of quasiparticle excitations described by the vertex operators
\begin{align}\label{appeq:vertex_43}
   \psi^{\dagger}_{\vec{n}_2} \sim e^{-i \vec{n}_2 \cdot \vec{\phi}} = \exp [-i (n_{2,1} \phi_1 + n_{2,2}\phi_2)]\,. 
\end{align}
The greater GF retains the integral representation
\begin{align}
\label{appeq:GF_43}
    \mathcal{G}^>(\tau) &\equiv \frac{-i}{2\pi a}  \int \mathcal{D}\rho\mathcal{D}\overline{\rho}  e^{iS[\rho,\overline{\rho}]} \notag \\ &\times e^{\frac{i n_{2,1}}{\sqrt{2}} [\phi_{1}(0, \tau) - \phi_{1}(0,0) - \overline{\phi}_{1} (0, \tau) - \overline{\phi}_{1} (0,0) ]} \notag \\ &\times e^{\frac{i n_{2,2}}{\sqrt{2}} [\phi_{2}(0, \tau) - \phi_{2}(0,0) - \overline{\phi}_{2} (0, \tau) - \overline{\phi}_{2} (0,0) ]}\,. 
   \end{align}
Integrating out the classical fields,  we obtain the following coupled equations for the quantum components:
 \begin{align}
K\partial_t \begin{pmatrix}
        \overline{\rho}_1 \\
        \overline{\rho}_2 
    \end{pmatrix} + \partial_x \begin{pmatrix}
        v_1 & u\\
        u & 3v_2
    \end{pmatrix}\begin{pmatrix}
        \overline{\rho}_1 \\
        \overline{\rho}_2 
    \end{pmatrix} = -j(x,t) \sigma_z \vec{n}_2.
\end{align}
Here, the $K$-matrix is given in Eq.~\eqref{appeq:K23}, $\vec{n}_2=(n_{2,1},n_{2,2})^T$, and $j(x,t)$ is given in Eq.~\eqref{appeq:source_j}. 
The $\sigma_z$ matrix on the right-hand-side follows from $\eta_1=-\eta_2=+1$ for the $\nu=2/3$ edge. 

By using the matrix $\Lambda$ in Eq.~\eqref{appeq:Lambda_23} and Fourier-transforming to frequency space, we decouple the equations:
\begin{align}
\label{eq:diag_eom_23_GF_SM}
    &-i\omega  \sigma_z\begin{pmatrix}
        \overline{\rho}_+ \\
        \overline{\rho}_- 
    \end{pmatrix} + \partial_x \begin{pmatrix}
         v_+ && 0\\
        0 && v_-
    \end{pmatrix} \begin{pmatrix}
        \overline{\rho}_+ \\
        \overline{\rho}_- 
    \end{pmatrix}  =-j(x,\omega)\begin{pmatrix}
       n_{2,+} \\ n_{2,-}
    \end{pmatrix},
\end{align}
where $j(x,\omega)$ is given in Eq.~\eqref{appeq:j_x_omega} 
and the rotated source components 
$n_{2,+}$ and $n_{2,-}$ are now given by
\begin{align}
    \begin{pmatrix}
        n_{2,+} \\ n_{2,-}
    \end{pmatrix} = \Lambda^T\sigma_z \vec{n}_2 = \begin{pmatrix}
       \cosh \gamma\, n_{2,1}+\frac{\sinh \gamma}{\sqrt{3}} n_{2,2} \\[0.2cm] -\sinh\gamma \,n_{2,1}- \frac{\cosh \gamma}{\sqrt{3}} n_{2,2}
    \end{pmatrix}.
\end{align} 
Comparing  Eq.~\eqref{eq:diag_eom_23_GF_SM} with the corresponding equation for the FCS, Eq.~\eqref{appeq:eoms_23_diag_SM}, we see that the substitution 
\begin{align} \label{appeq:FCStoGFs_43}
    \frac{\lambda }{2\pi} q_{\pm} \mapsto n_{2,\pm}\, 
\end{align}
allows us to directly apply the FCS results of Sec.~\ref{sec:FCS_counter_SM} 
to obtain the GFs. In this way, we obtain the following scattering phases (cf. the counting phases~\eqref{eq:phases_23_FCS_kid_SM} for $x_0=0$) entering the GFs,
\begin{subequations}
\label{eq:phases_23}
\begin{align}
    \delta_{1,\tau}(t) = \sum_{s=\pm}\sum_{m=0}^\infty \delta_{1,s,m}w_{\tau}(t,-t_{1,s,m}), \\
    \delta_{2,\tau}(t) = \sum_{s=\pm}\sum_{m=0}^\infty \delta_{2,s,m}w_{\tau}(t,-t_{2,s,m}).
\end{align}
\end{subequations}
Here, the window function $w_\tau(t_1,t_2)$ is given by Eq.~\eqref{appeq:window_function}, the time offsets $t_{1,s,m}$ and $t_{2,s,m}$ by Eq.~\eqref{eq:23_time_offsets}, and the phases $\delta_{1,s,m}$ and $\delta_{2,s,m}$ read
\begin{subequations}
\label{eq:gen_phases_23_SM}
\begin{align}
    &\delta_{1,+,m} =+2\pi n_{1,1} n_{2,+} t_Lr_L^mr^m_R,  \\ 
    &\delta_{1,-,m} = +2\pi n_{1,1}  n_{2,-}t_Lr_L^mr^{m+1}_R,\\
\label{eq:gen_phases_23_2_even_SM}
    &\delta_{2,-,m} =-\frac{2\pi n_{1,2}n_{2,-} t_R r_L^mr^m_R}{\sqrt{3}},\\\label{eq:gen_phases_23_2_odd_SM}
    &\delta_{2,+,m} = -\frac{2\pi n_{1,2} n_{2,+} t_R r_L^{m+1}r^{m}_R}{\sqrt{3}}.
\end{align}
\end{subequations}

Similarly to the GFs for the $\nu=4/3$ edge (see Sec.~\ref{sec:GFs_co} of the main text), we consider the GFs for the $\nu=2/3$ edge in the short- and long-length limits. In the short-length limit, $L \ll \bar{v} \tau$, the GFs take the form, 
\begin{align}
\label{appeq:G_43_short_length}
    \mathcal{G}^{\gtrless}(\tau)  &=\frac{\mp i}{2\pi a} \left(\frac{a}{ a \pm i v_{1} \tau}\right)^{n_{2,1}^2} \left(\frac{a}{ a \pm i v_{2} \tau}\right)^{n_{2,2}^2/3}\times  \left\{\overline{\Delta}[2\pi n_{1,1}n_{2,1}w_\tau(t,0)] \right\}^{1/n_{1,1}^2}  \times \left\{\overline{\Delta}[2\pi n_{1,2}n_{2,2}w_\tau(t,0)/3]  \right\}^{3/n_{1,2}^2}\,.
\end{align}
For the $\nu=2/3$, these GFs essentially probe the properties of the decoupled edge modes $\nu_1=1$ and $\nu_2=-1/3$. 

In the long-length limit, $L \gg \bar{v} \tau$, the GFs instead evaluate to
\begin{align}
\label{eq:G23_final_SM}
    &\mathcal{G}^{\gtrless}(\tau)  \simeq \frac{\mp i}{2\pi a} \left(\frac{a}{ a \pm i v_{+} \tau}\right)^{\alpha} \left(\frac{a}{ a \pm i v_{-} \tau}\right)^{\beta} \times  \prod_{s=\pm}\prod_{m=0}^\infty\left\{\overline{\Delta}[\delta_{1,s,m}w_\tau(t,0)]  \right\}^{1/n_{1,1}^2} \left\{\overline{\Delta}[\delta_{2,s,m}w_\tau(t,0)] \right\}^{3/n_{1,2}^2}\,,
\end{align}
where the phases $\delta_{1,s,m}$ and $\delta_{2,s,m}$ are given by Eq.~\eqref{eq:gen_phases_23_SM} and the two exponents are
\begin{subequations}
\label{appeq:alpha_beta_23_SM}
\begin{align}
    \alpha&=\frac{1}{n_{1,1}^2}\sum_{s=\pm}\sum_{m=0}^\infty \left(\frac{\delta_{1,s,m}}{2\pi}\right)^2=\frac{t_L^2}{1-r_L^2r_R^2}\left[n_{2,+}^2+r_R^2 n_{2,-}^2\right],\\
    \beta&=\frac{3}{n_{1,2}^2}\sum_{s=\pm}\sum_{m=0}^\infty \left(\frac{\delta_{2,s,m}}{2\pi}\right)^2=\frac{t_R^2}{1-r_L^2r_R^2}\left[n_{2,-}^2+r_L^2 n_{2,+}^2\right].
\end{align}
\end{subequations}
These exponents satisfy
\begin{align}
\label{eq:23sumrule_SM}
&\alpha+\beta = n_{2,+}^2+n_{2,-}^2 = (n_{2,1}^2+\frac{n_{2,2}^2}{3})\cosh(2\gamma)+\frac{2n_{2,1} n_{2,2}\sinh(2\gamma)}{\sqrt{3}}=\zeta,
\end{align}
thus adding up to the  scaling dimension $\zeta$ of the excitation~\eqref{appeq:vertex_43} on the $\nu=2/3$ edge. Note that this scaling dimension is interaction-dependent in view of the counterpropagating character of the edge modes~\cite{Wen_Topological_1995}. For vanishing interactions,  $\gamma=0$, we have that $\zeta=n_{2,1}^2+n_{2,2}^2/3$, as expected from the vertex operator~\eqref{appeq:vertex_43} and the $K$-matrix~\eqref{appeq:K23}.

We first consider the GFs~\eqref{eq:G23_final_SM} in the partial equilibrium situation, where the two distribution functions entering Eq.~\eqref{eq:G23_final_SM} take the equilibrium form~\eqref{eq:equilibriumdist} with $V=0$ at two separate temperatures $T_1$ and $T_{2}$. In this case, the action is Gaussian and the result can be obtained in a much simpler way, so that this serves to benchmark our general formalism and result. Using the results of Sec.~\ref{app:determinant_calcs}, we find that the GFs in Eq.~\eqref{eq:G23_final_SM} take the form 
\begin{align}
\label{appeq:43GF_partial_eq}
    \mathcal{G}^{\gtrless}(\tau)  =\frac{\mp i}{2\pi a} &\left(\frac{a}{ a \pm i v_{+} \tau}\right)^{\alpha} \left(\frac{\pi T_1 \tau}{\sinh(\pi T_1 \tau)}\right)^{\alpha} \times  \left(\frac{a}{ a \pm i v_{-} \tau}\right)^{\beta} \left(\frac{\pi T_{2} \tau}{\sinh(\pi T_{2} \tau)}\right)^{\beta}\,,
\end{align}
with exponents $\alpha$ and $\beta$ from Eq.~\eqref{appeq:alpha_beta_23_SM}. In the global equilibrium case, $T_1=T_2=T$, and using the relation between the exponents, $\alpha+\beta=\zeta$, we see that the temperature-dependent factor takes the usual equilibrium form as it should. 

We turn now to the case of our main interest, with the injected distribution functions in regions I and III being of the non-equilibrium, double-step form~\eqref{appeq:doublestep_43}.  The GFs evaluate then within the Szeg\H{o} approximation to
\begin{align} \label{eq:GFs_doubles_tep_23}
    &\mathcal{G}^{\gtrless}(\tau)  \simeq \frac{\mp i}{2\pi a} \left(\frac{a}{ a \pm i v_{+} \tau}\right)^{\alpha} \left(\frac{a}{ a \pm i v_{-} \tau}\right)^{\beta} \times \exp\Big[-
 \frac{\mathcal{T}_1|e V_1 \tau|}{2\pi n_{1,1}}\sum_{m=0}^\infty(1-e^{-i \delta_{1,m} \text{sgn}(eV_1\tau)})\Big] \nonumber \\ & \times \exp\Big[-
 \frac{\mathcal{T}_2|e V_2 \tau|}{2\pi n_{1,2}}\sum_{m=0}^\infty(1-e^{-i \delta_{2,m} \text{sgn}(eV_2\tau)})\Big]\,, 
\end{align}
where we used the condition of dilute injection, $\mathcal{T}_1,\mathcal{T}_2\ll 1$. 

\subsection{Transport properties}
\label{sec:transport_counter_SM}

With the non-equilibrium GFs calculated, we analyze here transport observables for tunneling between two $\nu=2/3$ edges, see Fig.~\ref{fig:QPC} of the main text. While both edges, labeled $u$ and $d$, can be driven out of equilibrium in various ways, we focus here on a setup analogous to that for $\nu=4/3$ in Sec.~\ref{sec:transport_co} in the main text: A non-equilibrium state is injected in the $\nu_2 = -1/3$ mode on the $u$ edge only, with the state produced by a ``minimal'' quasiparticle injection, 
$n_{1,2} = 1$, i.e., $e_{1,2}^*)=e/3$, and with $V_{2,u}=V$ and $\mathcal{T}_{2,u}=\mathcal{T}$. We further assume that the same type of minimal excitations tunnel at the QPC, corresponding to $\vec{n}_2=(n_{2,1}, n_{2,2}) = (0,1)$.  With these choices, we find that Fano factor 
is given by
\begin{align}
\label{appeq:Fano_43}
    F= \frac{e_2^*}{|e|} \cot(\pi \zeta) \cot[(1-2\zeta)\tan^{-1}(p)]\,, 
\end{align}
with the scaling dimension \eqref{eq:23sumrule} of the main text, and a modified dimensionless bias parameter $p$. Specifically, for sharp interfaces, 
the parameter $p$ evaluates to
\begin{align} \label{eq:dimensionlessparameter_23_SM}
    p = \frac{\text{sgn}(eV) \sum_{s=\pm}\sum_{m=0}^\infty\sin(\delta_{2,s,m})}{\sum_{s=\pm}\sum_{m=0}^\infty\left[1-\cos(\delta_{2,s,m})\right]}\,,
\end{align}
with the interaction-dependent phases $\delta_{2,\pm,m}$ given by Eq.~\eqref{eq:gen_phases_23_2_even_SM}-\eqref{eq:gen_phases_23_2_odd_SM}. We further note that the phases $\delta_{2,-,m+1} = 2\pi [\tanh(\gamma)^{2(m+1)}]/3$ and $\delta_{2,+,m} = -2\pi [\tanh(\gamma)]^{2(m+1)}/3$ for $m\geq0$ have equal magnitudes, but opposite signs. Consequently, their contributions to the numerator of
$p$ in Eq.~\eqref{eq:dimensionlessparameter_23_SM} cancel in a pairwise fashion. The only remaining contribution is that of the unpaired phase $\delta_{2,-,0} = 2\pi /3$, which leads to the following, simplified expression for $p$,
\begin{align} \label{eq:pcancellation}
   p =  \frac{\sqrt{3}\,\text{sgn}(eV)}{3+4\sum_{m=1}^{\infty} \sin^2[\pi \tanh(\gamma)^{2m}/3]} \,. 
\end{align}
Finally, for adiabatic interfaces, we obtain the bias parameter as
\begin{align}\label{eq:dimensionlessparameter_23_adiabatic_SM}
   p &=\text{sgn}(eV) \frac{\sin [2\pi |\nu_2| \cosh \gamma]+2\pi |\nu_2| (1-\cosh \gamma)}{1-\cos [2\pi \nu_2 \cosh\gamma]}\,.
\end{align}
We note that, in both Eq.~\eqref{eq:pcancellation} and~\eqref{eq:dimensionlessparameter_23_adiabatic_SM}, $p$ is an even function of the interaction parameter $\gamma$. 

\end{document}